\documentclass[traditabstract,a4]{aa}
\usepackage{graphicx,amsmath}
\usepackage{epsf}
\usepackage{txfonts}
\usepackage[hyphens]{url}
\usepackage{longtable}
\usepackage{lscape}
\usepackage[hyperindex,breaklinks=true, colorlinks, citecolor=blue]{hyperref}  
\usepackage{breakurl}
\usepackage{comment}
\usepackage{natbib}
\usepackage{upgreek}
\usepackage{float}
\usepackage[switch,pagewise]{lineno}
\usepackage{multirow}
\usepackage{caption}
\usepackage{lipsum}
\usepackage{threeparttable}
\usepackage{multirow}

\newcommand{\hLEt}[1]{}
\newcommand{\juan}[1]{{\bf #1}}
\renewcommand{\juan}[1]{#1}

\newcommand{\commentproof}[1]{{\bf \color{green}#1}}
\renewcommand{\commentproof}[1]{} 

\def\setsymbol#1#2{\expandafter\def\csname #1\endcsname{#2}}
\def\getsymbol#1{\csname #1\endcsname}

\def\Planck{\textit{Planck}}

\def\all2013resultspapers{\nocite{planck2013-p01, planck2013-p02, planck2013-p02a, planck2013-p02d, planck2013-p02b, planck2013-p03, planck2013-p03c, planck2013-p03f, planck2013-p03d, planck2013-p03e, planck2013-p01a, planck2013-p06, planck2013-p03a, planck2013-pip88, planck2013-p08, planck2013-p11, planck2013-p12, planck2013-p13, planck2013-p14, planck2013-p15, planck2013-p05b, planck2013-p17, planck2013-p09, planck2013-p09a, planck2013-p20, planck2013-p19, planck2013-pipaberration, planck2013-p05, planck2013-p05a, planck2013-pip56, planck2013-p06b, planck2013-p01a}}

\newbox\tablebox    \newdimen\tablewidth
\def\leaderfil{\leaders\hbox to 5pt{\hss.\hss}\hfil}
\def\tablenote#1 #2\par{\begingroup \parindent=0.8em
    \abovedisplayshortskip=0pt\belowdisplayshortskip=0pt
    \noindent
    $$\hss\vbox{\hsize\tablewidth \hangindent=\parindent \hangafter=1 \noindent
    \hbox to \parindent{$^#1$\hss}\strut#2\strut\par}\hss$$
    \endgroup}

\def\L2{\ifmmode L_2\else $L_2$\fi}

\def\DeltaT{\ifmmode \Delta T\else $\Delta T$\fi}
\def\deltat{\ifmmode \Delta t\else $\Delta t$\fi}
\def\fknee{\ifmmode f_{\rm knee}\else $f_{\rm knee}$\fi}
\def\Fmax{\ifmmode F_{\rm max}\else $F_{\rm max}$\fi}
\def\solar{\ifmmode{\rm M}_{\mathord\odot}\else${\rm M}_{\mathord\odot}$\fi}
\def\Msolar{\ifmmode{\rm M}_{\mathord\odot}\else${\rm M}_{\mathord\odot}$\fi}
\def\Lsolar{\ifmmode{\rm L}_{\mathord\odot}\else${\rm L}_{\mathord\odot}$\fi}
\def\inv{\ifmmode^{-1}\else$^{-1}$\fi}
\def\mo{\ifmmode^{-1}\else$^{-1}$\fi}
\def\sup#1{\ifmmode ^{\rm #1}\else $^{\rm #1}$\fi}
\def\expo#1{\ifmmode \times 10^{#1}\else $\times 10^{#1}$\fi}
\def\,{\thinspace}
\def\lsim{\mathrel{\raise .4ex\hbox{\rlap{$<$}\lower 1.2ex\hbox{$\sim$}}}}
\def\gsim{\mathrel{\raise .4ex\hbox{\rlap{$>$}\lower 1.2ex\hbox{$\sim$}}}}

\def\simprop{\mathrel{\raise .4ex\hbox{\rlap{$\propto$}\lower 1.2ex\hbox{$\sim$}}}}
\def\deg{\ifmmode^\circ\else$^\circ$\fi}
\def\pdeg{\ifmmode $\setbox0=\hbox{$^{\circ}$}\rlap{\hskip.11\wd0 .}$^{\circ}
          \else \setbox0=\hbox{$^{\circ}$}\rlap{\hskip.11\wd0 .}$^{\circ}$\fi}
\def\arcs{\ifmmode {^{\scriptstyle\prime\prime}}
          \else $^{\scriptstyle\prime\prime}$\fi}
\def\arcm{\ifmmode {^{\scriptstyle\prime}}
          \else $^{\scriptstyle\prime}$\fi}
\newdimen\sa  \newdimen\sb
\def\parcs{\sa=.07em \sb=.03em
     \ifmmode \hbox{\rlap{.}}^{\scriptstyle\prime\kern -\sb\prime}\hbox{\kern -\sa}
     \else \rlap{.}$^{\scriptstyle\prime\kern -\sb\prime}$\kern -\sa\fi}
\def\parcm{\sa=.08em \sb=.03em
     \ifmmode \hbox{\rlap{.}\kern\sa}^{\scriptstyle\prime}\hbox{\kern-\sb}
     \else \rlap{.}\kern\sa$^{\scriptstyle\prime}$\kern-\sb\fi}
\def\ra[#1 #2 #3.#4]{#1\sup{h}#2\sup{m}#3\sup{s}\llap.#4}
\def\dec[#1 #2 #3.#4]{#1\deg#2\arcm#3\arcs\llap.#4}
\def\deco[#1 #2 #3]{#1\deg#2\arcm#3\arcs}
\def\rra[#1 #2]{#1\sup{h}#2\sup{m}}

\def\dots{\relax\ifmmode \ldots\else $\ldots$\fi}
\def\WHzsr{\ifmmode $W\,Hz\mo\,sr\mo$\else W\,Hz\mo\,sr\mo\fi}
\def\mHz{\ifmmode $\,mHz$\else \,mHz\fi}
\def\GHz{\ifmmode $\,GHz$\else \,GHz\fi}
\def\mKs{\ifmmode $\,mK\,s$^{1/2}\else \,mK\,s$^{1/2}$\fi}
\def\muKs{\ifmmode \,\mu$K\,s$^{1/2}\else \,$\mu$K\,s$^{1/2}$\fi}
\def\muKRJs{\ifmmode \,\mu$K$_{\rm RJ}$\,s$^{1/2}\else \,$\mu$K$_{\rm RJ}$\,s$^{1/2}$\fi}
\def\muKHz{\ifmmode \,\mu$K\,Hz$^{-1/2}\else \,$\mu$K\,Hz$^{-1/2}$\fi}
\def\MJysr{\ifmmode \,$MJy\,sr\mo$\else \,MJy\,sr\mo\fi}
\def\MJysrmK{\ifmmode \,$MJy\,sr\mo$\,mK$_{\rm CMB}\mo\else \,MJy\,sr\mo\,mK$_{\rm CMB}\mo$\fi}
\def\microns{\ifmmode \,\mu$m$\else \,$\mu$m\fi}
\def\micron{\microns}
\def\muK{\ifmmode \,\mu$K$\else \,$\mu$\hbox{K}\fi}
\def\microK{\ifmmode \,\mu$K$\else \,$\mu$\hbox{K}\fi}
\def\muW{\ifmmode \,\mu$W$\else \,$\mu$\hbox{W}\fi}
\def\kms{\ifmmode $\,km\,s$^{-1}\else \,km\,s$^{-1}$\fi}
\def\kmsMpc{\ifmmode $\,\kms\,Mpc\mo$\else \,\kms\,Mpc\mo\fi}
\providecommand{\sorthelp}[1]{}

\def\Herschel{\textit{Herschel}}
\newcommand{\nh}{$N_{\textsc{H}}$}

\newcommand{\lognh}{$\log_{10}(N_{\textsc{H}}/\mbox{cm}^{-2})$}

\newcommand{\bperp}{$B_{\perp}$}
\newcommand{\bpara}{$B_{||}$}
\newcommand{\bvec}{$\vec{B}$} 
\newcommand{\meanbvec}{$\left<\vec{B}\right>$}

\newcommand{\healpix}{{\sc HEALPix}}

\providecommand{\sorthelp}[1]{}

\newcommand{\PlanckXXXV}{{PlanckXXXV}}
\newcommand{\meanphi}{{$\left<\phi\right>$}}

\begin{document}


\title{Using \Herschel\ and \Planck\ observations to delineate the role of magnetic fields in molecular cloud structure}
\titlerunning{Using \Herschel\ and \Planck\ to delineate the role of magnetic fields in molecular cloud structure}
\author{
Juan~D.~Soler$^{1}$
} 
\institute{
1. Max Planck Institute for Astronomy, K\"{o}nigstuhl 17, 69117, Heidelberg, Germany. {\tt soler@mpia.de}
}

\authorrunning{J.\,D.~Soler}

\date{Received 26 APR 2019 / Accepted 07 AUG 2019}

\abstract{
We present a study of the relative orientation between the magnetic field projected onto the plane of sky (\bperp) on scales down to 0.4\,pc, inferred from the polarized thermal emission of Galactic dust observed by \Planck\ at 353\,GHz, and the distribution of gas column density (\nh) structures on scales down to 0.026\,pc, derived from the observations by \Herschel\ in submillimeter wavelengths, toward ten nearby ($d\,<$\,450\,pc) molecular clouds.
\juan{
Using the histogram of relative orientation technique in combination with tools from circular statistics, we found that the mean relative orientation between \nh\ and \bperp\ toward these regions increases progressively from 0\deg, where the \nh\ structures lie mostly parallel to \bperp, with increasing \nh, in many cases reaching 90\deg, where the \nh\ structures lie mostly perpendicular to \bperp.}
We also compared the relative orientation between \nh\ and \bperp\ and the distribution of \nh, which is characterized by the slope of the tail of the \nh\ probability density functions (PDFs).
We found that the slopes of the \nh\ PDF tail are steepest in regions where \nh\ and \bperp\ are close to perpendicular.
This coupling between the \nh\ distribution and the magnetic field suggests that the magnetic fields play a significant role in structuring the interstellar medium in and around molecular clouds.
However, we found no evident correlation between the star formation rates, estimated from the counts of young stellar objects, and the relative orientation between \nh\ and \bperp\ in these regions.
}

\maketitle

\section{Introduction}\label{section:introduction}

The physical processes that regulate the formation of stars are one of the main open \juan{subjects of research} in contemporary astrophysics \citep[for a review see][]{mckee2007,dobbs2014,molinari2014}.
Interstellar magnetic fields are a long-standing candidate for the control of the star formation \juan{process} \citep[see, e.g.,][]{mouschovias2006}.
\juan{Observations indicate that in flux-freezing conditions, the interstellar magnetic fields, gas pressure, cosmic-ray pressure, and gravity are the most important forces on the diffuse gas} \citep{heilesANDcrutcher2005}.
It is likely that magnetic fields are responsible for reducing the star formation rate in dense molecular clouds (MCs) by a factor of a few by strongly shaping the interstellar gas \citep[][and references therein]{hennebelleANDinutsuka2019}.
However, observing the magnetic fields in and around MCs and determining their exact influence is a difficult task.

Observations of the Zeeman splitting in emission and absorption by species such as neutral hydrogen (H) and the hydroxyl (OH) and cyanide (CN) radicals provide the only direct detections of the magnetic field strength in the stellar medium \citep[see][for a review]{crutcher2012}.
The requirements of both high sensitivity and control of systematic effects considerably limit the coverage of the observations of the Zeeman splitting in MCs, however \citep{bourke2001,troland2008}.
Further information on the magnetic fields in MCs is provided by observations of the linear polarization, in extinction from background stars and in emission from interstellar dust grains, reveal the orientation of the interstellar magnetic field averaged along the line of sight (LOS) and projected onto the plane of the sky, \bperp, \citep{hiltner1949,davis1951,hildebrand1988,pattle2019}.

The magnetic field strength can be assessed indirectly from the linear polarization observations using the Davis-Chandrasekhar-Fermi method \citep[DCF,][]{davis1951a,chandrasekhar1953}, which combines estimations of the density, velocity dispersion, and polarization angle dispersion toward the studied regions \citep[see, e.g.,][]{houde2009,pattle2017a}.
An alternative to directly estimating the magnetic field strength is to statistically study the magnetic field orientation and its correlation with the observed column density, \nh, distribution.
Recent studies of numerical simulations of magnetohydrodynamic (MHD) turbulence indicate that there is a relation between the statistical trends of relative orientation between \nh\ and \bperp\ and the initial magnetization of the MCs \citep{soler2013,chen2016,solerANDhennebelle2017}.
The observational study of this trend toward ten nearby MCs reveals based on the \Planck\ observations that on spatial scales ranging from tens of parsecs to 0.2\,pc, turbulence is trans- or sub-Alfv\'{e}nic, that is, the magnetic fields are at least in equipartion with turbulence \citep[][from here on \PlanckXXXV]{planck2015-XXXV}.

In this work, we extend the relative orientation study presented in \PlanckXXXV\ to the \nh\ structures observed by the \Herschel\ satellite at higher resolution than that obtained by 
\Planck\ \citep{pitbratt2010}.  
Because of the difference in angular resolution between the linear polarization observations that are used to infer \bperp\ (10\arcmin\ FWHM at 353\,GHz) and the submillimeter observations that are used to determine \nh\ (36\parcs0 FWHM), the scientific question we address is what is the relation between the \nh\ structures and the larger-scale magnetic field.
This approach is not a substitute for the observation of polarized emission by dust and magnetic fields at higher angular resolutions, but rather uses the existing observations to explore the coupling between matter and magnetic fields across scales. 
This is central for our understanding of the formation of density structures in a magnetized medium


\juan{
We here determine whether the relative orientation trends reported in \PlanckXXXV\ are also present when we compare the cloud-scale \bperp\ observed by \Planck\ and the \nh\ structures observed by \Herschel.
We also investigate whether the distribution of \nh\ is related to the relative orientation between \nh\ and \bperp. 
Finally, we study whether the relative orientation between the \Planck\ \bperp and the \Herschel\ \nh\ structures is related to the star formation rates in the studied regions.
These three questions can only be addressed with the higher resolution \Herschel\ observations, which allow us to separate different regions within the MCs that were studied in \PlanckXXXV\ and estimate their internal \nh\ distribution. 
}

The first question addresses the anchoring of density structures by Galactic magnetic fields, which has been suggested as one of the observational proofs of the importance of the magnetic fields in MC formation \citep{li2014}. 
Recent studies of MHD simulations indicate that a strong magnetization preserves not only the field direction, but also the orientation of density structures across scales by creating a strong anisotropy in the flows that form and structure MCs \citep[see, e.g.,][]{hull2017,gomez2018,mocz2018}.
Recent observational studies of the relative orientation between the structures that were traced by nine molecular rotational emission lines and \bperp\ toward the Vela C region ($d$\,$\approx$\,$700$\,pc) indicate a connection between the structure of dense gas on small scales and the cloud-scale magnetic field \citep{fissel2019}.
We here therefore aim to evaluate whether the orientation of the cloud-scale magnetic field is related to the orientation of the density structures at smaller scales in nearby MCs.

The second question focuses on the probability distribution function (PDF) of \nh, which is a frequently used tool for describing the structure of MCs \citep[see, e.g.,][]{lombardi2008,kainulainen2009,goodman2009,schneider2015}. 
Theoretical studies indicate that the \nh\ PDFs of MCs are characterized by a log-normal peak and a power-law tail toward high \nh\, . This is interpreted as the effect of turbulent motions and gravity, respectively \citep[see, e.g.,][]{vasquez-semadeni1994,ballesteros-paredes2011,burkhart2018}.
Recent studies based on MHD simulations indicate that magnetic fields can affect \nh\ PDFs. Specifically, MCs with subcritical mass-to-flux ratios show significantly steeper power tails than supercritical MCs \citep{auddy2018}.
A proof of concept of the study of the relative orientation of \nh\ and \bperp\ and its relation with the \nh\ PDFs in observations can be found in \cite{soler2017}, where the authors considered the Serpens Main 2 region in the Aquila Rift.

Finally, the third question concentrates on the relation between the relative orientation of \nh\ and \bperp\  and the star formation rates (SFRs). This relation has been proclaimed an observational test of the role of magnetic fields as a primary regulator of star formation \citep{li2017}.
We use the reported SFR values evaluated from counts of young stellar objects (YSOs) \citep{lada2010,evans2014} and directly compare them to the \nh\ and \bperp\ relative orientation trends toward the studied regions.

Throughout this paper we assume that \bperp\ is well represented by the emission from the magnetically aligned dust grains \citep[for a review, see][]{andersson2015}.
This assumption is justified both by most recen observational evidence \citep{planck2014-XX} and by synthetic observations \citep{seifried2019} that indicate that the dust grains remain well aligned even at the highest column densities that are relevant for this study ($n$\,$>$\,10$^3$\,cm$^{-3}$). 

This paper is organized as follows.
We introduce the \Planck\ and \Herschel\ observations in Sect.~\ref{section:data}.
Section~\ref{section:method} describes how we implemented the method called the histogram of relative orientations (HRO), which we used for the systematic analysis of the relative orientation between \nh\ and \bperp.
We present the results of HRO analysis and study the relation between the relative orientations of \nh-\bperp\ and the shape of the \nh\ PDFs and SFRs in Sec.~\ref{section:results}
We discuss the physical implications of our main results in Sec.~\ref{section:discussion}
Finally, Sect.~\ref{section:conclusions} presents our conclusions and future prospects.
We reserve the technical details and additional information for a set of appendices.
Appendix~\ref{app:relativeorientation} presents complementary information on the HRO analysis.
Appendix~\ref{app:striationsTest} presents the HRO analysis of the \Herschel\ 250\micron\
maps toward the Musca and the IC5146 regions.
\juan{
Appendix~\ref{app:nhpdf} presents some additional comparisons of the relative orientation of \nh\ and \bperp\ and its relation with the \nh\ PDFs.
}
All the routines used in this paper, as well as other tools for the analysis of observations and simulations, are publicly available at \url{http://github.com/solerjuan/magnetar}.

\section{Observations}\label{section:data}

We present the study of ten nearby (d\,$<$\,450\,pc) MCs that have previously been studied in \PlanckXXXV.
These regions were partly covered in observations by ESA's \Herschel\ satellite.
We study the subregions using the data products described below.
For the sake of consistency, we use the same distances to the regions indicated in table\,1 of \PlanckXXXV\ and use Galactic coordinates in the analysis.
These selections do not affect our results.

\subsection{Polarization}

\Planck\ observed linearly polarized emission (Stokes $Q$ and $U$) in seven frequency bands from 30 to 353\,GHz over the whole sky.
In this study, we used the publicly available PR3 data from the High Frequency Instrument \citep[HFI,][]{lamarre2010} at 353\,GHz.
Toward MCs, the contribution of the polarized emission from the cosmic microwave background (CMB) is negligible in this frequency band. 
This \Planck\ map is therefore best suited for studying the polarized emission from interstellar dust grains \citep{planck2014-XIX}.

The maps of Stokes $Q$, $U$, their respective variances $\sigma_{QQ}$, $\sigma_{UU}$, and their covariance $\sigma_{QU}$ are initially at 4\parcm8 
resolution in \healpix\ format with a pixelization at $N_{\rm side}$\,$=$\,$2048$, which corresponds to an effective pixel size of 1\parcm72. 
For the sake of comparison with \PlanckXXXV\ and to increase the signal-to-noise ratio, we use the maps smoothed to a resolution of 10\arcmin\ FWHM.
The maps of \bperp\ inferred from these polarization observations are presented in the top panels of Figs.~\ref{fig:HOGpanel1}, \ref{fig:HOGpanel2}, \ref{fig:HOGpanel3}, and \ref{fig:HOGpanel4}.
Detailed maps of the \Herschel-observed regions are presented in Appendix~\ref{app:relativeorientation}. 

\subsection{Column density}

We used the publicly available column density maps presented in \cite{abreu-vicente2017}, which are based on the combination of the \Herschel\ data and \Planck\ observations. 
The \Herschel\ data correspond to the publicly available observations obtained with the Spectral and Photometric Imaging Receiver (SPIRE) in three bands centered on 250, 350, and 500\micron\ 
(17\parcs6, 23\parcs9, and 35\parcs2 FWHM resolution, respectively) and with the Photodetector Array Camera and Spectrometer  (PACS)\ in a band centered on 160\micron\ (13\parcs6 FWHM resolution).
The \Planck\ data in the corresponding bands were derived from the \Planck\ all-sky foreground dust emission model \citep{planck2013-p06b}.
The \Herschel\ data and \Planck\ observations in each band were cross-calibrated and combined in Fourier space using the feathering method \citep{csengeri2016}.
This method produces maps that improve the previous constant-offset corrections to the \Herschel\ maps by accounting for variations in the background emission levels.
The feathered data are convolved to the common 36\parcs0 FWHM resolution and are fit using a modified blackbody spectrum to infer the optical depth that is used as a proxy for \nh.
Further details are described in  \cite{abreu-vicente2017}.

The \Herschel\ observations toward the MCs studied in \PlanckXXXV\ cover objects in the Lynds Catalog of Dark Nebulae \citep[LDN,][]{lynds1962} or simply split the MCs into smaller arbitrary portions that facilitated observing them, for example, North and South in Perseus.
We separately analyzed each of the \Herschel-observed segments and combined them only in the regions where these segments correspond to a single well-known region, such as CrA, Perseus, Orion A, and Orion B.

\section{Histogram of relative orientations }\label{section:method}

We quantified the relative orientation between the column density structures and \bperp\ using the histogram of relative orientations \citep[HRO,][]{soler2013} technique. 
In the HRO, the \nh\ structures are characterized by their gradients, $\nabla N_{\rm H}$, which are by definition perpendicular to the iso-\nh\ curves. 
The gradient constitutes a vector field that we compared pixel by pixel to the \bperp\ orientation inferred from the \Planck\ 353-GHz polarization maps.

The angle $\phi$ between \bperp\ and the tangent to the \nh\ contours was evaluated using
\begin{equation}\label{eq:phi}
\phi\,=\,\arctan\left(|\nabla N_{\rm H}\times\hat{e}_{353}|, \nabla N_{\rm H}\cdot\hat{e}_{353}\right)
,\end{equation}
where the pseudo-vector $\hat{e}_{353}$ corresponds to the orientation of the \Planck\ 353\,GHz linear polarization defined by
\begin{equation}\label{eq:psi}
\psi\,=\,\frac{1}{2}\arctan\left(-U,Q\right).
\end{equation}
In Eq.~\eqref{eq:phi} as implemented, the norm  carries a sign when the range used for $\phi$ is between $-90$\deg\ and $90$\deg.
We computed the gradients using the Gaussian derivatives introduced in \PlanckXXXV\ with a kernel size equal to five pixels, thus establishing that the orientation of the \nh\ structures
is characterized on a scale of approximately 1\parcm17, given the pixel size of the \nh\ maps $\Delta l$\,$=$\,14\arcsec.

In this implementation, we report the results of the HRO analysis in terms of the histogram-dependent relative orientation parameter, $\xi$, introduced in equation~(4) of \PlanckXXXV. 
We also bypass the histogram step, however, by directly applying the tools of circular statistics introduced in \cite{jow2018}, that is, the projected Rayleigh statistic, $V$, and the mean relative orientation angle, \meanphi\ \citep[see][for general references in circular statistics]{batschelet1981}.
For the sake of completeness, we present the HROs and the maps of the individual subregions in Appendix~\ref{app:relativeorientation}.

As in \PlanckXXXV, the HRO analysis was performed in \nh\ bins with equal numbers of pixels, which guarantees comparable statistics.
The number of \nh\ bins was determined by requiring enough bins to resolve the highest \nh\ regions and at the same time maintaining enough pixels per \nh\ bin to obtain significant statistics.
The measurement errors in the values of $\xi$, $V$, and \meanphi\ were estimated using Monte Carlo realizations of the noise, quantified by the covariances of the intensity maps in the \Herschel\ bands and the \Planck\ Stokes parameters at 353\,GHz, following the procedure described in Appendix B of \PlanckXXXV.
As in \PlanckXXXV, we minimized the effect of noise by smoothing the \Planck\ maps to a 10\arcmin\ FWHM resolution and applying a 1\parcm17 FWHM derivative kernel to estimate the gradients.
Thus, the reported error bars around the values of $\xi$, $V$, and \meanphi\ correspond to the standard deviations, which are much larger than the estimated effect of the measurement noise.

Using $\xi$, $V$, and \meanphi\ as metrics to characterize the distribution of relative orientation angles $\phi_{k}$ allows for a more complete interpretation of the HRO results than in \PlanckXXXV.
The values of $\xi$ provide a description of the shape of the HROs.
The values of $V$ constitute an optimal statistical test independent of the HRO binning, and \meanphi\ allows evaluating the prevalence of relative orientations different to 0 or 90\deg,
as we detail below. 

\subsection{Projected Rayleigh statistic $V$}

The projected Rayleigh statistic ($V$) is a test of the nonuniformity and unimodal clustering of the angle distribution \citep[see, e.g., chapter 6 in][]{mardia2009directional}.
For a set of $N$ relative orientation vectors $\phi_{k}$, it is defined as
\begin{equation}\label{eq:V}
V\,\equiv\,2\frac{[\sum^{N}_{k}\cos(2(\phi_{k}-\phi_{0}))]^{2}}{\sum^{N}_{k}w_{k}},
\end{equation}
where $w_{k}$ corresponds to the statistical weights assigned to each angle $\phi_{k}$.
We selected the reference angle $\phi_0$\,$=$\,0\deg\ such that $V$\,$>$\,0 corresponds to $\phi_{k}$ clustered around 0\deg, or   \text{ }\nh\ contours mostly parallel to \bperp.
In the same way, $V$\,$<$\,0 corresponds to $\phi_{k}$ clustered around 90\deg, or\text{ } \nh\ contours mostly perpendicular to \bperp.

The projected Rayleigh ($V$) is primarily a hypothesis test, which means that if $|V|$ is small, there is no evidence that \nh\ and \bperp\ are oriented toward the reference angles 0 or 90\deg. 
If $|V|$ is relatively large, however, there must be some concentration around the reference angles.
The larger the value of $V$, the better chance there is of rejecting the null hypothesis of randomness \citep{batschelet1981}.
A useful analogy for the interpretation of $V$ is the random walk problem.
If we set all the weights $w_{k}$\,$=$\,1, Eq.~\eqref{eq:V} is equivalent to the displacement from the origin in a random walk if  steps of unit length were taken in the direction of each angle $2\phi_{k}$.
If the distribution of angles is uniform, then the expectation value of $V$ is $0$.

We evaluated the uncertainty that a single $V$ measurement reflects the dispersion of the relative angles following the description presented in \cite{jow2018}.
Assuming that the angles $\phi_{i}$ are independently and uniformly distributed, it follows that every $\cos 2\phi_{k}$ in Eq.~\eqref{eq:s_V} is independently and identically distributed, with a mean of $1/2$.
By the central limit theorem, it follows that in the limit of $N$\,$\rightarrow$\,$\infty$, $V$ follows a normal distribution with $\mu$\,$=$0 and $\sigma^{2}$\,$=$\,$1/2$.
The asymptotic limit of the $V$ distribution is therefore the standard normal distribution.

For a general distribution of angles, the variance in $V$ in the high $N$ limit is the variance of each $\cos 2\phi_{k}/\sqrt{1/2}$, which can be estimated as 
\begin{equation}\label{eq:s_V}
\sigma_{V}\,\equiv\,\left[\frac{2\sum^{N}_{k}\cos(2(\phi_{k})^{2}-V^{2}}{\sum^{N}_{k}w_{k}}\right]^{1/2}.
\end{equation}
We consider this value to be the uncertainty in the values of $V$.
\cite{jow2018} used Monte Carlo simulations to show that for uniformly distributed samples of $\phi_{k}$ the expectation value of $V$ converges to $0$ and $\sigma_{V}$\,$\rightarrow$\,$1$.
Thus, measurements of $V$\,$\gg$\,$1$ indicate a significant detection of parallel relative orientation, while measurements of $V$\,$\ll$\,$-1$ indicate a significant detection of perpendicular relative orientation.
Given our selection of a high signal-to-noise ratio in the emission and polarization observations, $\sigma_{V}$ is dominated by the measurement of a single $V$ in a particular MC rather that by the uncertainties in the observations.

\subsection{Mean relative orientation angle \meanphi}

If there is a concentration of the $\phi_{k}$ values, as can be inferred from the values of $V$, the mean relative orientation angle is defined as
\begin{equation}\label{eq:meanphi}
\left<\phi\right>\,\equiv\,\arctan\left(y,x\right),
\end{equation}
where
\begin{equation}\label{eq:components}
x\,\equiv\,\frac{\sum^{N}_{k}w_{k}\cos\phi_{k}}{\sum^{N}_{k}w_{k}} \mbox{ and } y\,\equiv\,\frac{\sum^{N}_{k}w_{k}\sin\phi_{k}}{\sum^{N}_{k}w_{k}}
.\end{equation}
Its associated standard deviation is defined as
\begin{equation}\label{eq:s_phi}
\sigma_\phi\,\equiv\,\left[-2\log(r)\right]^{1/2},
\end{equation}
where the mean resultant length, $r$, which is itself a measure of concentration in the angle distribution, is defined as
\begin{equation}\label{eq:mrl}
r\,\equiv\,(x^{2}+y^{2})^{1/2}.
\end{equation}
If the $\phi_{k}$ values are tightly clustered, $r$ is close to 1. 
If the $\phi_{k}$ values are broadly distributed, $r$ is close to 0.
A preferential relative orientation is only well defined if $r$ is greater than 0 and $\sigma_\phi$\,$\rightarrow$\,$\infty$ as $r$\,$\rightarrow$\,0.

\subsection{Statistical weights}

In this implementation, we accounted for the oversampling of $\phi$ values introduced by the pixel size of the maps in relation to the size of the beam using the statistical weights
\begin{equation}\label{eq:weights}
w_{k}\,=\,\left(\frac{\Delta l}{{\rm FWHM}_{\rm Pla}}\right)^{2},
\end{equation}
where $\Delta l$\,$=$\,14\arcsec\ is the pixel size of the \nh\ maps and ${\rm FWHM}_{\rm Pla}$\,$=$\,10\arcmin\ corresponds to the \Planck\ beam.

\begin{figure*}[ht!]
\vspace{1.0cm}
\centerline{
\includegraphics[width=0.33\textwidth,angle=0,origin=c]{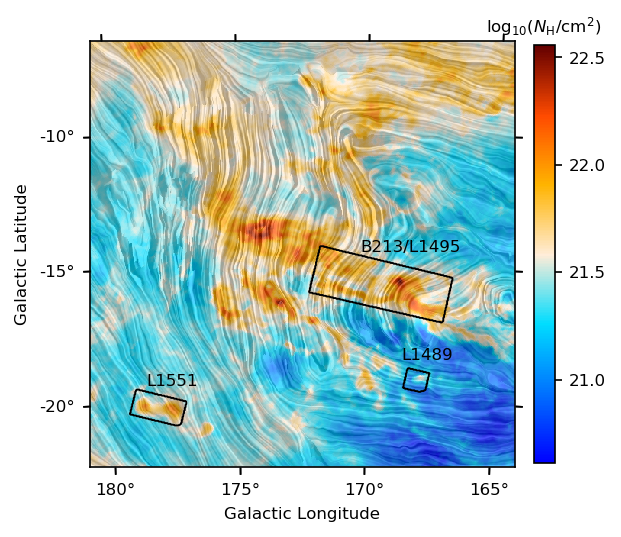}
\includegraphics[width=0.33\textwidth,angle=0,origin=c]{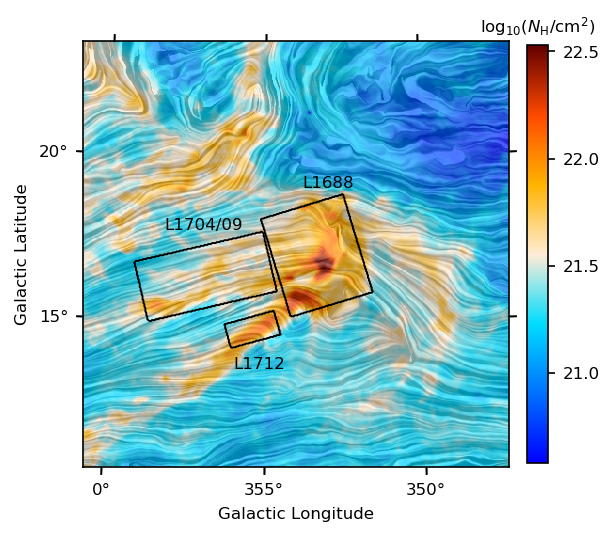}
\includegraphics[width=0.33\textwidth,angle=0,origin=c]{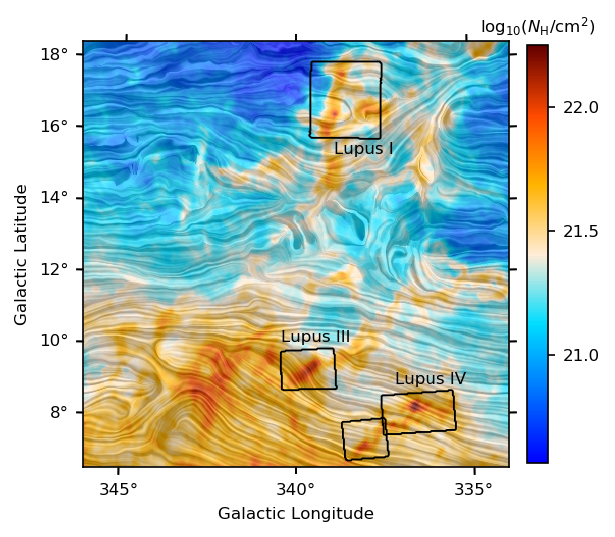}
}
\centerline{
\includegraphics[width=0.33\textwidth,angle=0,origin=c]{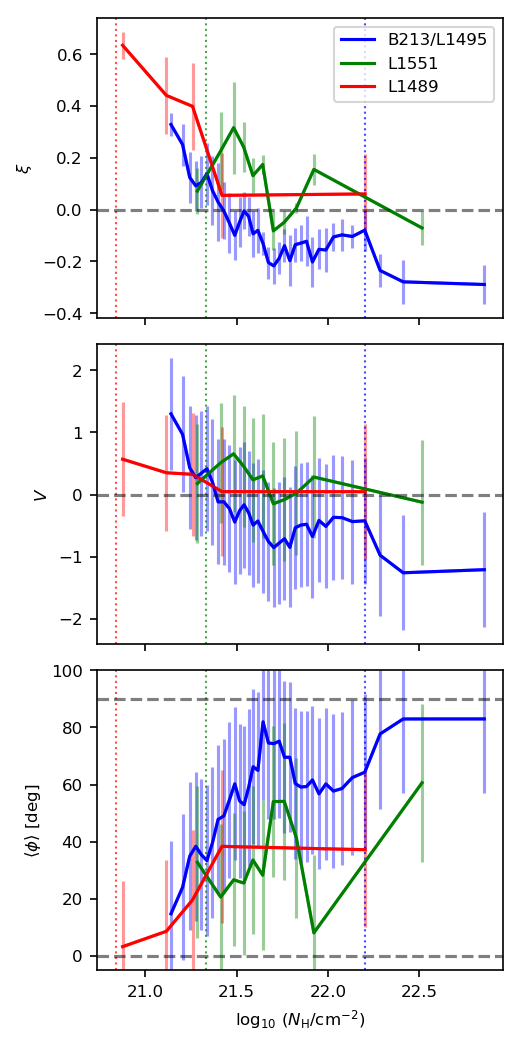}
\includegraphics[width=0.33\textwidth,angle=0,origin=c]{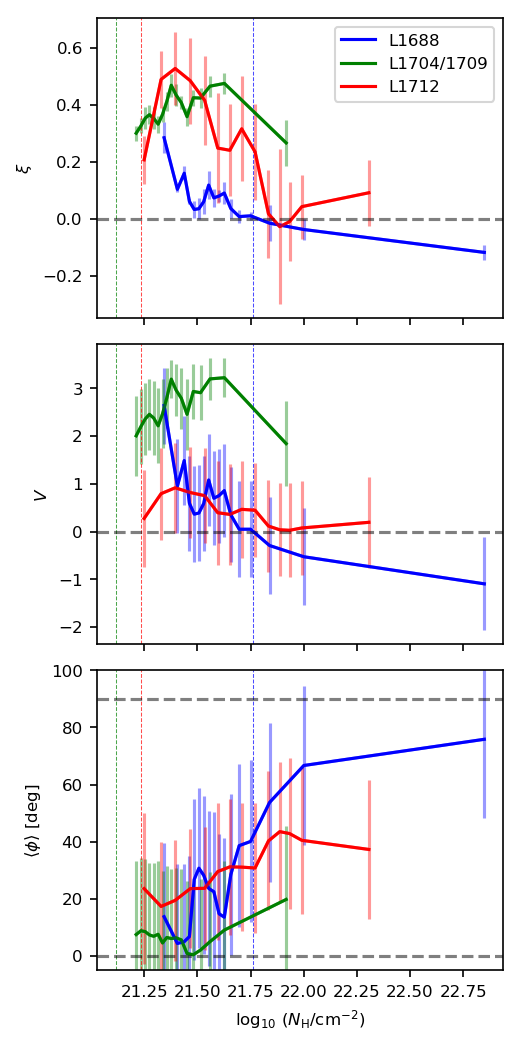}
\includegraphics[width=0.33\textwidth,angle=0,origin=c]{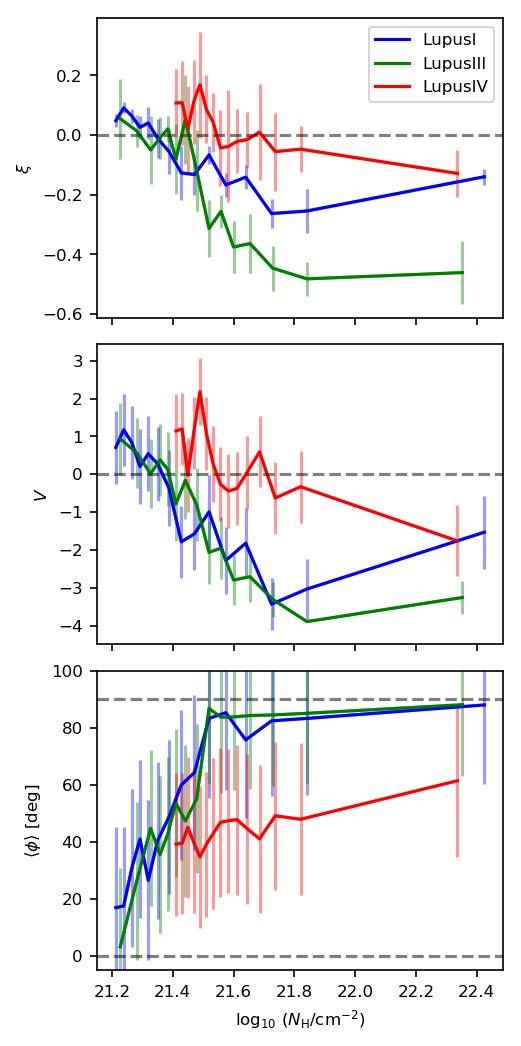}
}
\vspace{0.3cm}
\caption{\emph{Top}. Plane-of-the-sky magnetic field (\bperp) and column density (\nh) measured by 
\Planck\ toward the ten nearby molecular clouds studied in \PlanckXXXV.
The colors represent the column density inferred from the dust opacity at 353\,GHz.
The ``drapery'' pattern, produced using the LIC, indicates the orientation of magnetic field lines, which is orthogonal to the orientation of
the submillimeter polarization.
The rectangular frames indicate the regions where \Herschel\ observations are available.
Detailed maps of these \Herschel-observed regions are presented in Appendix~\ref{app:relativeorientation}.\\
\emph{Second row}.
Relative orientation parameter ($\xi$), a measure of the trends in the HRO introduced in \PlanckXXXV.
Values of $\xi$\,$>$\,0 correspond to histograms with many counts around 0\deg,  \text{i.e.}, \nh\ contours mostly parallel to \bperp.
Values of $\xi$\,$<$\,0 correspond to histograms with many counts around 90\deg,  \text{i.e.,} \nh\ contours mostly perpendicular to \bperp.
A selection of the HROs is presented in Appendix~\ref{app:relativeorientation}.\\
\emph{Third row}. 
Projected Rayleigh statistic ($V$), a statistical test of the nonuniformity and unimodal clustering in the distribution of relative orientation angles ($\phi_{k}$), defined in Eq.~\eqref{eq:V}.
Values of $V$\,$>$\,0 correspond to $\phi_{k}$ clustered around 0\deg, \text{i.e.,} \nh\ contours mostly parallel to \bperp.
Values of $V$\,$<$\,0 correspond to $\phi_{k}$ clustered around 90\deg, \text{i.e.,} \nh\ contours mostly perpendicular to \bperp.\\
\emph{Fourth row}. 
Mean relative orientation angle ($\left<\phi\right>$) defined in Eq.~\eqref{eq:meanphi}.
}
\label{fig:HOGpanel1}
\end{figure*}

\begin{figure*}[ht!]
\centerline{
\includegraphics[width=0.33\textwidth,angle=0,origin=c]{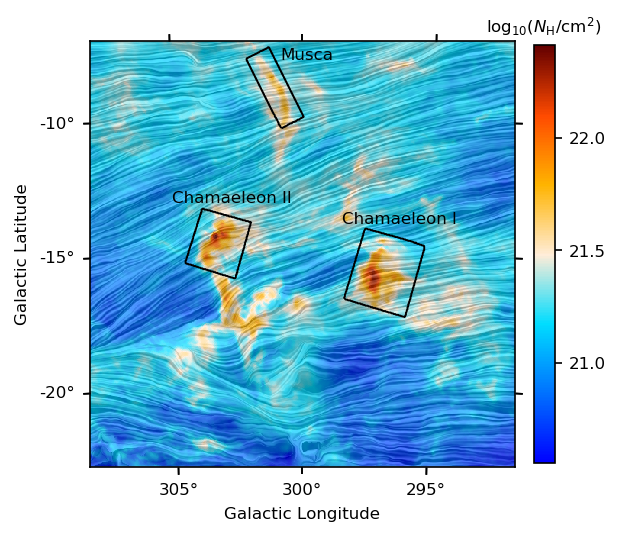}
\includegraphics[width=0.33\textwidth,angle=0,origin=c]{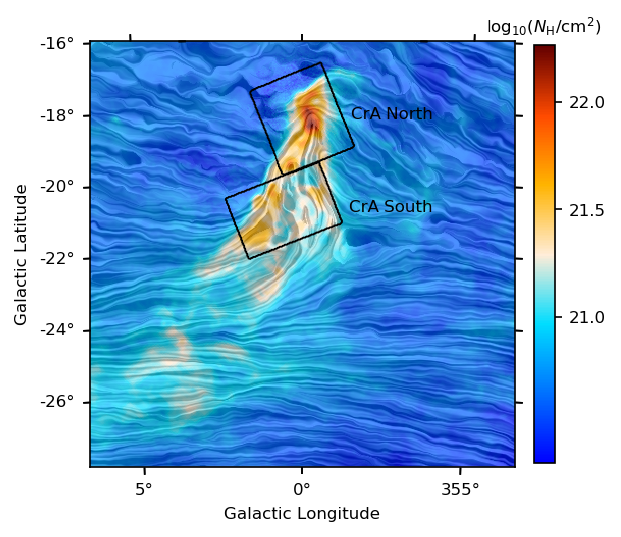}
\includegraphics[width=0.33\textwidth,angle=0,origin=c]{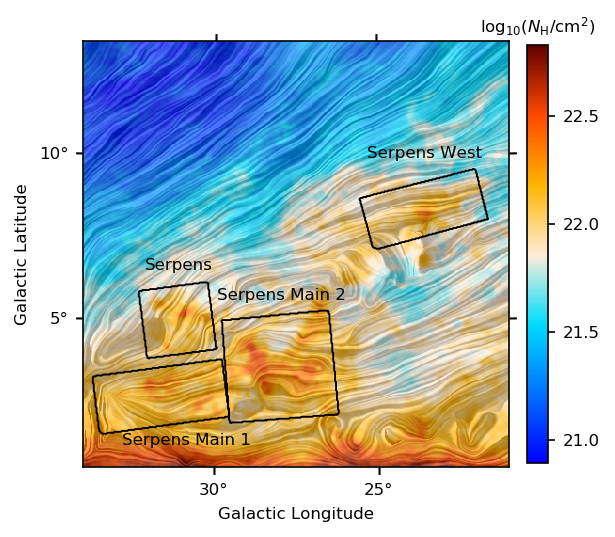}
}
\centerline{
\includegraphics[width=0.33\textwidth,angle=0,origin=c]{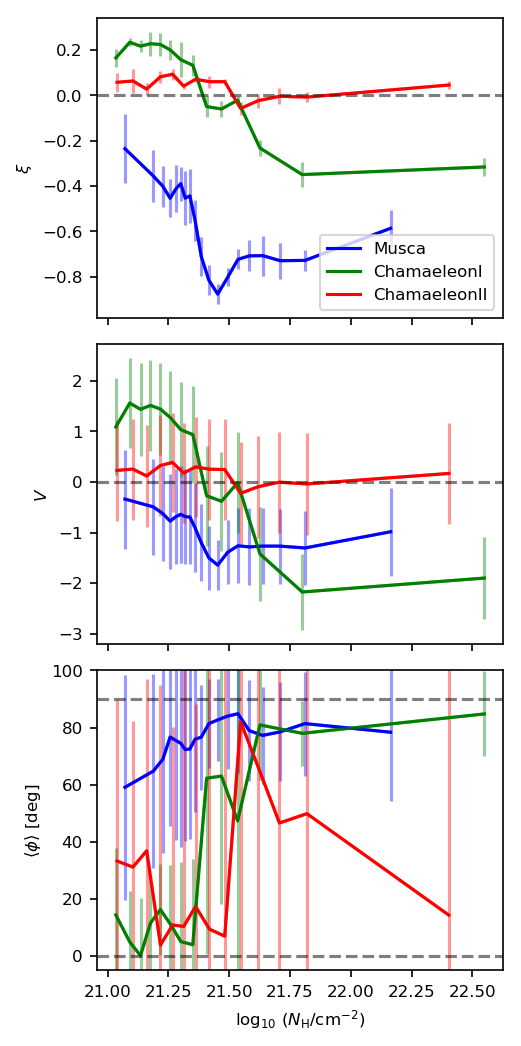}
\includegraphics[width=0.33\textwidth,angle=0,origin=c]{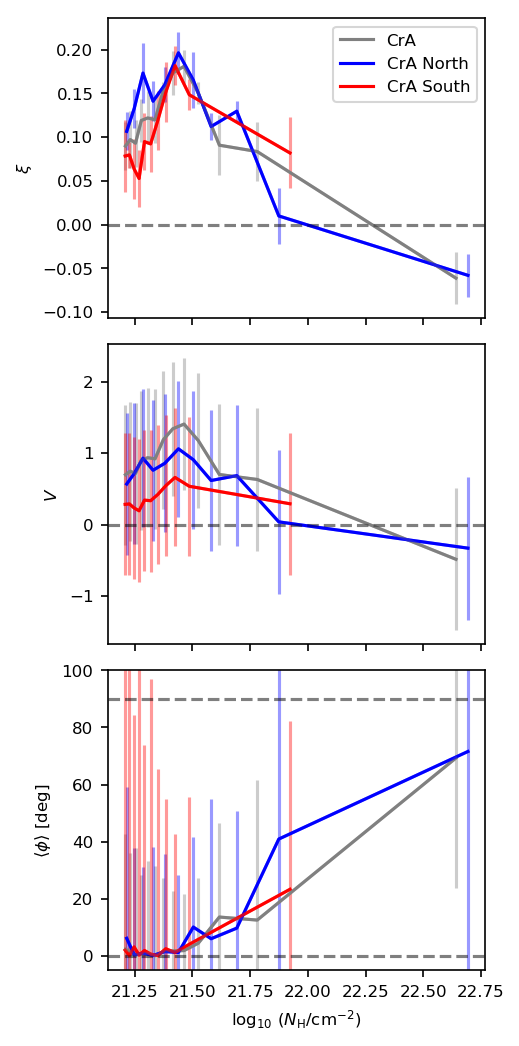}
\includegraphics[width=0.33\textwidth,angle=0,origin=c]{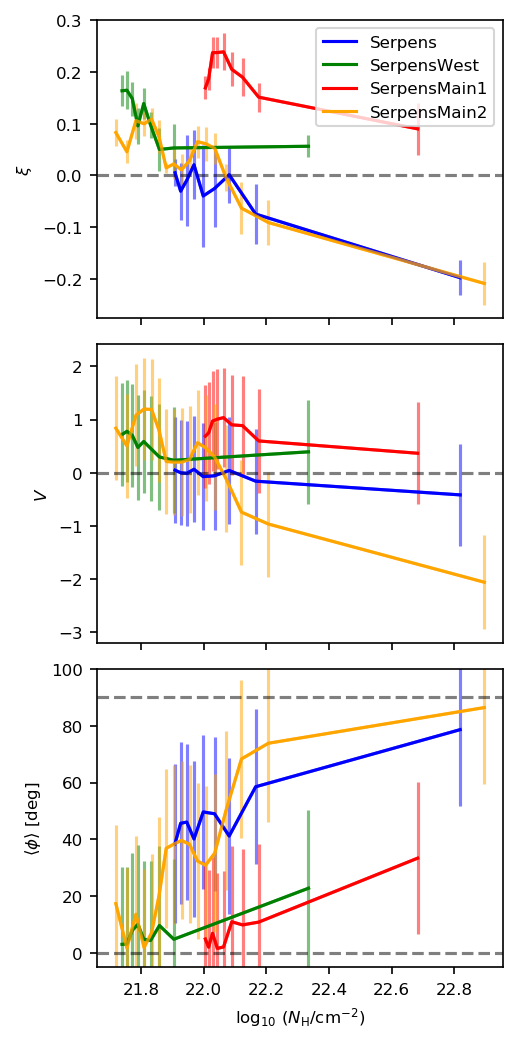}
}
\caption{Same as Fig.~\ref{fig:HOGpanel1} for Chamaeleon-Musca, CrA, and the Aquila Rift.}
\label{fig:HOGpanel2}
\end{figure*}

\begin{figure*}[ht!]
\centerline{
\includegraphics[width=0.33\textwidth,angle=0,origin=c]{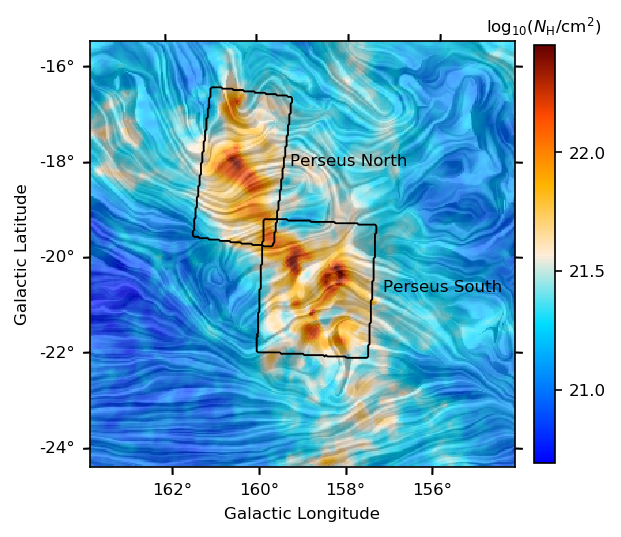}
\includegraphics[width=0.33\textwidth,angle=0,origin=c]{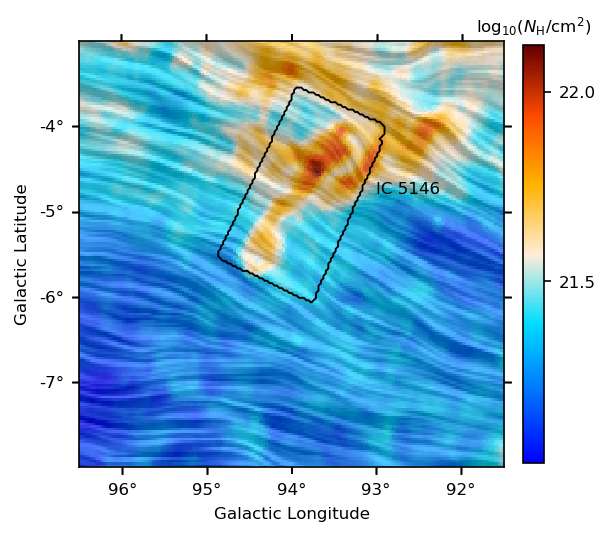}
\includegraphics[width=0.33\textwidth,angle=0,origin=c]{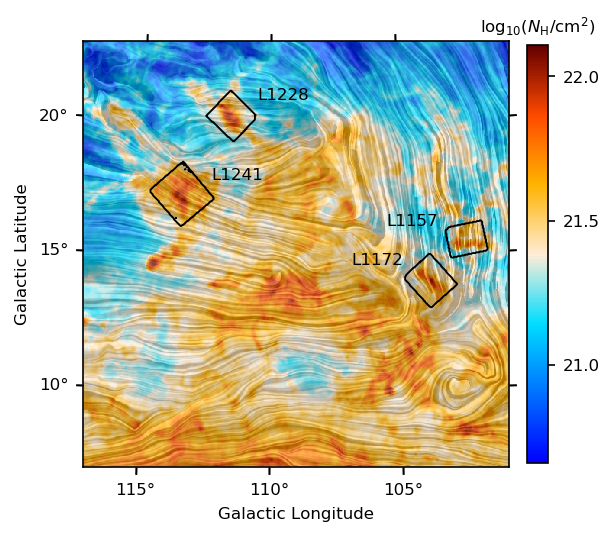}
}
\centerline{
\includegraphics[width=0.33\textwidth,angle=0,origin=c]{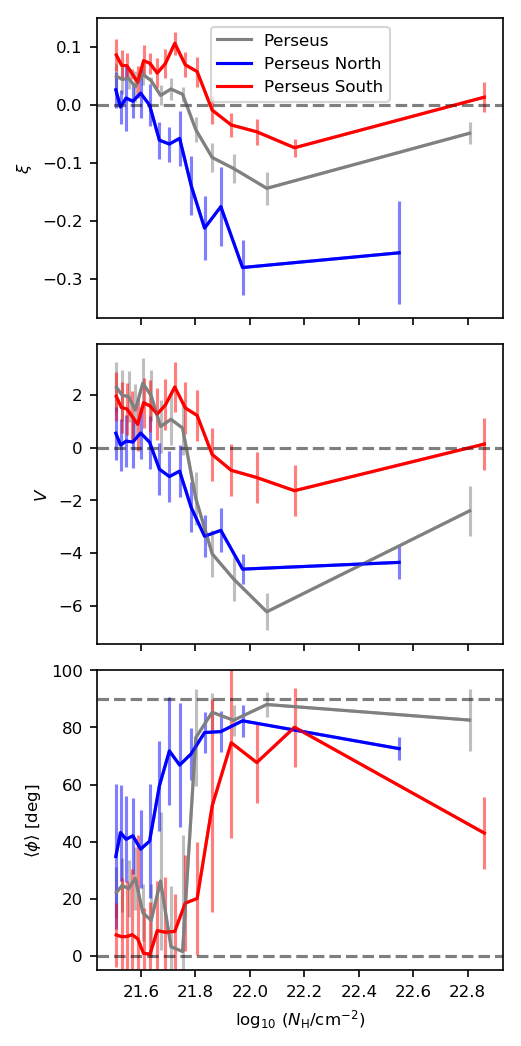}
\includegraphics[width=0.33\textwidth,angle=0,origin=c]{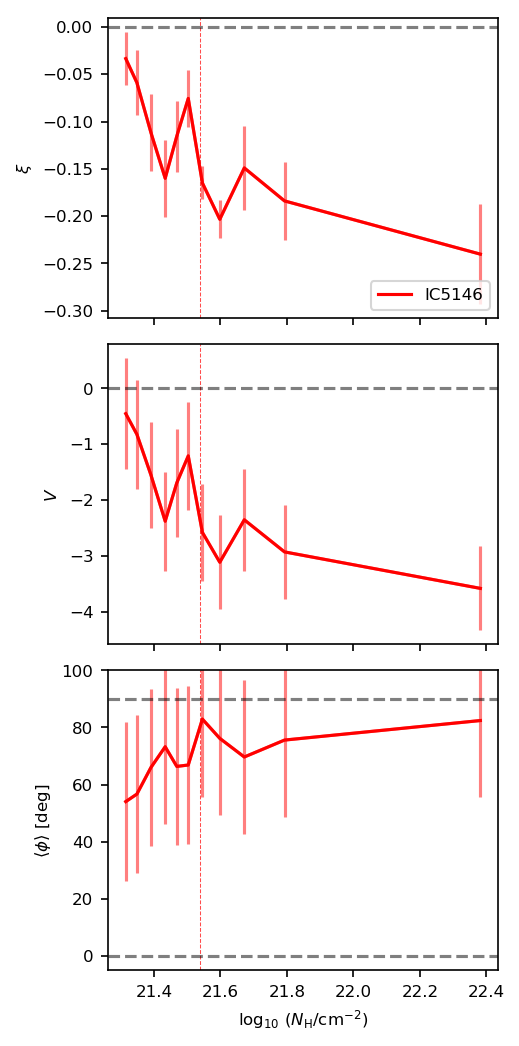}
\includegraphics[width=0.33\textwidth,angle=0,origin=c]{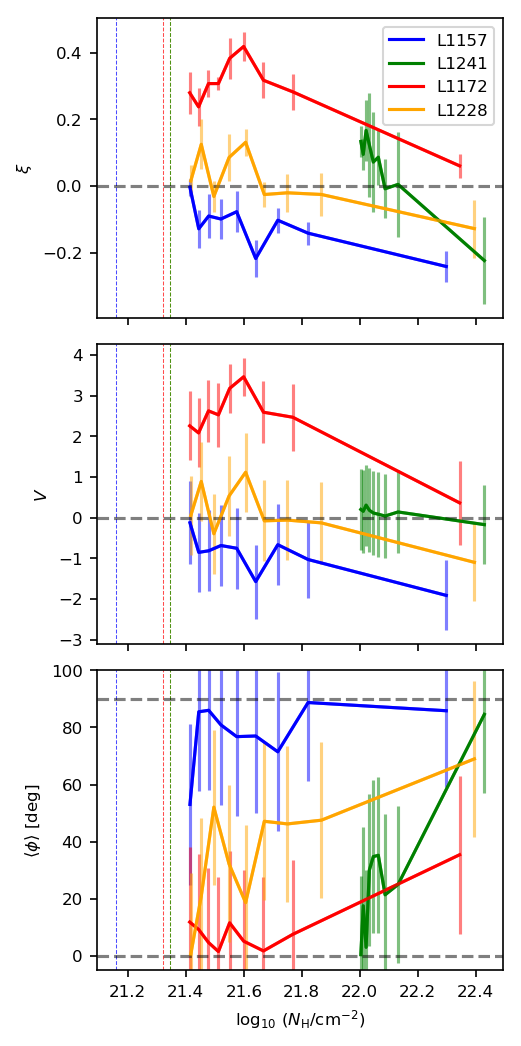}
}
\caption{Same as Fig.~\ref{fig:HOGpanel1} for Perseus, IC5146, and Cepheus.}
\label{fig:HOGpanel3}
\end{figure*}

\begin{figure*}[ht!]
\centerline{
\includegraphics[width=0.33\textwidth,angle=0,origin=c]{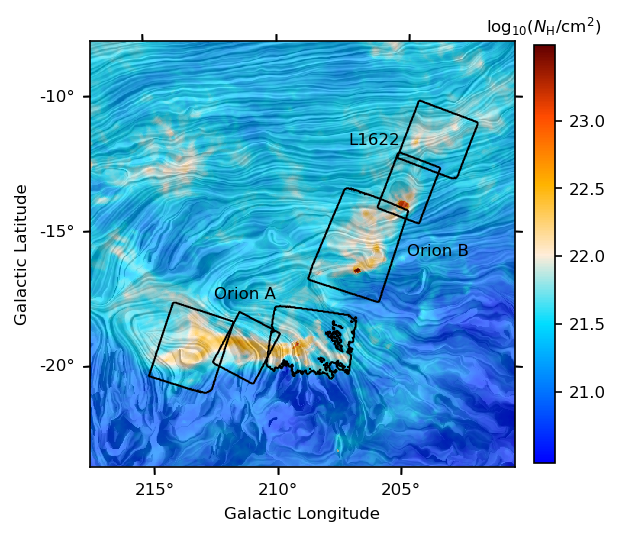}
\includegraphics[width=0.4\textwidth,angle=0,origin=c]{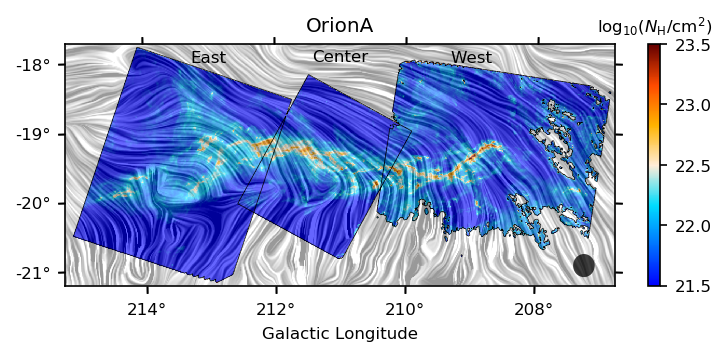}
}
\centerline{
\hspace{-20.0mm}
\includegraphics[width=0.33\textwidth,angle=0,origin=c]{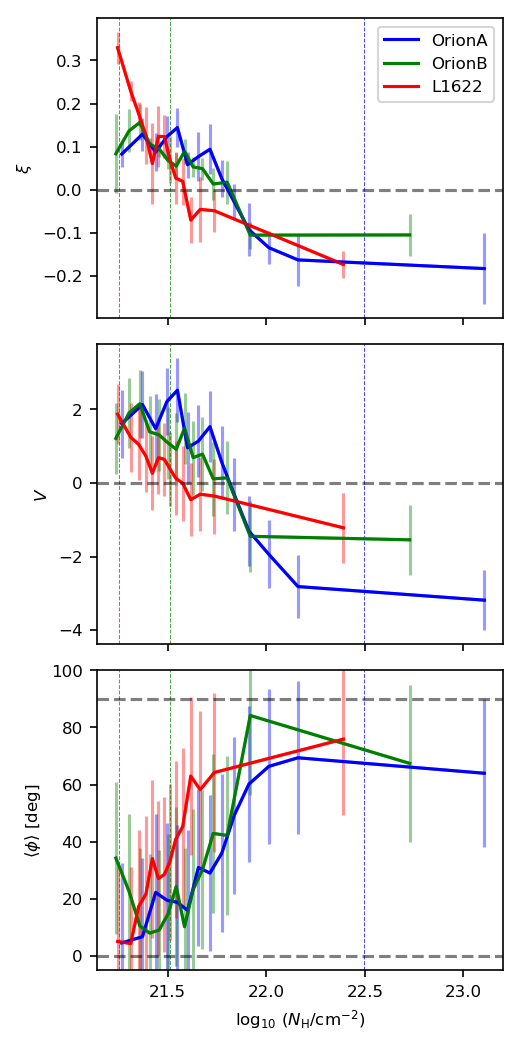}
\includegraphics[width=0.33\textwidth,angle=0,origin=c]{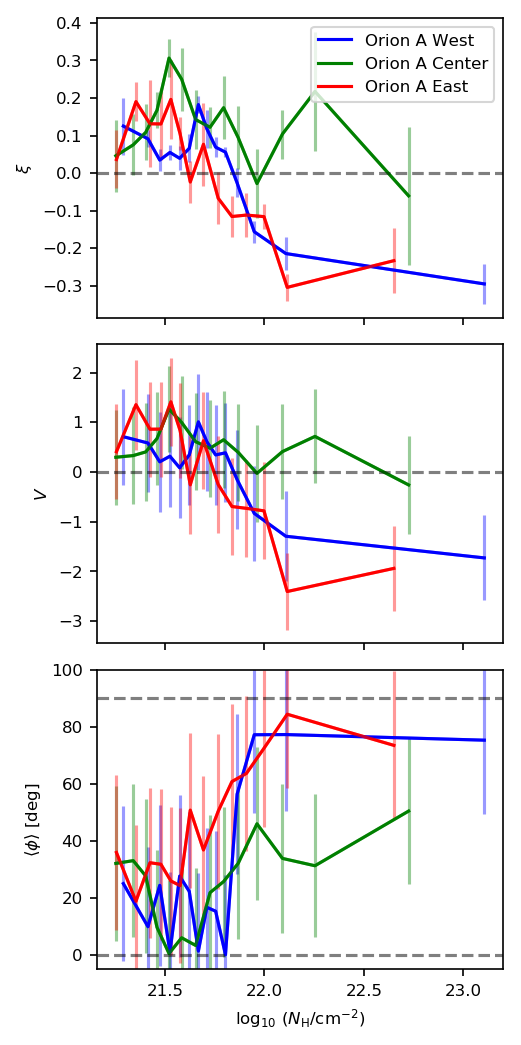}
}
\caption{Same as Fig.~\ref{fig:HOGpanel1} for Orion.}
\label{fig:HOGpanel4}
\end{figure*}

\section{Results}\label{section:results}

\subsection{Relative orientation between \nh\ and \bperp}

The general results of the HRO analysis of the ten MCs are presented in the three lower panels of Fig.~\ref{fig:HOGpanel1}, Fig.~\ref{fig:HOGpanel2}, Fig.~\ref{fig:HOGpanel3}, and Fig.~\ref{fig:HOGpanel4}.
These panels correspond to the mean resultant length, $r$, the mean relative orientation angle, $\left<\phi\right>$, and the projected Rayleigh statistic, $V$, introduced in Sec.~\ref{section:method}.
Because some of the subregions correspond to previously studied objects presented in the literature, we comment on the HRO results for each of them.
This division does not imply that these subregions correspond to a single astronomical object and is made for the sake of convenience in the discussion of our results.

\subsubsection{Taurus}

The largest subregion covered by the \Herschel\ observations toward Taurus corresponds to the elongated Barnard Dark Nebulae B213 and the dark cloud L1495, which was studied in \citep{palmeirim2013}.
Previous studies based on near-infrared polarization observations showed that \bperp\ is mostly perpendicular to B213 \citep{chapman2011}.
The HRO analysis toward this region indicates that the trend in the relative orientation between \nh\ and \bperp\ corresponds to the archetypical behavior identified in 
\PlanckXXXV\ for the Taurus region, that is, \nh\ and \bperp\ are mostly parallel at the lowest \nh\ and mostly perpendicular at the highest \nh. 
This is evident in the transition from $V$\,$>$\,0 to $V$\,$<$\,0 and \meanphi\ from 0\deg\ to 90\deg\ with increasing \nh. 

The HRO analysis toward the L1551 region reveals that the smooth increment in \meanphi\ with increasing \nh\ is interrupted around \lognh\,$\approx$\,21.7\,cm$^{-2}$.
This behavior is similar to that found in the L1489 dark cloud subregion, where the smooth increase of \meanphi\ from 0\deg\ is interrupted around \lognh\,$\approx$\,21.5\,cm$^{-2}$.
In principle, the behavior of \meanphi\ in L1551 and L1489 can be interpreted as the effect of the projection of the magnetic field and the density structure onto the plane of the sky.
The relative orientation between the 2D projections of two vectors that are perpendicular in 3D is likely to be less than 90\deg\ for many of the possible configurations of the vectors and the plane of projection.
However, the low values of $V$ indicate that the relative orientation angle distributions are not significantly unimodal and centered on 0\deg\ or 90\deg, thus preventing us from drawing further conclusions from \meanphi\ alone.

\subsubsection{Ophiuchus}

The largest subregion in Ophiuchus corresponds to the L1688 dark cloud, an exhaustively studied region of low-mass star formation \citep{wilking2008}.
This region shows a similar behavior to that observed in the B213/L1495 region in Taurus, that is, a clear transition of \meanphi\ from 0\deg\ to 90\deg\ and $V$\,$>$\,0 to $V$\,$<$\,0 with increasing \nh. 
This transition from \nh\ and \bperp\ mostly parallel at the lowest \nh\ and mostly perpendicular at the \lognh\,$>$\,21.8 is much clearer than that found in \PlanckXXXV\ studies of the whole Ophiuchus region.

The other two subregions in Ophiuchus correspond to the dark clouds that are often called streamers. These trail L1688 to the southeast (L1712) and the east (L1704 and L1709).
The HRO analysis toward the L1712 region reveals that \meanphi\ is roughly between 20 and 40\deg, although $\xi$\,$\approx$\,0 and $V$\,$\approx$\,0.
Toward L1704/L1709, the HRO analysis reveals that \nh\ and \bperp\ are mostly parallel, \meanphi\,$\approx$\,0\deg.
In sum, the region with lowest \nh\ values (L1704/L1709) shows \nh\ and \bperp\ mostly parallel, the region with intermediate \nh\ values (L1712) has a transition from mostly parallel to roughly 40\deg, and finally, the region with the highest \nh\ values (L1688) shows \nh\ and \bperp\ changing from mostly parallel to mostly perpendicular with increasing \nh.

\subsubsection{Lupus}

 Lupus I and Lupus III show a clear transition of $V$\,$>$\,0 to $V$\,$<$\,0 and \meanphi\ from 0\deg\ to 90\deg\ with increasing \nh.
In the Lupus IV subregion, which is located at lower Galactic latitude than the other two subregions, $V$\,$\approx$\,0 in for \lognh\,$<$\,21.8, and it is only $V$\,$<$\,0 in the highest \nh\ bin, where \meanphi\ is approximately 60\deg, as shown in Fig.~\ref{fig:HOGpanel1}.

\subsubsection{Chamaeleon-Musca}

Musca is the quintessential example of a well-defined elongated molecular cloud with a homogeneous \bperp\ perpendicular to it, as first revealed by starlight polarization observations \citep{pereyra2004}.
The HRO analysis confirms this visual assessment with clearly defined negative values of $\xi$ and $V$ and \meanphi\ very close to 90\deg, as shown in Fig.~\ref{fig:HOGpanel2}.
However, the HOG analysis does not reveal a significant trend of \nh\ structures being parallel to \bperp\ for \lognh\,$>$\,21.0.
This is most likely the result of the smoothing of the \Herschel\ observations to a common resolution to produce the \nh\ maps, which washes out most of the striations that are clearly visible at 250\micron\ 
in its native resolution.
Further HRO analysis of the \Herschel\ 250\micron\
maps and \bperp, presented in Appendix~\ref{app:striationsTest}, shows a clear transition of $V$\,$>$\,0 to $V$\,$<$\,0 and \meanphi\ from 0\deg\ to 90\deg\ with increasing \nh.
This quantitatively confirms the relative orientation assessment reported in \cite{cox2016}.

The two Chamaeleon clouds covered by the \Herschel\ observations, for which we used the designations in \cite{luhman2008}, show dissimilar behaviors.
Chamaeleon~I (Cham~I) shows a clear transition from $V$\,$>$\,0 to $V$\,$<$\,0 and \meanphi\ from 0\deg\ to 90\deg\ with increasing \nh, very much like the one found throughout the entire Chamaeleon-Musca region in \PlanckXXXV.
In contrast, Chamaeleon~II (Cham~II) presents values of $\xi$ and $V$ close to 0, which indicates that the distribution of relative orientation angles does not clearly peak on either 0\deg\ or 90\deg,\ as can be inferred from the \meanphi\ values shown in Fig.~\ref{fig:HOGpanel2}.

Cham~I and Cham~II have similar sizes and masses, but the SFR, based on counting YSOs identified by their infrared excess, is higher in Cham I by roughly a factor of six \citep{evans2014}.
Furthermore, \cite{alvesdeoliveira2014} identified clear differences between the two Cham I and Cham II clouds based on the \Herschel\ observations.
On the one hand, Cham~I has formed stars, but it seems to have arrived at the end of its star formation. 
Morphologically, it appears as a ridge that is embedded in low-density gas with striations.
On the other hand, Cham~II has a clumpy structure with ongoing star formation, and faint striations are only marginally observed.

Toward Cham~II the values of $V$\,$\approx$\,0 indicate that the distribution of relative orientation angle is neither unimodal nor centered on 0\deg\ or 90\deg.
Still, the values of \meanphi\ are close to 0\deg\ in the lowest \nh\ bins, increase up to 60\deg\ around \lognh\,$\approx$\,21.6, and then decrease to roughly 20\deg.
This behavior is one of the two exceptions that we found to the general increment of \meanphi\ with increasing \nh.

\subsubsection{Corona Australis}

The HRO study of the Corona Australis (CrA), presented in Fig.~\ref{fig:HOGpanel2}, indicates that throughout the entire region, the \nh\ structures are predominantly parallel to \bperp\ over most \nh\ values and are only perpendicular in the last \nh\ bin, which corresponds to \lognh\,$>$\,22.
However, the significance of the relative orientation in terms of $\xi$ and $V$ toward CrA is not very high. 
It is mostly due the correlation introduced by the \Planck\ beam and not by the contrast in the histograms shown in Fig.~\ref{fig:CrAHROs}. 

To facilitate the discussion, we divided the CrA region into two portions.
CrA North includes the Cr-A, Cr-B, Cr-C, and Cr-D, designations introduced in \cite{nutter2005}. 
These are the densest regions within CrA and contain most of the gravitationally bound cores \citep{bresnahan2018}.
CrA South includes \nh\ structures with cometary appearance with fewer starless cores, candidate prestellar cores, and confirmed prestellar cores than CrA North.
The HRO analysis reveals significant differences between the two regions.
On the one hand, CrA North shows a transition from $\xi$ and $V$\,$>$\,0 to $\xi$ and $V$\,$<$\,0 with increasing \nh\ and with \meanphi\ around 0\deg\ at roughly \lognh\,$<$\,21.7 and then increasing to approximately 70\deg.
Although the statistical significance of the $\xi$ and $V$\,$<$\,0 are low, the HROs presented in Fig.~\ref{fig:CrAHROs} further confirm that they represent a substantial change in the distribution of $\phi$ toward that portion of the CrA region.
On the other hand, CrA South shows $\xi$ and $V$\,$>$\,0 and \meanphi\ around 0\deg\ for all \nh, which only reaches a maximum value \lognh\,$=$\,21.03.

\subsubsection{Aquila Rift}

The HRO analysis of Serpens Main 2 \citep{bontemps2010,konyves2015}, which has been presented as a proof of concept in \citep{soler2017} and is shown here in Fig.~\ref{fig:HOGpanel2}, reveals a clear transition from \nh\ and \bperp\ mostly parallel at the lowest \nh\ and mostly perpendicular at the highest \nh, which is evident in the change of \meanphi\ from 0\deg\ to 90\deg\ and $V$\,$>$\,0 to $V$\,$<$\,0 with increasing \nh\ in the rightmost panels of Fig.~\ref{fig:HOGpanel2}.
This confirms the trends found by estimating \bperp\ based on near-infrared polarimetry and its qualitative comparison with the \nh\ structure in the region \citep{sugitani2011}.

A similar behavior is found in the Serpens dark cloud \citep{eiroa2008}, where the significance in the values of $V$ is lower, a smaller cloud has less independent measurements of $\phi$, but the values of \meanphi\ are clearly close to 90\deg\ in the highest \nh\ bin.
Toward the Serpens Main 1 and the Serpens West regions, the trends in $\xi$, $V$, and \meanphi\ reveal that \nh\ and \bperp\ are mostly parallel, although the values of $V$\,$\approx$\,0 indicate that the significance of these values is low, mostly due the correlation introduced by the \Planck\ beam and not by the contrast in the histograms shown in Fig.~\ref{fig:AquilaRiftHROs}. 

\subsubsection{Perseus}

The HRO study of Perseus \citep{bally2008a}, presented in Fig.~\ref{fig:HOGpanel3}, indicates that over the whole region the relative orientation trends are consistent with those found in \PlanckXXXV,\, that is, from $\xi$\,$>$\,0 and $V$\,$>$\,0 to $\xi$\,$<$\,0 and $V$\,$<$\,0 with increasing \nh.
To facilitate discussion, we divided Perseus into two portions, following the two \Herschel\ maps toward Perseus. 
Perseus North includes B5 (L1471), which is one of the most extensively studied dark clouds in the sky, and IC 348. 
Perseus South includes B1, the dark nebulae L1448 and 1455, and NGC 1333, which is currently the most active region of star formation in Perseus \citep{sadavoy2014}.

The main difference between Perseus North and Perseus South is the significance in the values of $\xi$ and $V$, which are roughly a factor for two higher in Perseus North.
This difference is not simply the result of the larger number of independent $\phi$ observations, as shown by the normalized quantity $\xi$, but it reflects a significant difference in the distribution of its values in each \nh\ bin, as further illustrated in Fig.~\ref{fig:Perseus-IC5146HROs}.
Perseus North and South both show the transition from $\xi$\,$>$\,0 and $V$\,$>$\,0 to $\xi$\,$<$\,0 and $V$\,$<$\,0, which corresponds to \nh\ and \bperp\ being mostly parallel at the lowest \nh\ and mostly perpendicular at the highest \nh.
However, in the highest \nh\ bin of Perseus South, corresponding to 22.2\,$<$\,\lognh\,$<$\,23.5, $\xi$ and $V$ return to 0 and \meanphi\ is roughly 50\deg.
The values of $\xi$ and $V$ indicate that the distribution of $\phi$ is neither unimodal nor centered on 0 or 90\deg.
This behavior is the second of the exceptions that we found to the general trends discussed in Sec.~\ref{section:discussion}

\subsubsection{IC5146}

The IC5146 region is composed of a bulbous dark cloud at the end of a long dark streamer extending to the northwest, where a dark cloud complex lies \citep{herbig2008}.
The HRO toward that region, presented in Fig.~\ref{fig:HOGpanel3}, shows clearly defined negative values of $\xi$ and $V$ and \meanphi\ very close to 90\deg, particularly at \lognh\,$>$\,21.6.
As in the case of Musca, we repeated the HRO analysis of IC5146 using the \Herschel\ 250\micron\ 
map in Appendix~\ref{app:striationsTest}, which revealed that the lower resolution in the \nh\ maps washes out structures that are clearly parallel to \bperp\ and would complete the clear transition from $\xi$\,$>$\,0 and $V$\,$>$\,0 to $\xi$\,$<$\,0 and $V$\,$<$\,0 with increasing \nh.

\subsubsection{Cepheus}

The region of the Cepheus flare \citep{kun2008} studied in \PlanckXXXV\ corresponds to the northeast portion of the region covered in the \cite{yonekura1997} $^{13}$CO survey.
Although in \PlanckXXXV\ this region was analyzed without considering individual clouds, the \Herschel\ observations toward Cepheus cover four regions around dark cloud associations: 
L1241 (Cepheus 2  in \cite{evans2014}), L1172/74 (Cepheus 3), L1157/1152/1155/1147 (Cepheus 4), and L1228 (Cepheus 5). 
We therefore present the HRO analysis for each of these separate regions in Fig.~\ref{fig:HOGpanel3}.

The HRO analysis toward L1228 shows a clear transition from $V$\,$>$\,0 and \meanphi\,$\approx$\,0\deg\ to $V$\,$<$\,0 and \meanphi\,$\approx$\,90\deg\ with increasing \nh, that is, from \nh\ and \bperp\ being mostly parallel at the lowest \nh\ and mostly perpendicular at the highest \nh.
The analysis of L1241 shows a similar trend, with less significance and mostly at \lognh\,$>$\,22.0. 
In contrast with L1228, which is the region with the highest SFR in Cepheus, L1241 shows very little evidence of current star formation \citep{kirk2009}.

Toward L1172, \nh\ and \bperp\ appear to be mostly parallel across all \nh\ values, but the trends show that $\xi$ and $V$ tend to 0 and \meanphi\ increasing progressively from 0\deg\ to approximately 30\deg.
In contrast, toward L1157, \nh\ and \bperp\ appear to be perpendicular across all \nh\ values.
The SFRs, based on counting YSOs, indicate that L1157 is currently forming eight times more stars than L1172 \citep{evans2014}.

\subsubsection{Orion}

The HRO study of the Orion A, Orion B, and L1622, presented in Fig.~\ref{fig:HOGpanel4}, indicates that the relative orientation trends are consistent with those found in \PlanckXXXV,\, that is, from $\xi$\,$>$\,0 and $V$\,$>$\,0 to $\xi$\,$<$\,0 and $V$\,$<$\,0, or \nh\ and \bperp\ changing from mostly parallel to mostly perpendicular, with increasing \nh.
The high values of $\xi$ and $V$ indicate that these relative orientation trends are very significant, although the values of \meanphi\ only reach approximately 70\deg\ in the highest \nh\ bins of the three regions.
The similarity in the relative orientation in the three regions is clear in the values of $\xi$, $V$, and \meanphi\ in the range 21.0\,$\lesssim$\,\lognh\,$\lesssim$\,21.8 and it extends to the highest \nh\ bin, although the maximum \nh\ values are lower for L1622 and Orion~B than for Orion~A.

The relative orientation in the highest \nh\ bins in Orion~A, which indicate that \nh\ and \bperp\ are mostly perpendicular, is consistent with the relative orientation reported at smaller scales toward the eastern end of the region, in an area of roughly 6\arcmin\,$\times$\,6\arcmin\ around the Kleinmann-Low nebula \citep[KL,][]{KleinmannANDLow1967} and the Becklin-Neugebauer object \citep[BN,][]{BecklinANDNeugebauer1967}, using observations by the SCUBA-2 Polarimeter (POL-2) on the James Clerk Maxwell Telescope (JCMT) at 850\micron\
\citep[14\parcs1 resolution,][]{pattle2017} and the High-Resolution Airborne Wideband Camera-Plus (HAWC+) on board the Stratospheric Observatory for Infrared Astronomy (SOFIA) at 53, 89, 154, and 214\micron\ \citep[4\parcs9, 7\parcs8, 13\parcs6, and 18\parcs2 resolution, respectively][]{chuss2019}.
This indicates that the hourglass configuration of \bperp\ reported in \cite{pattle2017} and \cite{chuss2019} in arcminute scales is not decoupled from the large-scale \bperp\ configuration, which is perpendicular in degree scales to the integral-shaped filament that hosts the KL-BN region.

The size of the region allows us to further divide the HRO analysis of Orion A into three subregions that we present in the rightmost panels of Fig.~\ref{fig:HOGpanel4}.
The western region contains the KL-BN region and most of the YSOs in Orion A \citep{grossschedl2019}, the center region contains most of the L1641 dark cloud \citep{polychroni2013}, and finally, the eastern region contains the L1647 dark cloud. 
Although the western and eastern regions have known differences in terms of the slope of their \nh\ distribution and number of Class 0 protostars \citep{stutz2015}, their behavior in terms of the relative orientation between \nh\ and \bperp\ is very similar. It changes from mostly paralllel to mostly perpendicular with increasing \nh.
In contrast, the center region does not show a very clear preferential relative orientation, as inferred from $\xi$ and $V$\,$\approx$\,0, although the \meanphi\ values indicate a tendency for the relative orientation to increase from roughly 0\deg\ to 50\deg\ with increasing \nh.

\begin{figure*}[ht!]
\centerline{
\includegraphics[width=0.33\textwidth,angle=0,origin=c]{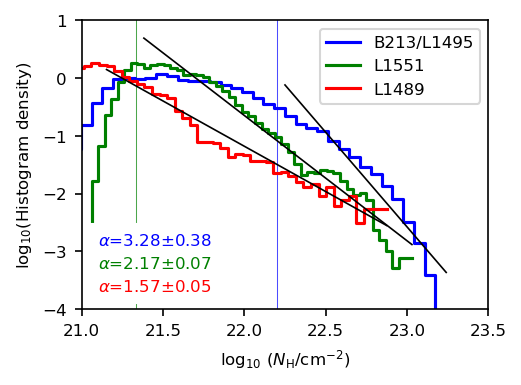}
\includegraphics[width=0.33\textwidth,angle=0,origin=c]{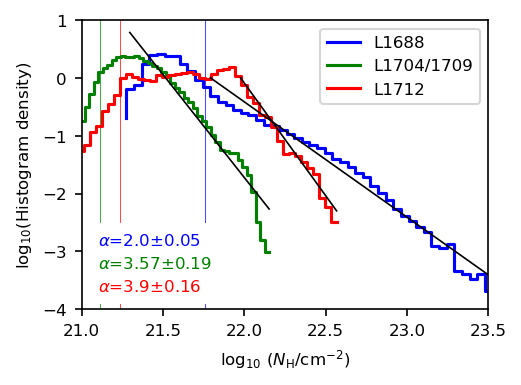}
\includegraphics[width=0.33\textwidth,angle=0,origin=c]{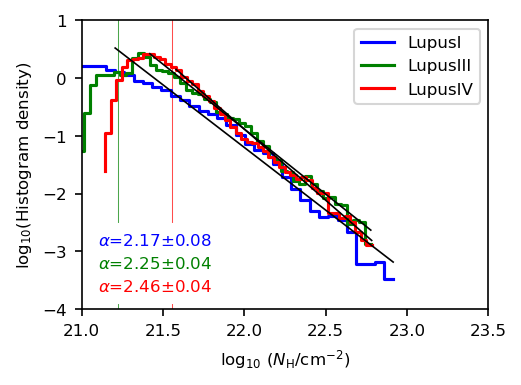}
}
\centerline{
\includegraphics[width=0.33\textwidth,angle=0,origin=c]{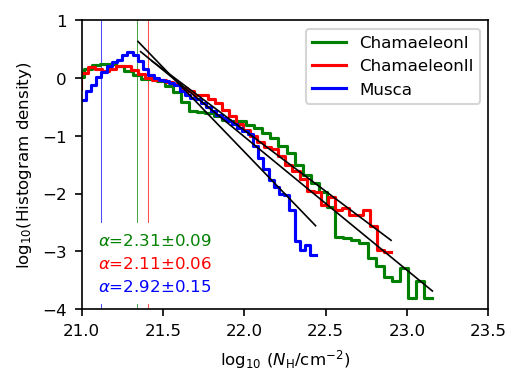}
\includegraphics[width=0.33\textwidth,angle=0,origin=c]{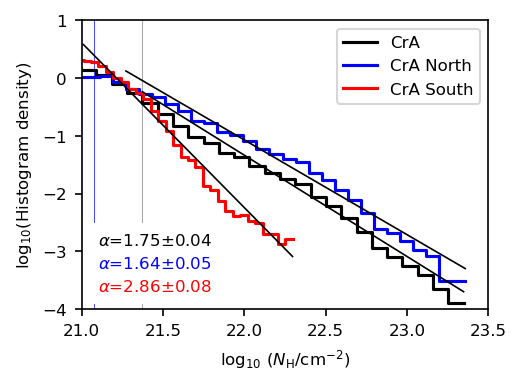}
}
\centerline{
\includegraphics[width=0.33\textwidth,angle=0,origin=c]{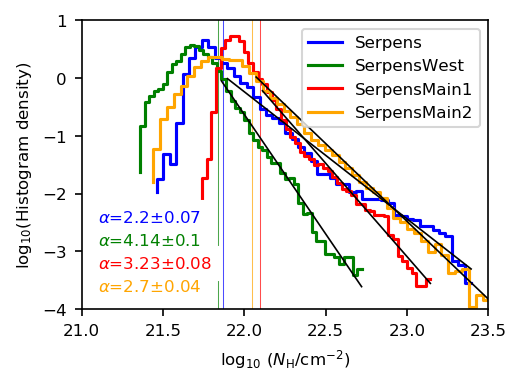}
\includegraphics[width=0.33\textwidth,angle=0,origin=c]{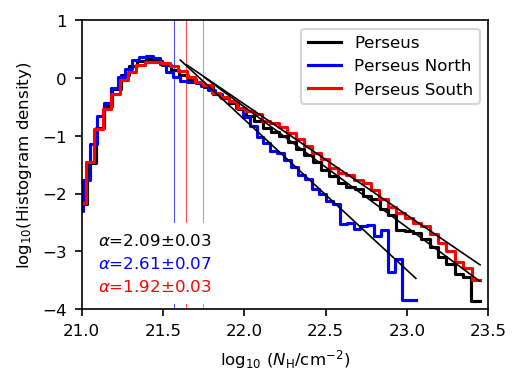}
}
\centerline{
\includegraphics[width=0.33\textwidth,angle=0,origin=c]{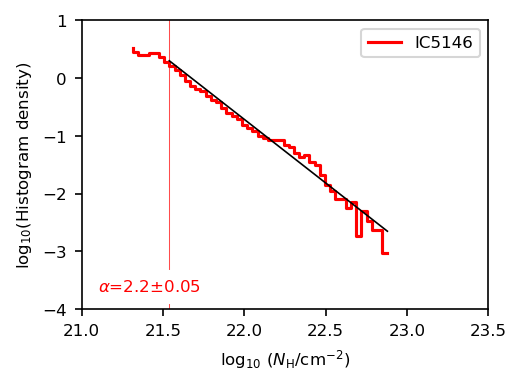}
\includegraphics[width=0.33\textwidth,angle=0,origin=c]{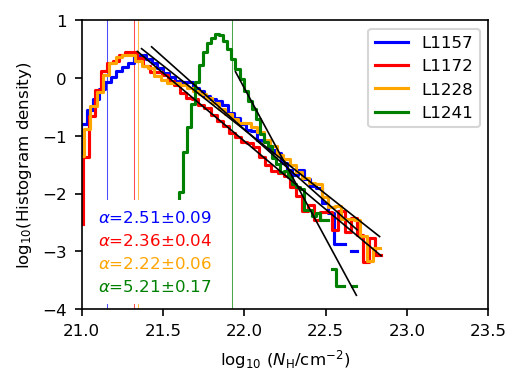}
\includegraphics[width=0.33\textwidth,angle=0,origin=c]{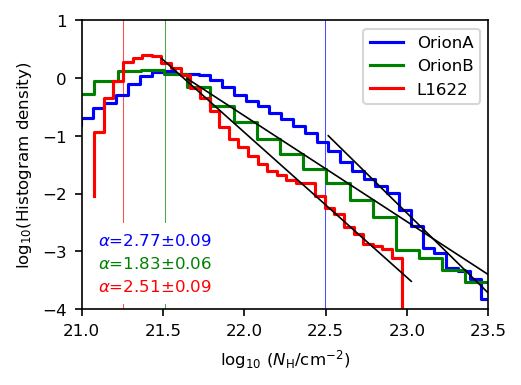}
}
\caption{Probability density functions of \nh.
The black lines correspond to the fits of Eq.~\eqref{eq:nhpdf} to the \nh\ PDFs within the last closed contour of the subregions that are indicated by the vertical lines.
}
\label{fig:NHPDFs}
\end{figure*}

\subsection{Relative orientation of \nh\ and \bperp\ and column density PDFs}

Following the considerations presented in \cite{ossenkopf2016} and \cite{alves2017}, we focused our analysis of the \nh\ PDF on the area within the last closed contour in each of the subregions.
With this selection, the \nh\ PDF can be reasonably modeled as
\begin{equation}\label{eq:nhpdf}
\log_{10}({\rm PDF}) = C - \alpha\log_{10}(N_{H}),
\end{equation}
as illustrated in Fig.~\ref{fig:NHPDFs}.
Although there are deviations from this linear behavior in some regions, the values of the slope $\alpha$ provide a general characterization of the \nh\ that we can compare to the trends in the relative orientation of \nh-\bperp\ .

\begin{figure*}[ht!]
\centerline{
\includegraphics[width=0.5\textwidth,angle=0,origin=c]{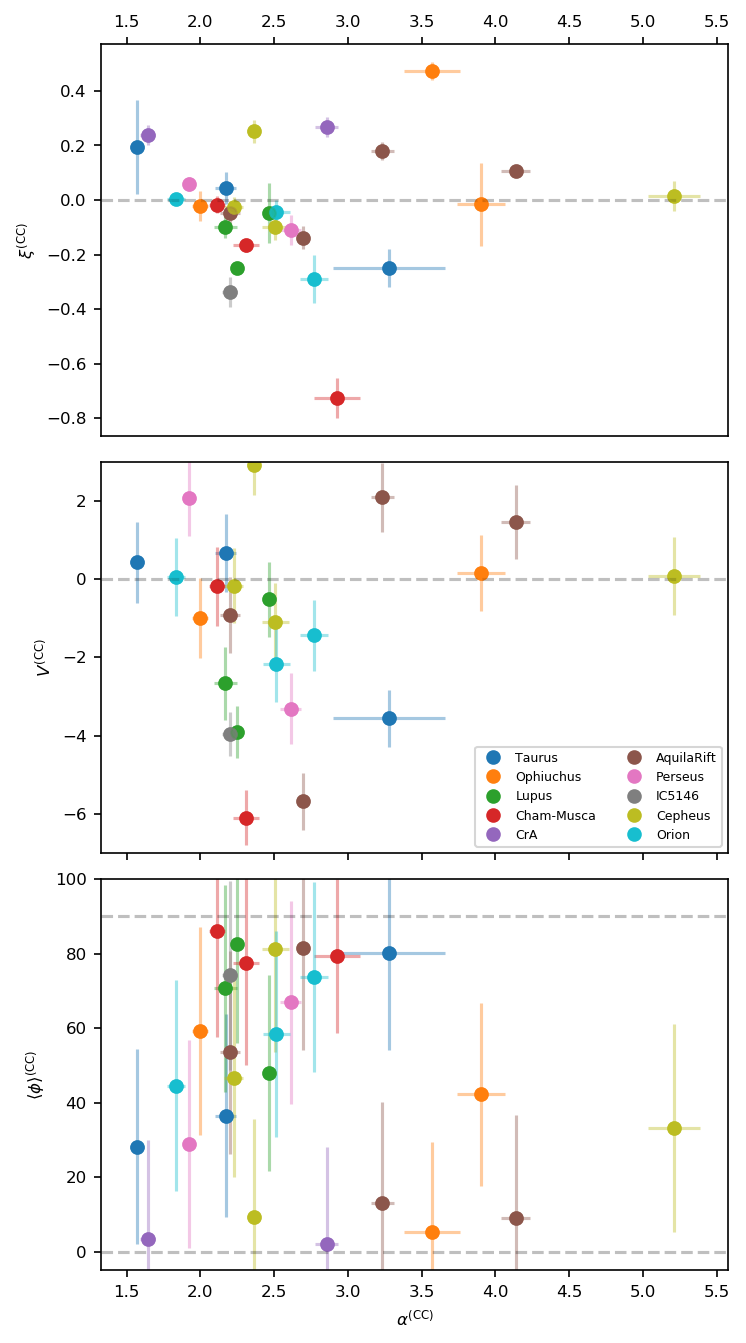}
\includegraphics[width=0.5\textwidth,angle=0,origin=c]{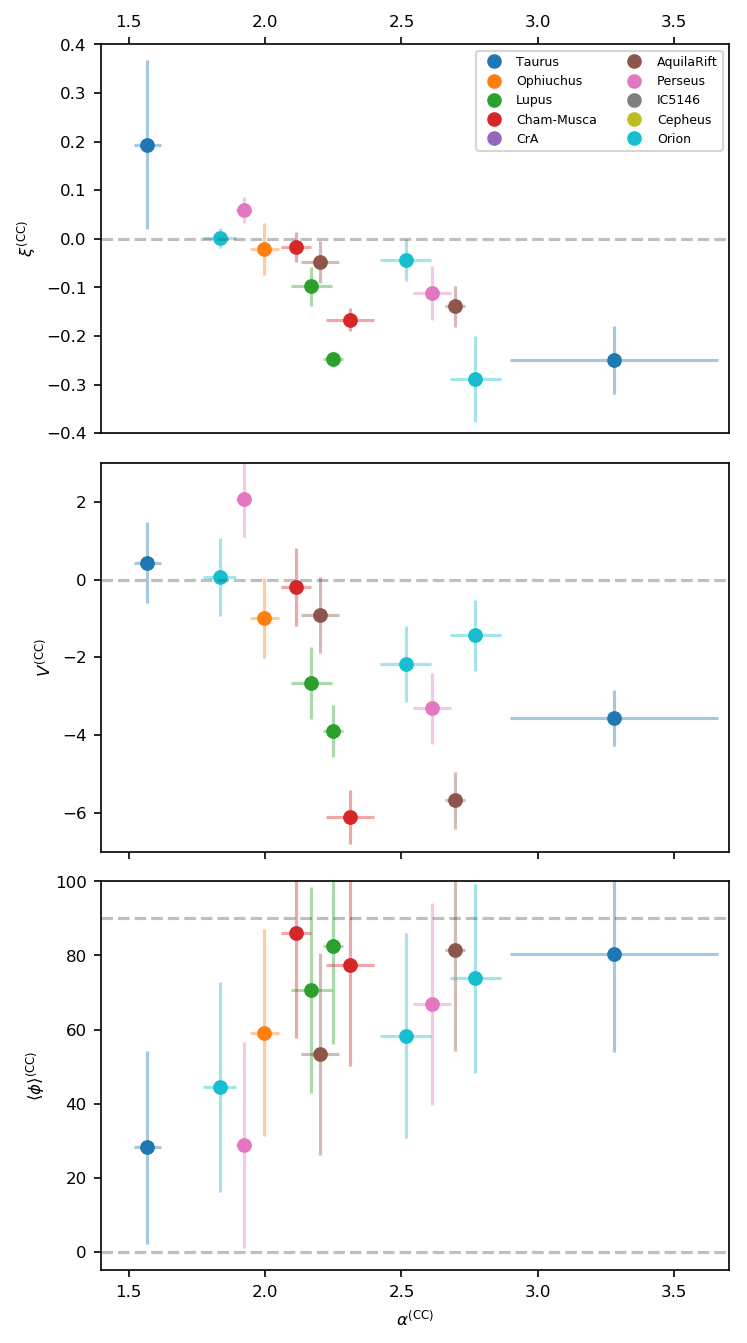}
}
\caption{Relative orientation parameter ($\xi$, top), projected Rayleigh statistic ($V$, middle), and mean relative orientation angle ($\left<\phi\right>$, bottom) as a function of the slope of the \nh\ distribution ($\alpha$, Eq.~\ref{eq:nhpdf}) within the last closed contour of the studied regions before (left) and after (right) the selection criterion $|\left<\psi\right>^{({\rm in})}-\left<\psi\right>^{({\rm out})}|$\,$>$\,20\deg, where $\left<\psi\right>^{({\rm in})}$ and $\left<\psi\right>^{({\rm out})}$ are the mean \bperp\ orientation angles inside and outside the largest closed \nh\ contour, respectively.
}
\label{fig:GeneralTrendsNHPDF}
\end{figure*}

The left-hand side of Fig.~\ref{fig:GeneralTrendsNHPDF} shows a comparison of the values of $\xi$, $V$, and \meanphi\, evaluated within the last closed contours, and values of $\alpha$ in all of the studied MC subregions.
At first glimpse, this figure reveals a general lack of correlation between the values $\alpha$ and the metrics of the relative orientation.
The MCs with the steepest \nh\ PDFs ($\alpha$\,$>$\,$3.2$) show values of $\xi$, $V$, and \meanphi\ that are consistent with \nh\ and \bperp\ being mostly parallel.
Most of the MCs show slopes $3.2$\,$>$\,$\alpha$\,$<$\,$1.8,$ and within this range lie clear examples of clouds that are predominantly perpendicular, such as Musca ($\alpha$\,$\approx$\,$2.9$, \meanphi\,$\approx$\,79\deg\,$\pm$\,20\deg) and CrA South ($\alpha$\,$\approx$\,$2.9$, \meanphi\,$\approx$\,1\deg\,$\pm$\,26\deg).
The MCs with the shallowest \nh\ PDFs ($1.8$\,$>$\,$\alpha$) also show values of $\xi$, $V$, and \meanphi\ that are consistent with \nh\ and \bperp\ being mostly parallel.
However, the MCs with $\alpha$\,$<$\,$3.2$ suggest a general trend: the values of $\xi$ and $V$ appear to progressively decrease with increasing $\alpha$ and change from mostly positive to mostly negative, which is reflected in the increase in \meanphi\ from 0 to 90\deg\ with increasing $\alpha$.

We further investigated the correlation between the relative orientation of \nh-\bperp\  and the \nh\ PDFs by applying a selection criterion that accounts for the potential effect of the polarization background in the \bperp\ that we observed toward the MCs. 
For this purpose, we evaluated the mean \bperp\ orientation angle within and outside the largest closed \nh\ contour, $\left<\psi\right>^{({\rm CC})}$ and $\left<\psi\right>^{({\rm off})}$, respectively.
In principle, MCs where \bperp\ is dominated by the polarization background, here defined as the polarization signal outside of the largest closed \nh\ contour, have relatively low values of $|\left<\psi\right>^{({\rm CC})}-\left<\psi\right>^{({\rm off})}|$. 
This excludes these MCs and minimizes the effect of the background.
A selection of MCs based on $|\left<\psi\right>^{({\rm CC})}-\left<\psi\right>^{({\rm off})}|$ automatically excludes MCs where \bperp\ within and outside the largest closed \nh\ contour are possibly parallel, such as Musca, but it is a sensible initial standard that does not require any further information or modeling of the spatial distribution of dust along the line of sight.
We note that this criterion is more effective and simple than a similar approach based on the evaluation of the polarization fraction ($p$\,$=$\,$P/I$), which is more sensitive to the degeneracy between the orientation of the mean magnetic field with respect to the LOS and the amplitude of the turbulent component of the field.

We selected MCs where $|\left<\psi\right>^{({\rm CC})}-\left<\psi\right>^{({\rm off})}|$\,$>$\,20\deg\ and report their values of $\xi$, $V$, and \meanphi\ in comparison with the values of $\alpha$ on the right-hand side of Fig.~\ref{fig:GeneralTrendsNHPDF}.
With the exception of the L1241 in Cepheus, the selected MCs show a clear trend of decreasing $\xi$ and $V$ values with increasing \nh\ PDF slopes, that is, the MCs with the steepest \nh\ PDF tend to be those with \nh\ and \bperp\ close to perpendicular and those with the shallowest tend to be those with \nh\ and \bperp\ close to parallel. 
This tendency is well represented by the L1489 ($\alpha$\,$\approx$\,$1.5$, \meanphi\,$\approx$\,28\deg\,$\pm$\,26\deg) and the B213/L1495 ($\alpha$\,$\approx$\,$3.1$, \meanphi\,$\approx$\,79\deg\,$\pm$\,20\deg) regions in Taurus, the South ($\alpha$\,$\approx$\,$1.9$, \meanphi\,$\approx$\,29\deg\,$\pm$\,28\deg) and North ($\alpha$\,$\approx$\,$2.6$, \meanphi\,$\approx$\,28\deg\,$\pm$\,26\deg) regions in Perseus, and Orion A ($\alpha$\,$\approx$\,$2.0$, \meanphi\,$\approx$\,74\deg\,$\pm$\,25\deg), L1622 ($\alpha$\,$\approx$\,$2.4$, \meanphi\,$\approx$\,58\deg\,$\pm$\,27\deg), and Orion B ($\alpha$\,$\approx$\,$2.7$, \meanphi\,$\approx$\,44\deg\,$\pm$\,28\deg) regions in Orion.

The results in Fig.~\ref{fig:GeneralTrendsNHPDF} indicate that each cloud as a whole, if we define the cloud as the area within the last closed contour, is not exclusively parallel or perpendicular to \bperp\ but its found in a range of \meanphi\ between 0 to 90\deg.
This constitutes a critical difference with respect to the results of the studies of the cloud elongation, characterized by its autocorrelation, and the mean field direction \citep{li2014}.
Given that the gradient method for characterizing the relative orientations includes the information from the \nh\ orientation and \bperp\ within the last closed contour, it is less sensitive to that particular selection, which is an arbitrary boundary determined by the edges of the observed area.

\subsection{Relative orientation and SFRs}

\begin{figure*}[ht!]
\centerline{
\includegraphics[width=0.5\textwidth,angle=0,origin=c]{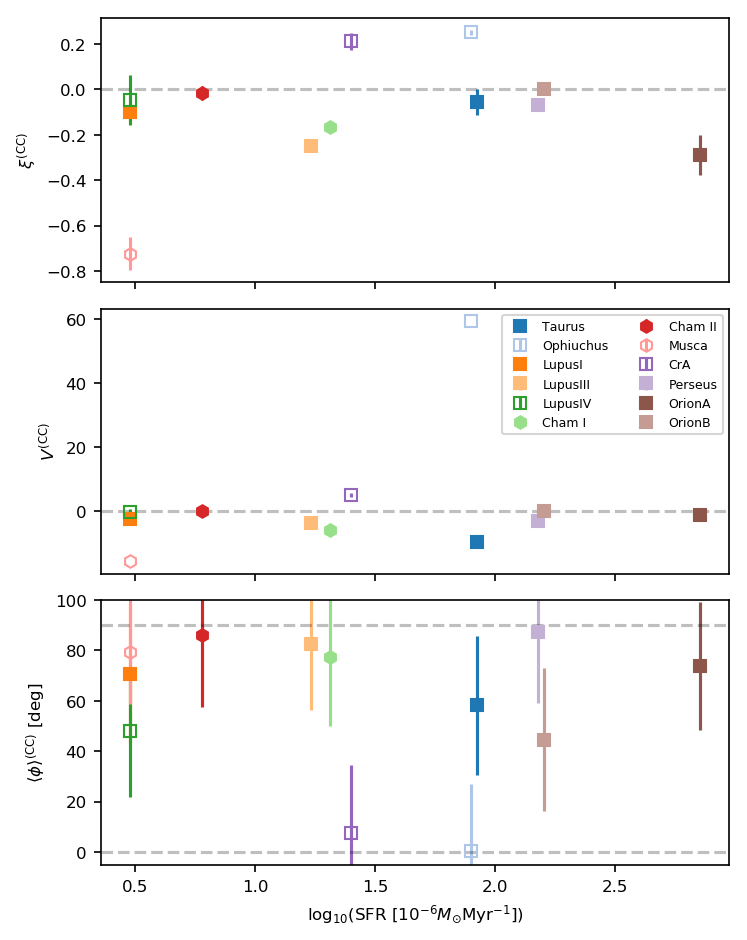}
\includegraphics[width=0.5\textwidth,angle=0,origin=c]{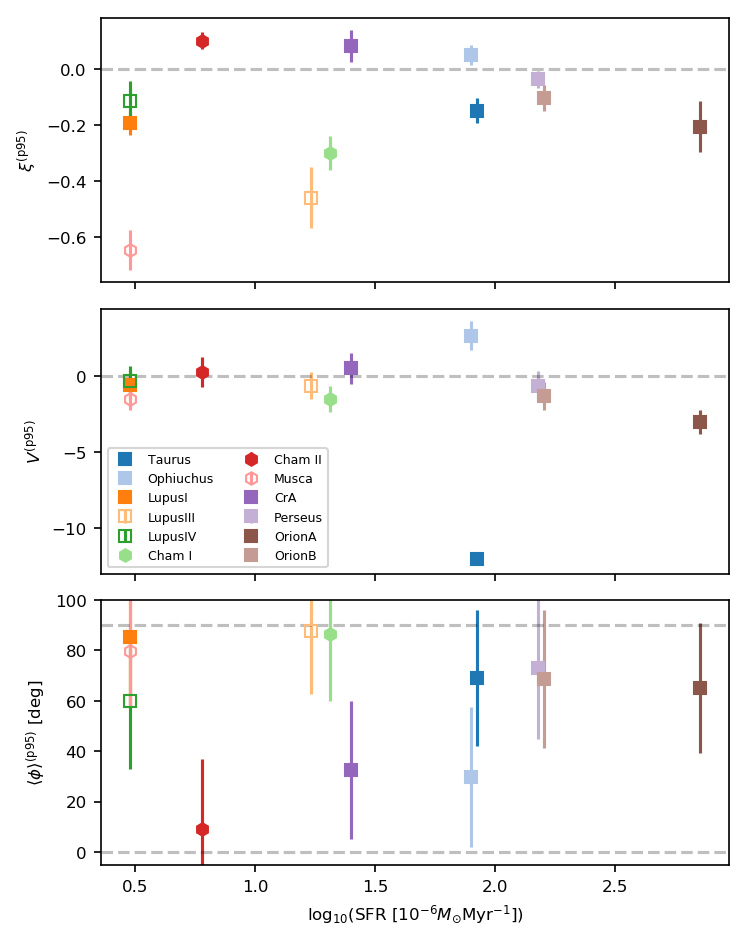}
}
\caption{Relative orientation parameter ($\xi$, top), projected Rayleigh statistic ($V$, middle), and mean relative orientation angle (\meanphi, bottom) within the largest close contours (left) and above the 95th percentile of column density (right) in each region plotted against the SFRs evaluated from the counts of YSOs reported in \cite{lada2010} and \cite{evans2014}, \juan{represented by the squares and hexagons, respectively.}
\juan{The filled and open markers correspond to regions where $|\left<\psi\right>^{({\rm in})}-\left<\psi\right>^{({\rm out})}|$\,$>$\,20\deg\ and $<$\,20\deg, respectively.}
}
\label{fig:GeneralTrendsSFR}
\end{figure*}

We used the reported values of the SFRs evaluated from YSO counts \citep{lada2010,evans2014} and directly compared them to the values of $\xi$, $V$ and \meanphi\ in the regions we studied, as presented in Fig.~\ref{fig:GeneralTrendsSFR}.
We considered two portions of each region to estimate the relative orientation: (i) the last closed contour, and (ii) the contour defined by the 95th percentile of \nh.
The first selection aims to evaluate the \nh-\bperp\ relative orientation in the largest well-delineated object within the limits of the \Herschel\ map.
The second selection aims to evaluate the \nh-\bperp\ relative orientation in the highest \nh\ portion of each \Herschel\ map, which in principle corresponds to the regions that have accumulated matter more effectively and potentially are more prone to star formation. 

When the largest closed contours are considered, we found that the region with the highest SFR per unit of mass (Orion A) shows \meanphi\ close to 90\deg, that is, the \nh\ structures are close to perpendicular to \bperp.
Regions such as Taurus, Ophiuchus, Perseus, and Orion B show relatively similar SFRs, but different relative orientations: \nh\ and \bperp\ are preferentially parallel (Ophiuchus), are preferentially perpendicular (Perseus), or have other angles (Taurus and Orion B).
We also found that some of the regions with the lowest SFRs (Cham II, Musca, and Lupus) show \meanphi\ values close to 90\deg, but another region with low SFR (Lupus IV) presents \meanphi\,$\approx$\,45\deg. 
In sum, however, we do not find an evident correlation between the relative orientation between \nh\ and \bperp\ and the SFRs in the regions we studied.

The lack of correlation between the relative orientation of \nh-\bperp\  and the SFRs is also present in the highest \nh\ portions of the studied clouds, which we selected using the 95th percentile of \nh.
With this selection, most of the regions show \meanphi\ values that are consistent with preferentially perpendicular \nh\ and \bperp\   in clouds with both high and low SFRs.
As expected from the results of the HRO analysis, Cham II shows a change in \meanphi\ from 90\deg to 0\deg, which is an exception to the general trend of  \meanphi\ changing from 0\deg to 90\deg\ with increasing \nh.

\begin{figure*}[ht!]
\centerline{
\includegraphics[width=0.5\textwidth,angle=0,origin=c]{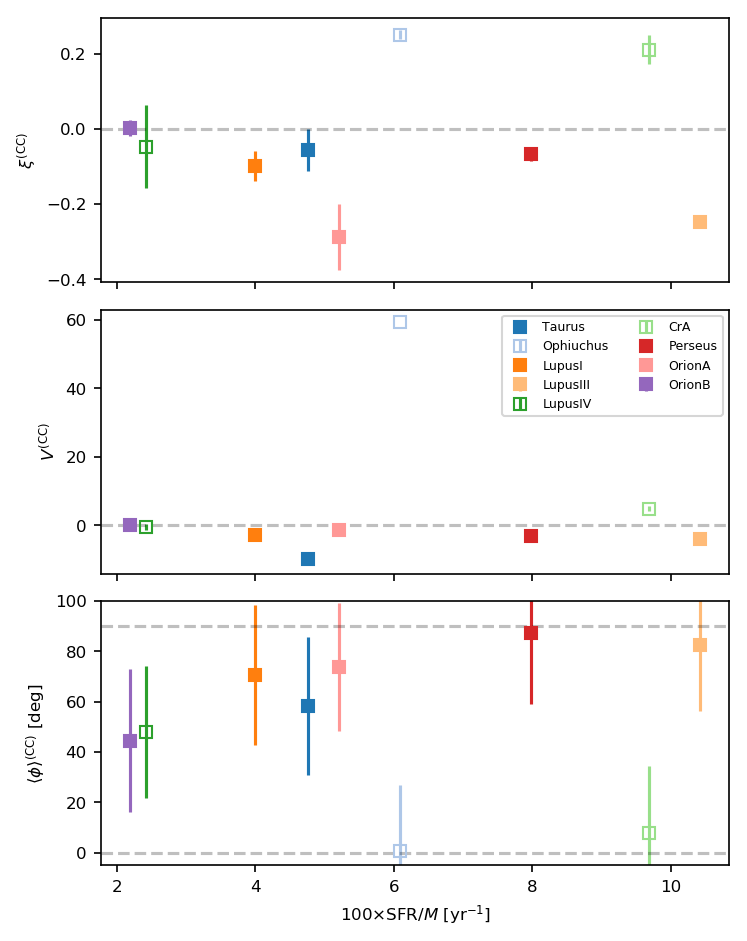}
\includegraphics[width=0.5\textwidth,angle=0,origin=c]{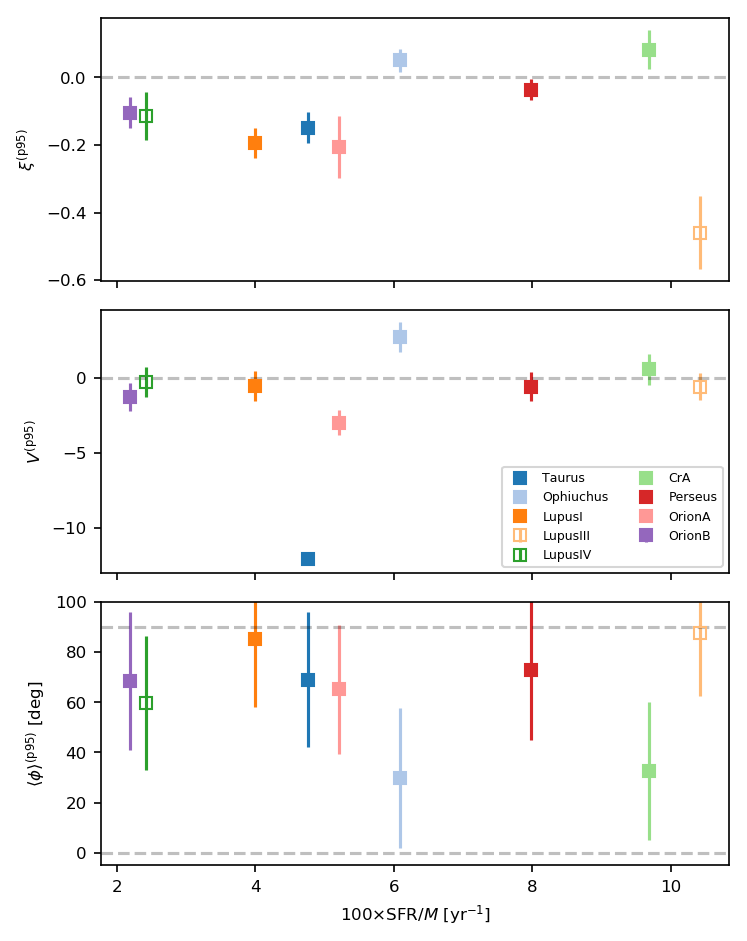}
}
\caption{Relative orientation parameter ($\xi$, top), projected Rayleigh statistic ($V$, middle), and mean relative orientation angle (\meanphi, bottom) within the largest close contours (left) and above the 95th percentile of column density (right) in each region plotted against the ratio of the SFR to cloud mass.
\juan{The filled and open squares correspond to regions where $|\left<\psi\right>^{({\rm in})}-\left<\psi\right>^{({\rm out})}|$\,$>$\,20\deg\ and $<$\,20\deg, respectively.}
}
\label{fig:GeneralTrendsSFRoverM}
\end{figure*}

To facilitate comparison with \cite{li2017}, we also compared the ratio of the SFR to the cloud mass with the values of $\xi$, $V,$ and \meanphi\ in the regions we studied, as presented in Fig.~\ref{fig:GeneralTrendsSFRoverM}. 
Our comparison is based exclusively on the regions reported in \cite{lada2010}, where the masses are all calculated within the $A_{V} > 8$ contours.
This selection does not imply any loss of generality and is made to maintain a consistent definition of the masses using published values.
Figure~\ref{fig:GeneralTrendsSFRoverM} shows that both for the last closed contours and for the 95th percentile of \nh, there is no evidence of a clear relation between the relative orientation of \nh-\bperp\  and the ratio of the SFR to cloud mass.

\section{Discussion}\label{section:discussion}

\subsection{General trends in the relative orientation of \nh\ and \bperp}\label{subsection:generaltrends}

If we can point to a common denominator in this HRO study, it is that there is a general trend for $\xi$ and $V$ to decrease and \meanphi\ increase from 0\deg\ with increasing \nh, illustrated in Fig.~\ref{fig:HOGpanel1}, \ref{fig:HOGpanel2}, \ref{fig:HOGpanel3}, and \ref{fig:HOGpanel4}. 
There are two exceptions that are worth mentioning, however: Cham II (leftmost panel of Fig.~\ref{fig:HOGpanel2}) and Perseus South (leftmost panel of Fig.~\ref{fig:HOGpanel3}), where the highest \nh\ bins show values of $V$ that are less negative and \meanphi\ lower than in the preceding \nh\ bins. 
There are observational reasons, for example, the orientation with respect to the LOS, and physical conditions, for example, the tangling of the magnetic field at higher densities, that can be used to explain both of these exceptions, as we further discuss in Sec.~\ref{section:discussion}.
Nevertheless, it is still remarkable that such a heterogeneous set of MCs for the most part show avsimilar behavior in the relative orientation between the 36\arcsec-scale \nh\ structures and the 10\arcmin-scale \bperp.

The change in $V$ from positive to negative and \meanphi\ from 0 to 90\deg\ is particularly clear in some of the objects that have previously been identified as archetypes of the paradigm of less dense and magnetically-aligned filaments feeding dense filaments perpendicular to the magnetic field \citep{andre2014}, for example, B213 in Taurus, Musca, and Serpens Main 2 in Aquila Rift.
Here, we have identified that this trend is generally followed in the studied MCs using a method that does not require the identification of filaments.
The column densities where the transition from mostly parallel to mostly perpendicular is found are close to those where the change of the slopes in the \nh\ distributions \citep{kainulainen2009} and the scaling of the LOS magnetic field (\bpara) with increasing \nh, inferred from the Zeeman observations, have been reported previously \citep{crutcher2012}.
In clouds where the mostly parallel portion of the general trend is missing, for example, Musca and IC5146, we have shown that the increase in angular resolution obtained when the \nh\ maps are replaced with the higher resolution 250\micron\ Herschel observations recovers the mostly parallel to mostly perpendicular \nh\ and \bperp\ trend, as shown in Appendix~\ref{app:striationsTest}.

\subsection{Density structure within MCs anchored in the cloud-scale magnetic fields}

We first observe about the relative orientation trends between the \nh\ structures and \bperp, reported in Figs.~\ref{fig:HOGpanel1} to \ref{fig:HOGpanel4}  that the trend in the relative orientation between \nh\ and \bperp\ found in \PlanckXXXV\ is also found in many of the subregions observed by \Herschel.
Notable examples are the B213 region in Taurus, L1688 in Ophiuchus, Lupus I and III, Chamaeleon I, Serpens Main 2, Perseus North, Orion A and B, and L1622, all of which show a progressive change in relative orientation with increasing \nh\, from mostly parallel to mostly perpendicular.
However, the interpretation of these results is different from that in \PlanckXXXV.

In this case, the relation corresponds to the coupling between different scales, where the orientation of the \nh\ structures bears a relation to \bperp\ on a physical scale that is an order of magnitude larger. 
Insofar as this correlation exists, there is a strong argument that the magnetic field plays a significant role in the distribution of the density structures inside MCs.
It suggests that the strong magnetization in the diffuse gas, inferred from the Zeeman H{\sc i} observations \citep{heilesANDcrutcher2005} and the anisotropy of H{\sc i} structures with respect to \bperp\ \citep{clark2014,clark2019}, also affects the density structure of these MCs.
If the matter and the magnetic fields are strongly coupled, as is most likely the case in the diffuse ISM, it is unlikely that the accumulation of the parcels of gas that eventually become MCs is made perpendicular to the magnetic field because of the effect of Lorentz force \cite[see discussion in][]{solerANDhennebelle2017}.
Numerical simulations of MC formation indeed indicate that the magnetic field delays the formation of dense molecular gas by imposing an anisotropy in the direction of the mass flows \citep{ntormousi2017,girichidis2018}.
It is therefore not surprising that we find the densest parts of the MC perpendicular to the magnetic field.

The magnetic fields are not strong enough to fully determine the orientation of the \nh\ structures within some MCs, however.
Inevitably, gravity may dominate the magnetic fields at least in portions of the MCs, as is evident by the presence of prestellar objects, either by accretion of matter from the larger scales or by the loss of magnetic flux by a diffusion process, such as ambipolar difussion or magnetic reconnection.
It is tempting to predict that the trends of the relative orientation of \nh-\bperp\ we reported here extend to the structures at smaller scales, but we find at least two examples of regions where the \nh\ structures and \bperp\ are mostly perpendicular at high \nh\ but less perpendicular at the highest \nh: Cham II and the southern portion of Perseus.
These examples further confirm that the relative orientation is not necessarily inherited from the largest scales and  is subject to change depending on the local conditions, as observed in \cite{zhang2014}.

Numerical simulations of MHD turbulence indicate that only a very strong magnetic field can preserve the relative orientation with respect to the large-scale field from cloud scales (tens of parsecs) to disk scales and to an hourglass-shaped field below 1000\,au scales \citep{hull2017}.
Just as the parsec-scale magnetic field is not directly inherited from the Galactic magnetic field, as is evident in the Orion-Eridanus superbubble \citep{soler2018}, it is premature to extrapolate the relative orientation between \nh\ structures and \bperp\ at one particular scale to a smaller scale.

\subsection{Relation of the relative orientation to the distribution of \nh\ within MCs}\label{sec:DiscussionNHPDFs}

After the selection based on the polarization orientation angle, it is clear that the clouds with the shallowest \nh\ PDFs appear to be those where \meanphi\ is closer to 0\deg, while the clouds with the steepest \nh\ PDF tails appear to be those where \meanphi\ is closer to 90\deg, as illustrated in Fig.~\ref{fig:GeneralTrendsNHPDF}.
This trend is exactly opposite to what would be expected for the gravitational collapse of a magnetized MC.
If the magnetic field is dynamically important, a shallow \nh\ PDF tail, which is interpreted as the effect of gravitational collapse, should appear after the accumulation of matter along the field lines, whose final product are density structures that are mostly perpendicular to the field.
By the same token, a steep \nh\ PDF tail, which is associated with regions that are dominated by turbulence rather than by gravitational collapse, should correspond to relatively early stages of the accumulation of matter along the field lines, when density structures tend to be mostly parallel to the field, where the matter flows are restricted by the Lorentz force.
However, these expectations are based on a laminar description of a fluid that is eminently turbulent.

Numerical simulations of MHD turbulence indicate that an increasing magnetic field strength results in narrower density PDFs with steeper tails \citep{collins2012,molina2012}.
More recent studies of MHD turbulence in self-gravitating clouds indicate that the slope of the \nh\ PDF tail is significantly steeper in magnetically subcritical clouds than in supercritical clouds \citep{auddy2018}.
These numerical experiments suggest that the magnetic fields act as a density cushion in turbulent gas, shaping the cloud by preventing the gas from reaching very high densities. 
Thus, it is plausible that the regions where \nh\ is mostly perpendicular to \bperp\ and the \nh\ PDFs are the steepest correspond to magnetically subcritical clouds. 

The progressive increase in the values of \meanphi (decrease in $\xi$ and $V$) with increasing \nh\ PDF tail slope is singular.
This behavior, which is not exclusive of the last-closed-contour selection, as illustrated in Appendix~\ref{app:nhpdf}, suggests that the gravitational collapse and the turbulence are not exclusively responsible for the distribution of \nh\ and that the magnetic field plays an active role in structuring these MCs.
Detailed studies of this trend, using MHD simulations of MCs with different magnetizations and turbulence levels at different evolutionary stages, are beyond the scope of this work, but are instrumental for characterizing the broad range of physical conditions that are responsible for the trends in Fig.~\ref{fig:GeneralTrendsNHPDF}.

\subsection{Relation of the relative orientation to star formation}\label{sec:DiscussionStarFormation}

\begin{figure*}[ht!]
\centerline{
\includegraphics[width=1.0\textwidth,angle=0,origin=c]{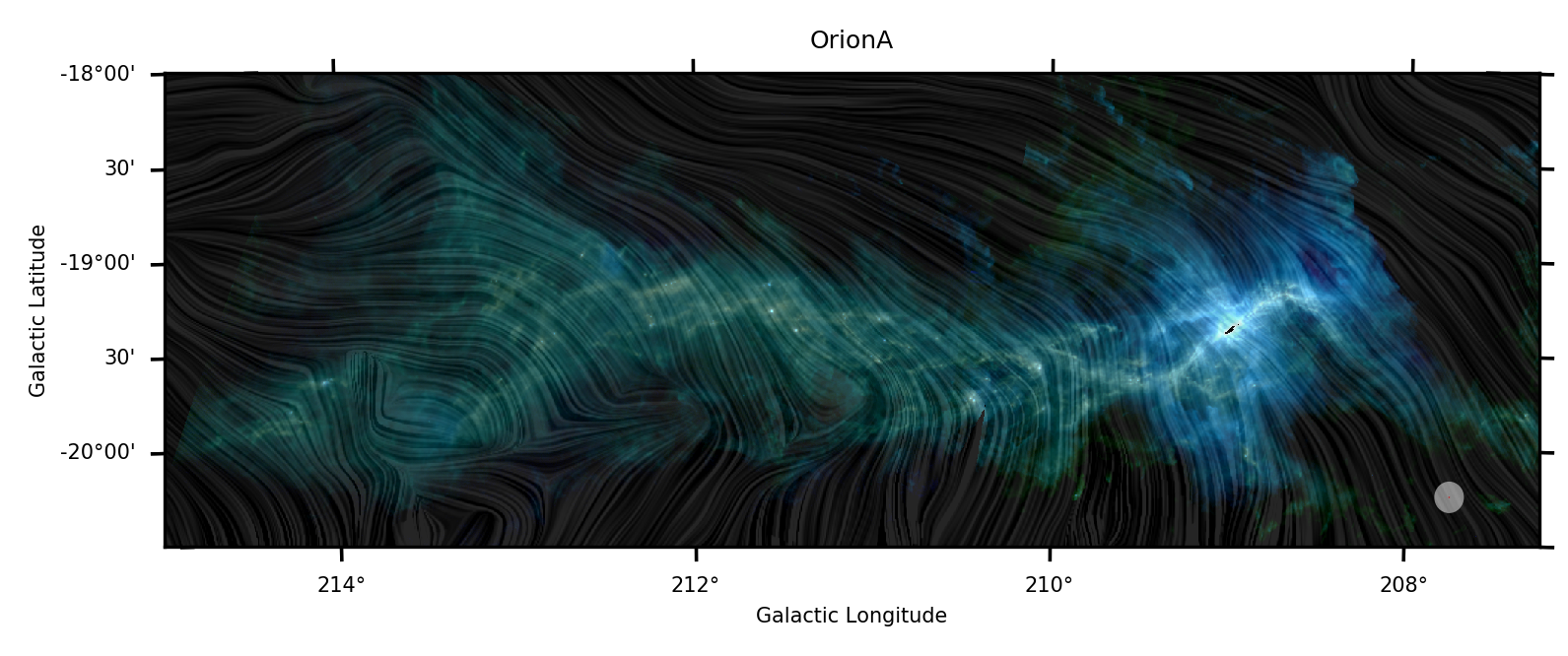}
}
\caption{Emission toward the Orion~A region observed by \Herschel\ at 160 (blue), 250 (green), and 500\micron\ (red) and \bperp\ (drapery pattern) inferred from the \Planck\ 353\,GHz polarization observations.
The gray disk represents the size of the \Planck\ beam.
The red dot represents the size of the \Herschel\ 500\micron\ beam.
}
\label{fig:OrionMultifreq}
\end{figure*}

The correlation between the relative orientations of \nh-\bperp\  and the \nh\ distributions makes it tempting to extrapolate this trend to the scales of star formation.
However, the lack of correlation between the relative orientations of \nh\ and \bperp\  and the SFRs, illustrated in Fig.~\ref{fig:GeneralTrendsSFR} and Fig.~\ref{fig:GeneralTrendsSFRoverM}, is not entirely unexpected.
Without an unusually strong magnetic field, the coupling between the matter and the fields at 0.026\,pc scales is on the one hand subject to a broad range of physical conditions that affect star formation at smaller scales, for example, protostellar outflows, turbulence induced by the gravitational collapse, or simply a change in the balance of magnetic to gravitational forces.
On the other hand, the observed clouds are not all observed in the same period of the evolution and the SFR derived from the YSO counts only offer one snapshot of their evolution.
It is possible that a cloud such as Musca or Lupus I has been formed by flows along the magnetic field and has effectively accumulated matter perpendicular to the field but is in an early stage of its evolution, thus showing only a few YSOs.
It is also possible that one of these clouds was unable to accumulate enough matter from its surrounding to globally form as many stars as Orion~A, for instance, and is in a late stage of its evolution, confined by pressure and gravity, but supported by the magnetic field against gravitational collapse.
Even the eastern and western end of Orion~A present the same relative orientation trends, but their star formation activity is very different, as illustrated in the rightmost panel of Fig.~\ref{fig:HOGpanel4} or directly in Fig.~\ref{fig:OrionMultifreq}.

These results call into question the interpretation of other analyses that are premised on the idea that the density structures are anchored in the large-scale magnetic field, which determines the elongation of the MCs, and sets their star formation activity.
Based on this expectation, \cite{li2017} proposed that magnetic fields are a primary regulator of the SFRs.
We found, however, that the densest portions of the regions with highest SFRs are mostly perpendicular to \bperp,\ as are the densest portions of some of the regions with the lowest SFRs, as illustrated in Fig.~\ref{fig:GeneralTrendsSFR}.
This lack of correlation with the relative orientation of \nh-\bperp\ is also found in the ratios of the SFR to cloud mass presented in Fig.~\ref{fig:GeneralTrendsSFRoverM}.

The main difference between our results and those reported in \cite{li2017} lies in the definition of the relative orientation. 
In \cite{li2017}, the authors define a single structre elongation in \nh\ , estimated using the \nh\ correlation function, and a single mean \bperp by averaging the polarization observations.
In the HRO, each cloud is characterized by the \nh\ contours, which are compared to the local \bperp\ orientation. This accounts for the internal \nh\ structure of the regions and the fact that they may not correspond to a single object anchored by the large-scale field.
In this sense, the HRO provides greater statistical significance and a more complete description of the \nh\ structure.

Although the results of our analysis contradict the conclusions of \cite{li2017}, they do not discard the importance of the magnetic fields in the star formation process.
Magnetic fields can globally reduce the SFRs in MCs by controlling the accumulation of the material that is available for gravitational collapse and by their influence on the stellar feedback that disperses this material back into the ISM \citep[see, e.g., the discussions in][]{maclow2017,rahner2019}.
Further studies of the dynamics of nearby MCs and the magnetic fields in more distant MCs are necessary, however, to find conclusive evidence of these effects.

\subsection{Projection onto the plane of the sky}

The observed quantities \nh\ and \bperp\ are the result of the integration along the LOS and the projection onto the plane of the sky. 
The observed relative orientation may therefore not reflect the orientation of the density structures and the 3D magnetic field (\bvec).
One of the main consequences of projection is that two vectors that are parallel in 3D are almost always projected parallel in 2D, but two perpendicular vectors in 3D are only projected as perpendicular if they are in a plane that is almost parallel to the plane of the sky, as discussed in detail in Appendix C of \PlanckXXXV\ and section 4.1 of \cite{zhang2014}.
This would imply that if the density and \bvec\ have no preferred orientations in 3D, it is very unlikely to observe an elongated \nh\ structure that is almost perpendicular to \bperp.
 And yet we see clear examples of this configuration in Taurus, Lupus~I, Musca, IC5146, and Orion~A.
 
However, it is clear that the density and \bvec\ are not entirely isotropic in the vicinity of the Sun, and toward the nearby clouds we may be observing the \nh\ and \bperp\ configuration in the walls of the cavity of hot ionized gas that extends all around the Sun, known as the Local Bubble \citep[see][and references therein]{alves2018}, the expanding supernova remnant known as the Orion-Eridanus superbubble \citep[see][and references therein]{soler2018}, and other nearby superbubbles \citep[see, e.g.,][]{bally2008}.
These magnetized superbubbles would generate anisotropies in the mean direction of the magnetic field (\meanbvec) that would favor the observations of \bvec\ close to parallel to the plane of the sky. 
This would explain the prevalence of dense \nh\ structures perpendicular to \bperp.
Demonstrating this hypothesis is beyond the scope of this work, however. 

The observations toward regions where \meanbvec\ is close to parallel to the LOS should show higher dispersions of \bperp\ than regions where \meanbvec\ is close to parallel to the plane of the sky \citep[see, e.g.,][]{chen2019}.
However, given that the dispersion of \bperp\ can also be the result of the tangling of the field caused by turbulence, it is difficult to determine whether the difference in the relative orientation trends between the clouds presented in this study are exclusively the result of the mean field orientation with respect to the LOS \citep[see][for a discussion of this effect]{planck2016-XLIV}.
Comparisons of \bperp\ inferred from starlight polarization and from the \Planck\ 353\,GHz observations suggest that in regions like Taurus and Musca, \bvec\ is mostly in the plane of the sky, while in regions like Lupus I and the Pipe Nebula,  \meanbvec\ is either oriented close to the LOS or high levels of field tangling caused by turbulence \citep{soler2016}.
Breaking this degeneracy between turbulence and the orientation of \meanbvec\ would require additional information, such as that provided by Zeeman observations and Faraday rotation \citep{tahani2018}. 
This is the subject of forthcoming works \citep[][Sullivan et al. in preparation]{tahani2019submitted}.

\section{Conclusions}\label{section:conclusions}

We presented a study of the relative orientation between the column density structures, derived from the \Herschel\ observations at 36\parcs0 resolution, and the plane-of-the-sky magnetic field orientation, inferred from the \Planck\ linear polarization observations at 10\arcmin\ resolution, toward portions of ten nearby molecular clouds. 
This is an extension of the analysis presented in \cite{planck2015-XXXV}.

Our study indicates that the relative orientation trends found in \PlanckXXXV, that is, that \nh\ and \bperp\ change progressively from mostly parallel to mostly perpendicular with increasing \nh, globally persist in the \nh\ structures that are observed by \Herschel\ at higher angular resolution.
The transition from mostly parallel to mostly perpendicular is particularly significant toward a set of subregions that includes B213 in Taurus, L1688 in Ophiuchus, Lupus~I, Chamaeleon I, Musca, the northern portion of CrA, Serpens and Serpens Main 2 in the Aquila Rift, Perseus, IC5146, Orion~A, Orion~B, and L1622.
This common behavior is notable given that the clouds we considered represent a relatively heterogeneous set of physical conditions and environments in the vicinity of the Sun. 
This provides observational evidence of the coupling between the cloud-scale magnetic fields and the \nh\ structure within the MCs.

We also studied the relation between the relative orientation of \nh-\bperp\  and the distribution of \nh\ within the MCs, which is characterized by the \nh\ PDFs within the last closed \nh\ contours.
We found that the regions with the steepest \nh\ PDF tails are those where \nh\ and \bperp\ are close to perpendicular.
We also found that regions with shallower PDF tails show a broad range of global relative orientations of \nh\ and \bperp\ , with the shallowest being those regions where \nh\ and \bperp\ are closer to parallel.
These results provide observational evidence of the effects of the magnetic fields in the \nh\ distribution based on MHD simulations \citep[see, e.g.,][]{collins2012,molina2012,auddy2018}.
However, further numerical studies are necessary to determine the physical mechanisms that shape these nearby MCs and properly account for observational effects, such as the mean orientation of the field with respect to the LOS.

Finally, we investigated the relation between the relative orientation of \nh-\bperp\ and the SFRs and the SFR to cloud-mass ratios and found no significant trends.
These results directly disagree with the studies that link the MC elongation and its orientation with respect to the mean orientation of \bperp\ with the aforemention quantities \citep{li2017}.
Our results do not exclude the magnetic field as an important agent in setting the SFRs, however, but call for further studies of their role in the dynamics of the gas reservoirs of these and other star-forming \juan{regions} and its influence on the dispersion of that reservoir through stellar feedback.

We presented a study that uses the most recent estimates of \nh\ and \bperp\ toward nearby MCs that have become available through observations made with the ESA \Planck\ and \Herschel\ satellites.
Still, our conclusions are limited by the difference in angular resolution between the intensity and polarization observations.
This gap is expected to be bridged in the future by  balloon-borne experiments, such as the Next-Generation Balloon-borne Large Aperture Submillimeter Telescope \citep[BLAST-TNG,][]{dober2014}, and proposed satellite missions, such as the Space Infrared Telescope for Cosmology and Astronomy (SPICA) satellite \citep{SPICA2019}, which will yield extended polarization maps that will reveal further details of the magnetic fields within the most nearby regions of star formation.

\bibliographystyle{aa}
\bibliography{35779arxiv.bbl}

\begin{acknowledgements}
JDS acknowledges funding from the European Research Council under the Horizon 2020 Framework Program via the ERC Consolidator Grant CSF-648505. 
Part of the discussions that lead to this work took part under the program Milky-Way-Gaia of the PSI2 project funded by the IDEX Paris-Saclay, ANR-11-IDEX-0003-02.
JDS thanks the various people who helped with their encouragement and conversation; notably, Henrik Beuther, Andrea Bracco, Fran\c{c}ois Boulanger, Laura Fissel, Patrick Hennebelle, Mordecai MacLow, Peter Martin, Marc-Antoine Miville-Desch\^{e}nes, and Douglas Scott.
JDS also thanks the anonymous reviewer for the thorough review and highly appreciates the comments and suggestions that significantly contributed to improving the quality of this work.

The development of \Planck\ has been supported by: ESA; CNES and CNRS/INSU-IN2P3-INP (France); ASI, CNR, and INAF (Italy); NASA and DoE (USA); STFC and UKSA (UK); CSIC, MICINN, JA, and RES (Spain); Tekes, AoF, and CSC (Finland); DLR and MPG (Germany); CSA (Canada); DTU Space (Denmark); SER/SSO (Switzerland); RCN (Norway); SFI (Ireland); FCT/MCTES (Portugal); and PRACE (EU).

\Herschel\ SPIRE has been developed by a consortium of institutes led by Cardiff Univ. (UK) and including Univ. Lethbridge (Canada); NAOC (China); CEA, LAM (France); IFSI, Univ. Padua (Italy); IAC (Spain); Stockholm Observatory (Sweden); Imperial College London, RAL, UCL-MSSL, UKATC, Univ. Sussex (UK); Caltech, JPL, NHSC, Univ. Colorado (USA). 
This development has been supported by national funding agencies: CSA (Canada); NAOC (China); CEA, CNES, CNRS (France); ASI (Italy); MCINN (Spain); SNSB (Sweden); STFC
(UK); and NASA (USA). 

\Herschel\ PACS has been developed by a consortium of institutes led by MPE (Germany) and including UVIE (Austria); KUL, CSL, IMEC (Belgium); CEA, OAMP (France); MPIA (Germany); IFSI, OAP/AOT, OAA/CAISMI, LENS, SISSA (Italy); IAC (Spain). 
This development has been supported by the funding agencies BMVIT (Austria), ESA-PRODEX (Belgium), CEA/CNES (France), DLR (Germany), ASI (Italy), and CICT/MCT (Spain).

This research has made use of data from the \Herschel\ Gould Belt survey (HGBS) project.
The HGBS is a \Herschel\ Key Programme jointly carried out by SPIRE Specialist Astronomy Group 3 (SAG 3), scientists of several institutes in the PACS Consortium (CEA Saclay, INAF-IFSI Rome and INAF-Arcetri, KU Leuven, MPIA Heidelberg), and scientists of the \Herschel\ Science Center (HSC).

\end{acknowledgements}

\appendix

\section{Histograms of the relative orientation}\label{app:relativeorientation}

As its name indicates, the method histograms of relative orientation (HRO) characterizes the distribution of the relative orientation angles $\phi$ as defined in Eq.~\eqref{eq:phi}.
For the sake of discussion,  we have reported the trends in the HROs in the main text using the derived quantities that describe the shape of the distribution, as is the case of $\xi$ and $V$, or its average values, as is the case of \meanphi.
Here we present some representation HROs that correspond to the lowest, one intermediate, and to the highest column density (\nh) bins with an equal number of $\phi$ measurements.
These \nh\ bins are derived from the inversion of the \nh\ distribution and aim to have comparable error bars. 

\begin{figure*}[ht!]
\centerline{
\includegraphics[width=0.3\textwidth,angle=0,origin=c]{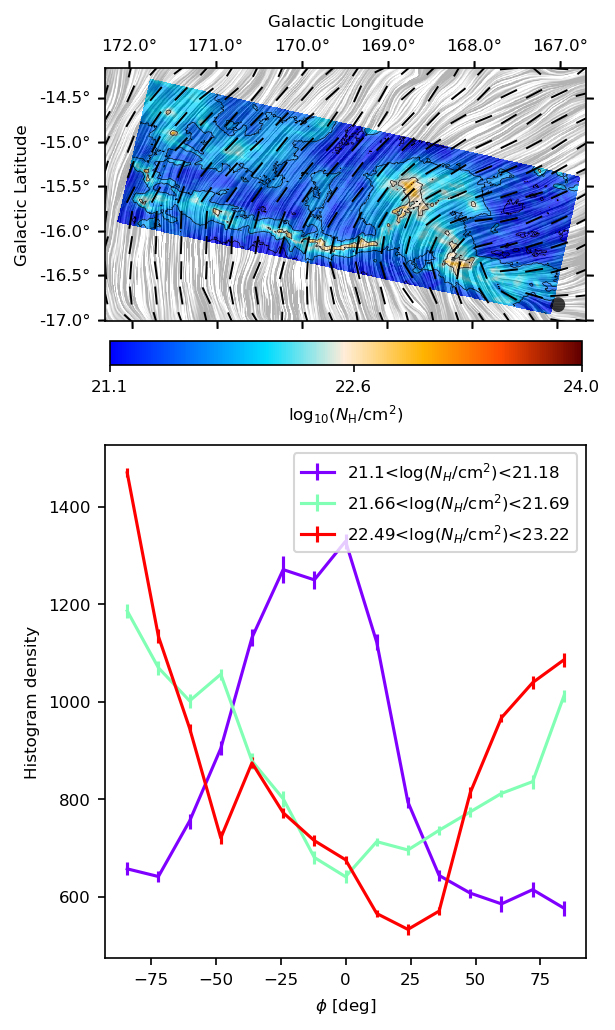}
\includegraphics[width=0.3\textwidth,angle=0,origin=c]{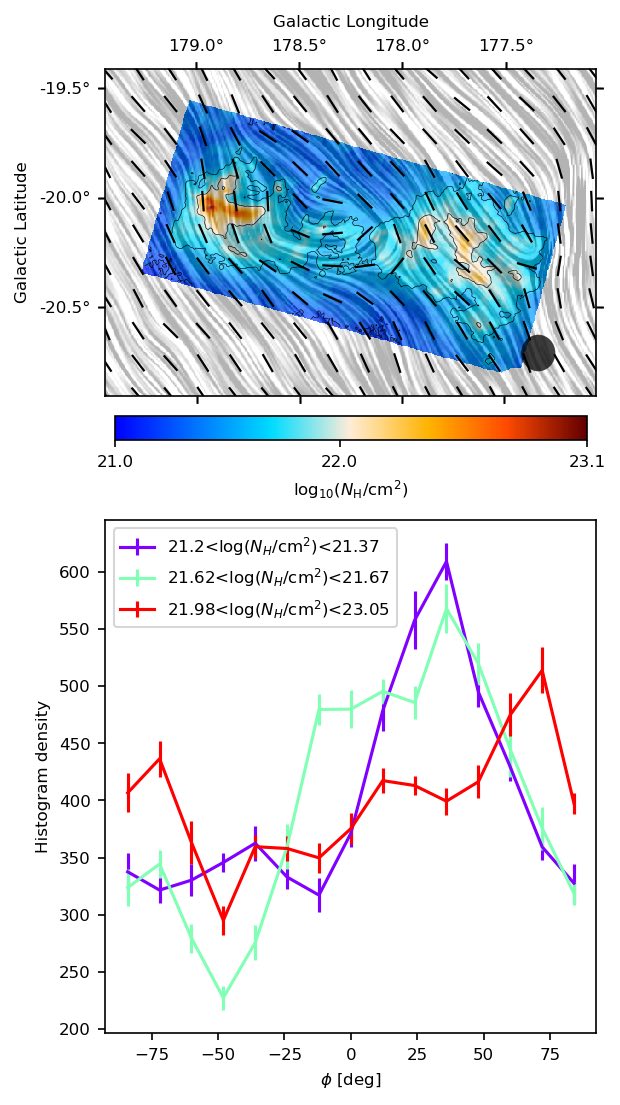}
\includegraphics[width=0.3\textwidth,angle=0,origin=c]{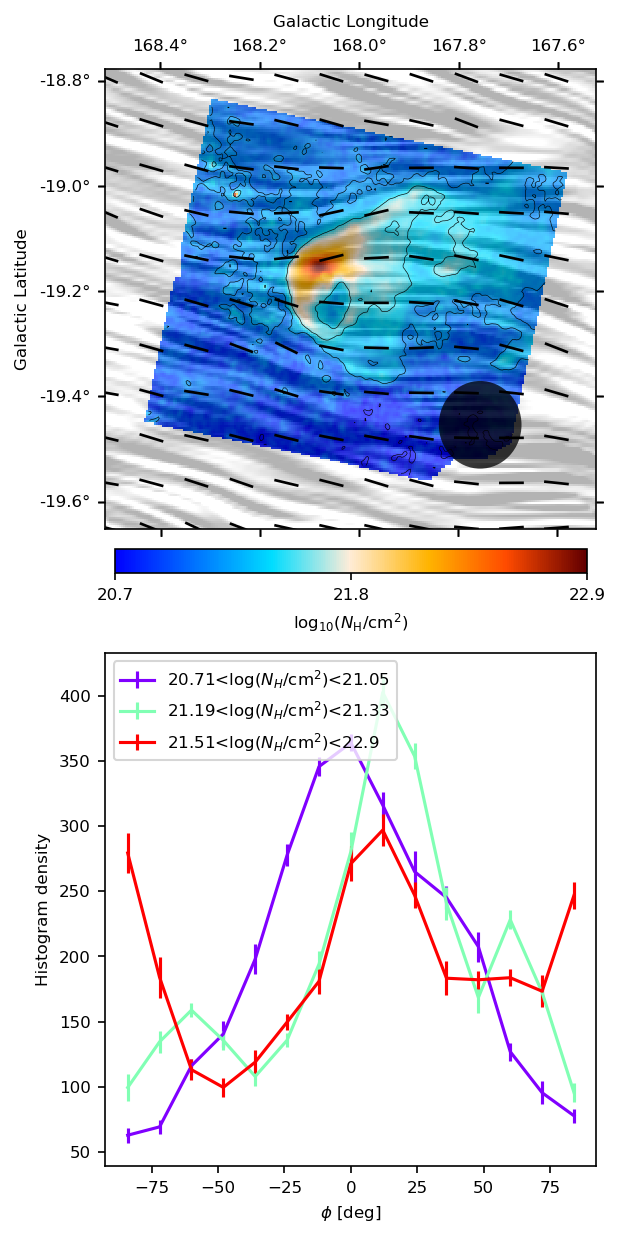}
}
\vspace{-2.0mm}
\caption{\emph{Top}. Column density (\nh) and magnetic field (\bperp) toward the Taurus subregions B213/L1495 (left), L1551 (center), and L1489 (right). 
The colors represent \nh\ derived from the \Herschel\ data and \Planck\ observations.
The drapery pattern represents \bperp, as inferred from the \Planck\ 353\,GHz observations.
The black circle represents the size of the \Planck\ beam.
\emph{Bottom.} Histograms of relative orientations (HROs) for three representative \nh\ bins.}
\label{fig:TaurusHROs}
\end{figure*}

\begin{figure*}[ht!]
\centerline{
\includegraphics[width=0.3\textwidth,angle=0,origin=c]{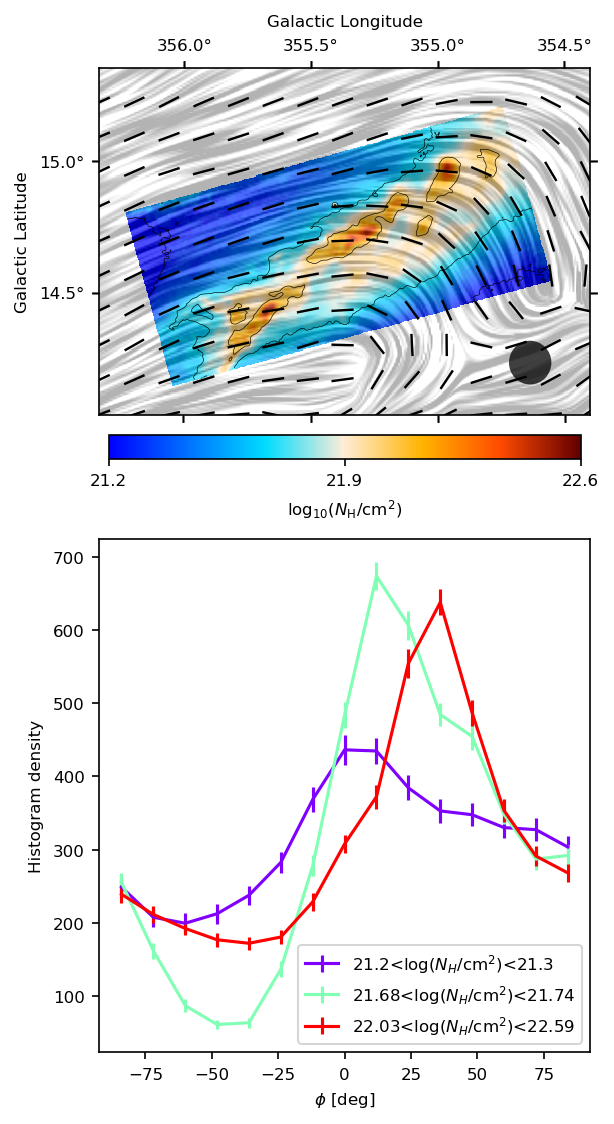}
\includegraphics[width=0.3\textwidth,angle=0,origin=c]{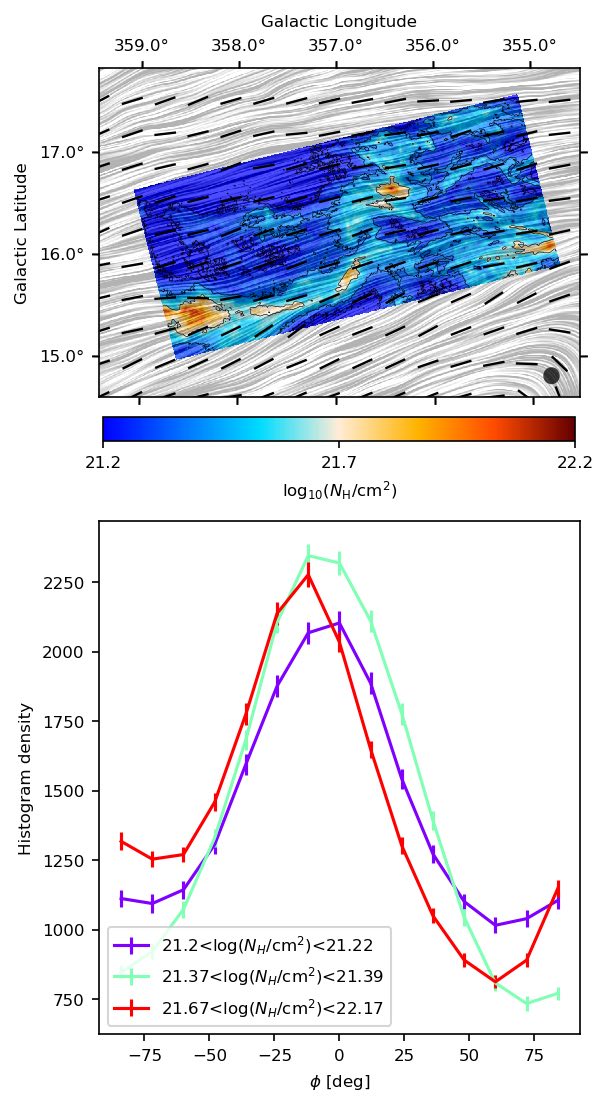}
\includegraphics[width=0.3\textwidth,angle=0,origin=c]{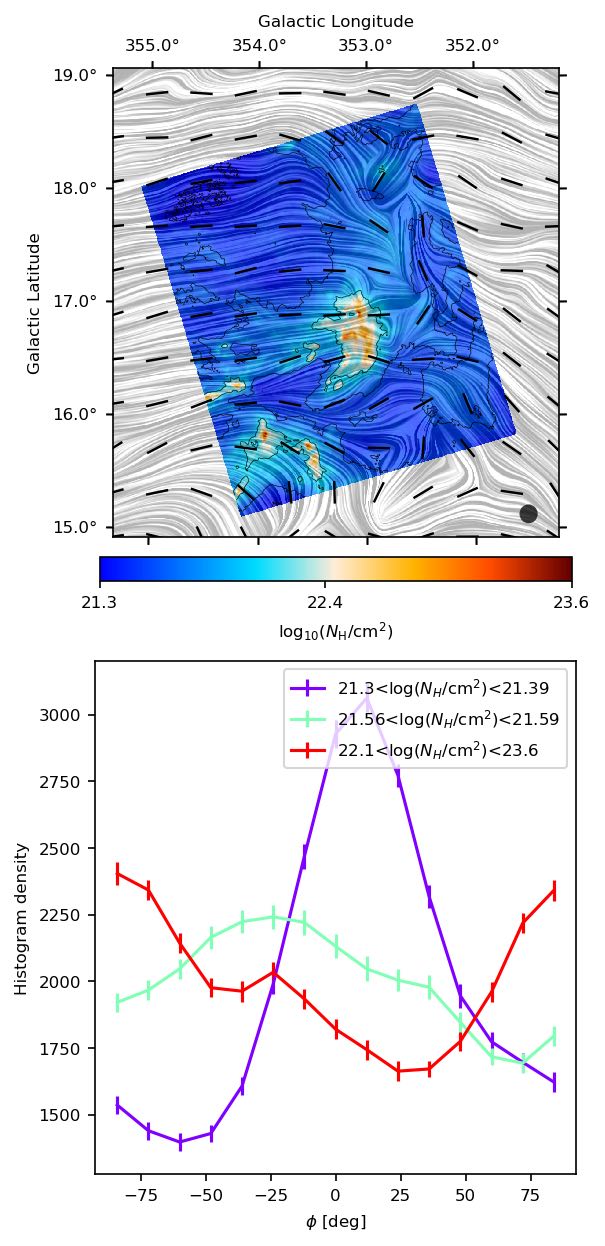}
}
\vspace{-2.0mm}
\caption{Same as Fig.~\ref{fig:TaurusHROs} for the Ophiuchus subregions L1712 (left), L1704/L1409 (center), and L1688 (right).}
\label{fig:OphiuchusHRO}
\end{figure*}

\begin{figure*}[ht!]
\centerline{
\includegraphics[width=0.3\textwidth,angle=0,origin=c]{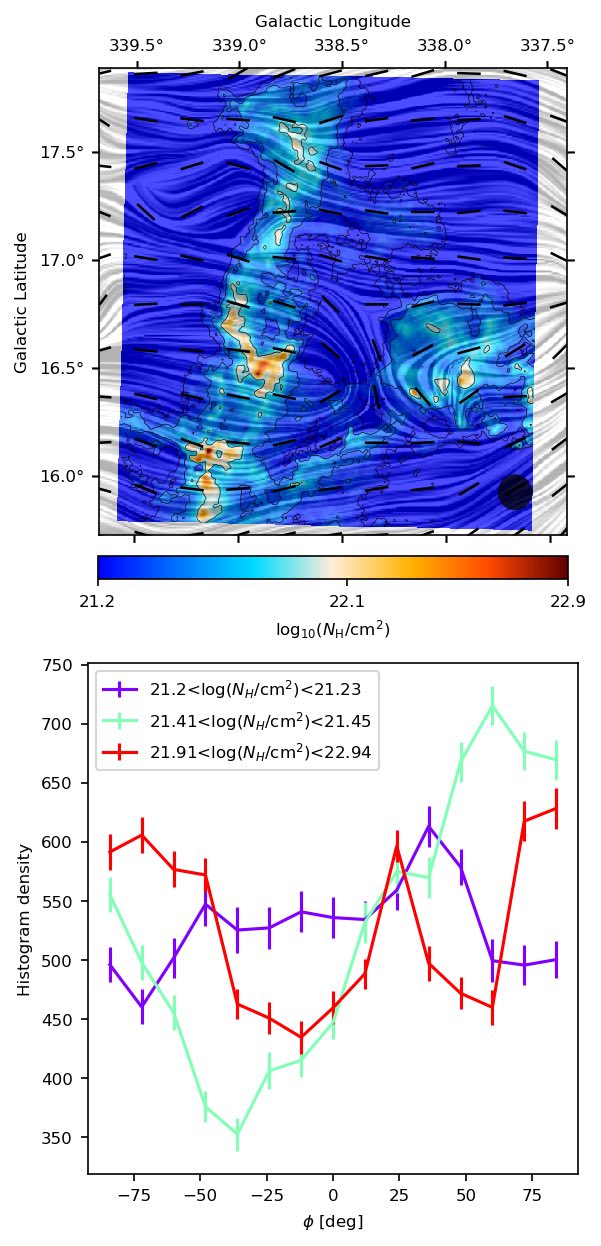}
\includegraphics[width=0.3\textwidth,angle=0,origin=c]{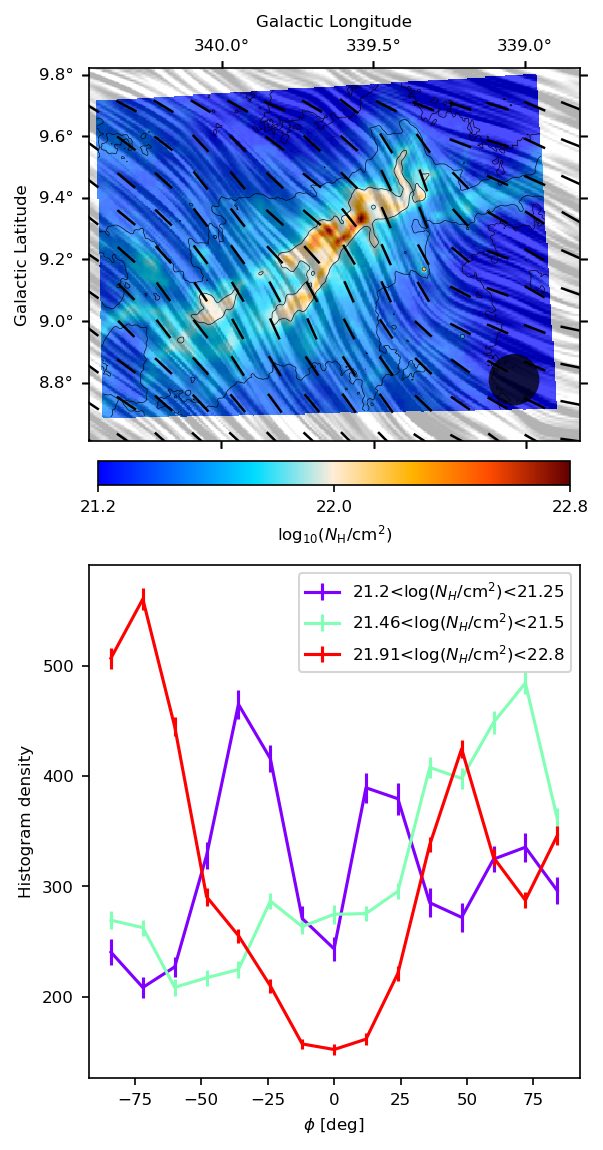}
\includegraphics[width=0.3\textwidth,angle=0,origin=c]{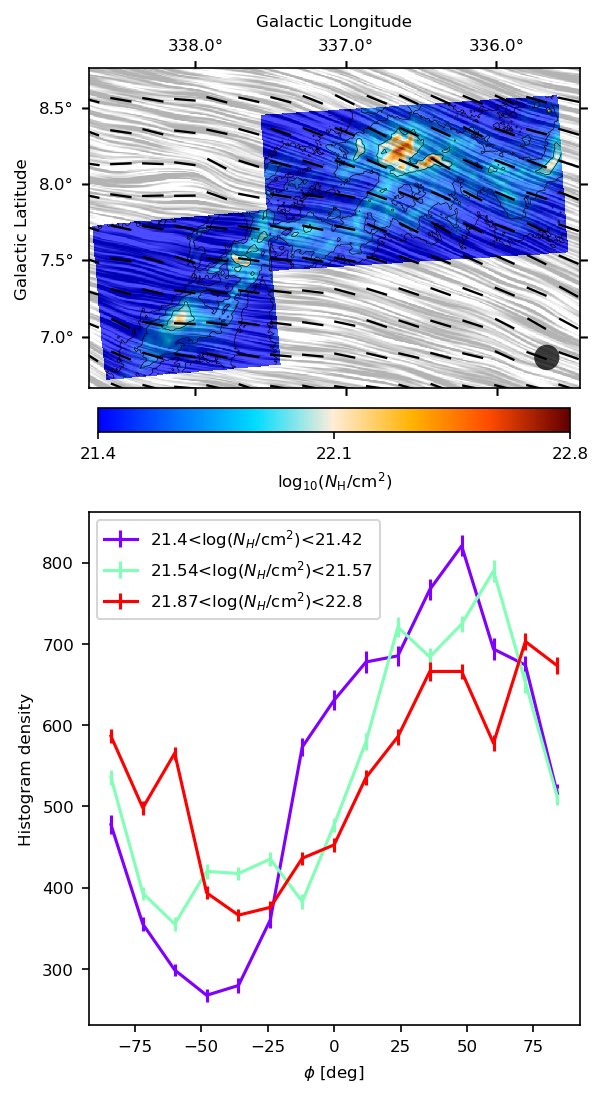}
}
\vspace{-2.0mm}
\caption{Same as Fig.~\ref{fig:TaurusHROs} for the Lupus\,I (left), Lupus\,III (center), and Lupus\,IV (right) subregions.}
\label{fig:LupusMaps}
\end{figure*}

\begin{figure*}[ht!]
\centerline{
\includegraphics[width=0.3\textwidth,angle=0,origin=c]{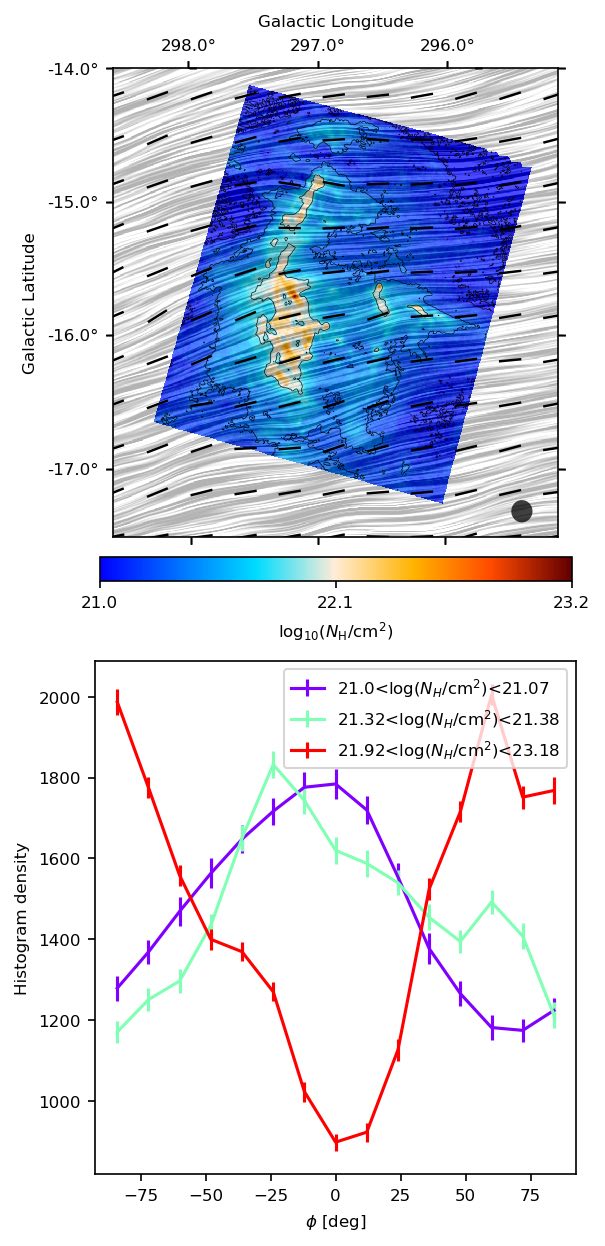}
\includegraphics[width=0.3\textwidth,angle=0,origin=c]{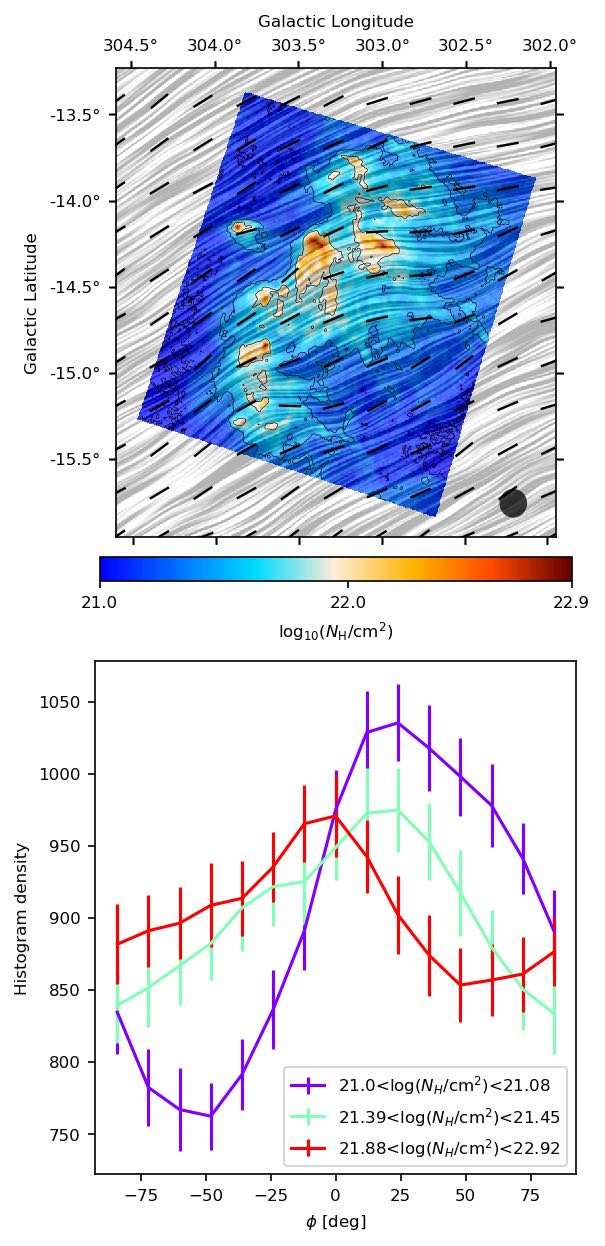}
\includegraphics[width=0.3\textwidth,angle=0,origin=c]{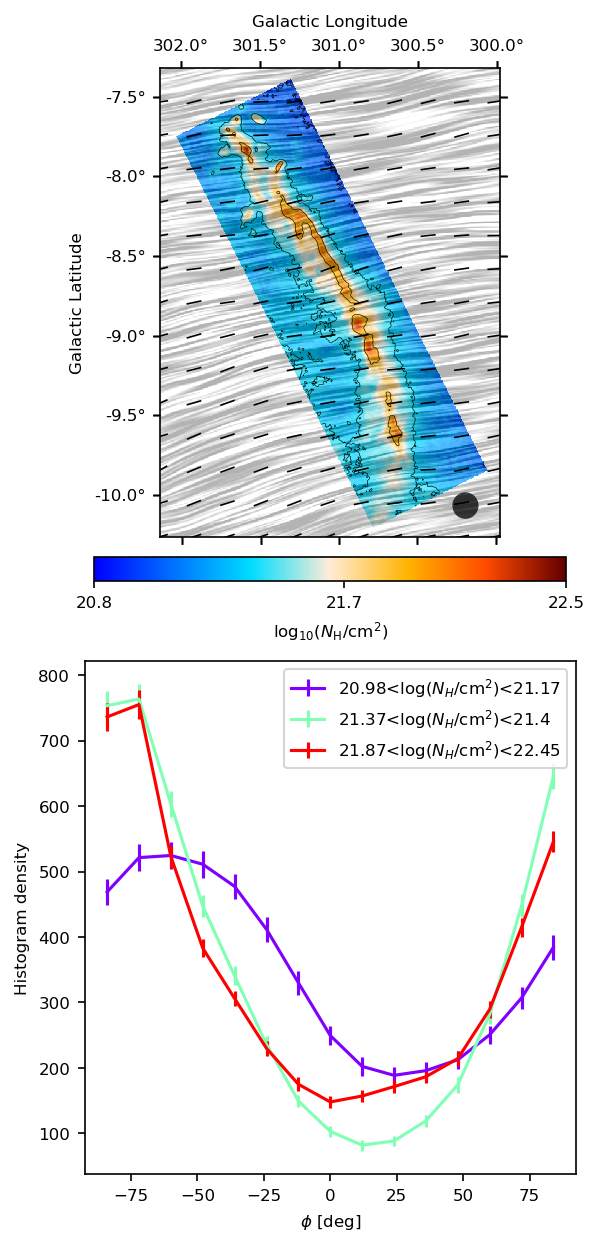}
}
\vspace{-2.0mm}
\caption{Same as Fig.~\ref{fig:TaurusHROs} for the Chamaeleon\,I (left), Chamaeleon\,II (center), and Musca (right) subregions.}
\label{fig:Cham-MuscaHROs}
\end{figure*}

\begin{figure*}[ht!]
\centerline{
\includegraphics[width=0.3\textwidth,angle=0,origin=c]{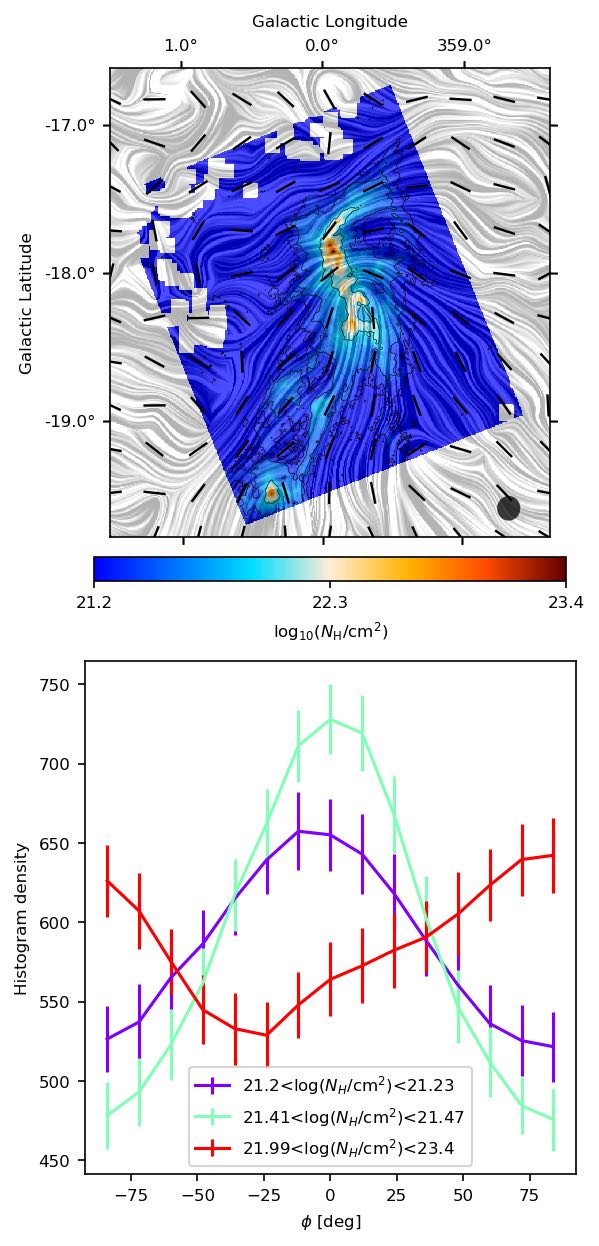}
\includegraphics[width=0.3\textwidth,angle=0,origin=c]{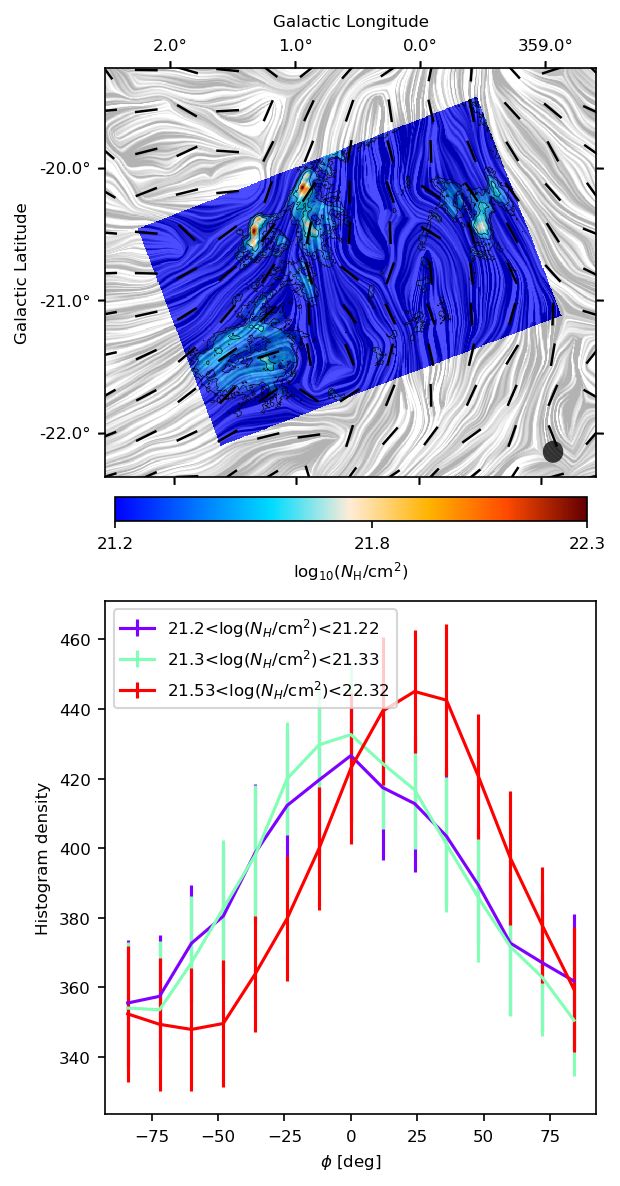}
}
\caption{Same as Fig.~\ref{fig:TaurusHROs} for the subregions CrA North (left) and CrA South (right).}
\label{fig:CrAHROs}
\end{figure*}

\begin{figure*}[ht!]
\centerline{
\includegraphics[width=0.3\textwidth,angle=0,origin=c]{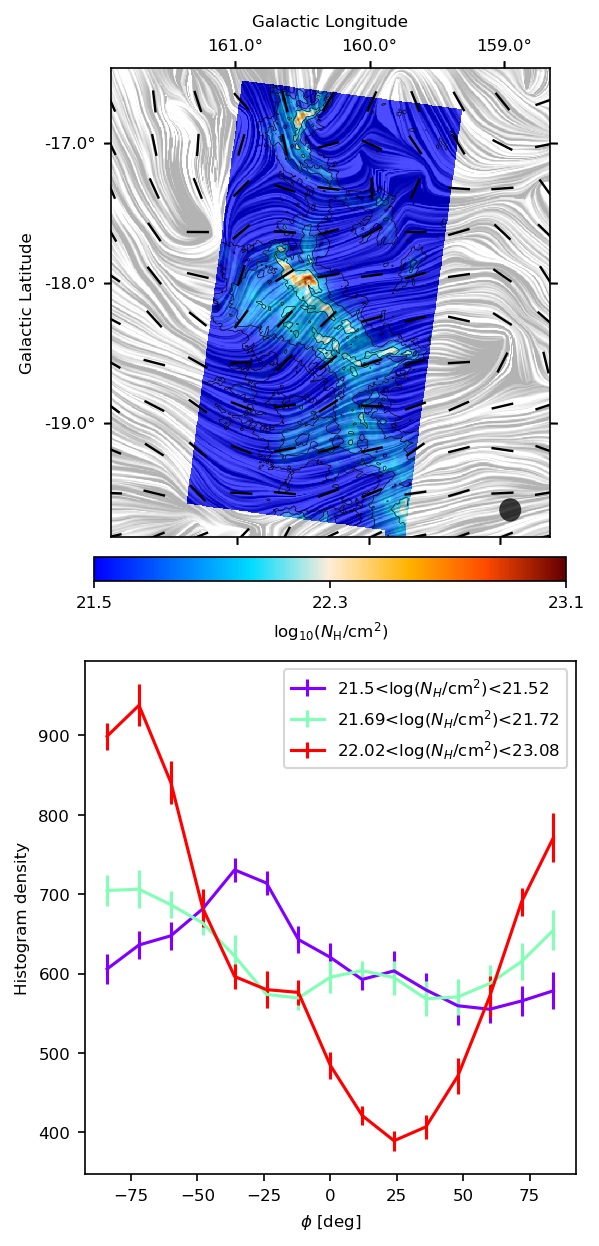}
\includegraphics[width=0.3\textwidth,angle=0,origin=c]{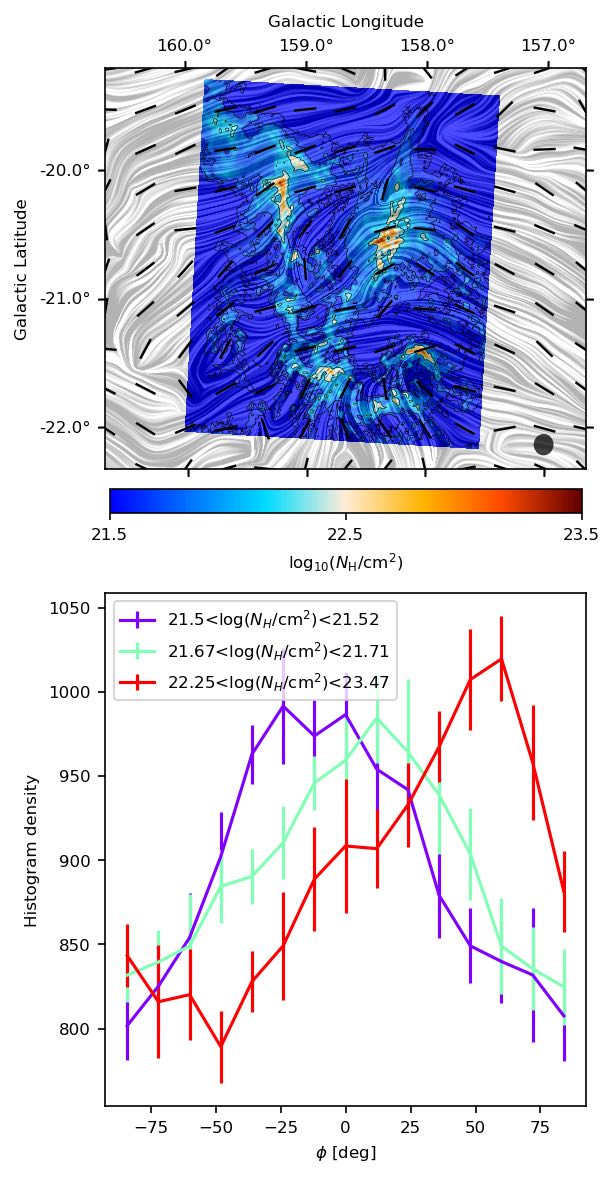}
\includegraphics[width=0.3\textwidth,angle=0,origin=c]{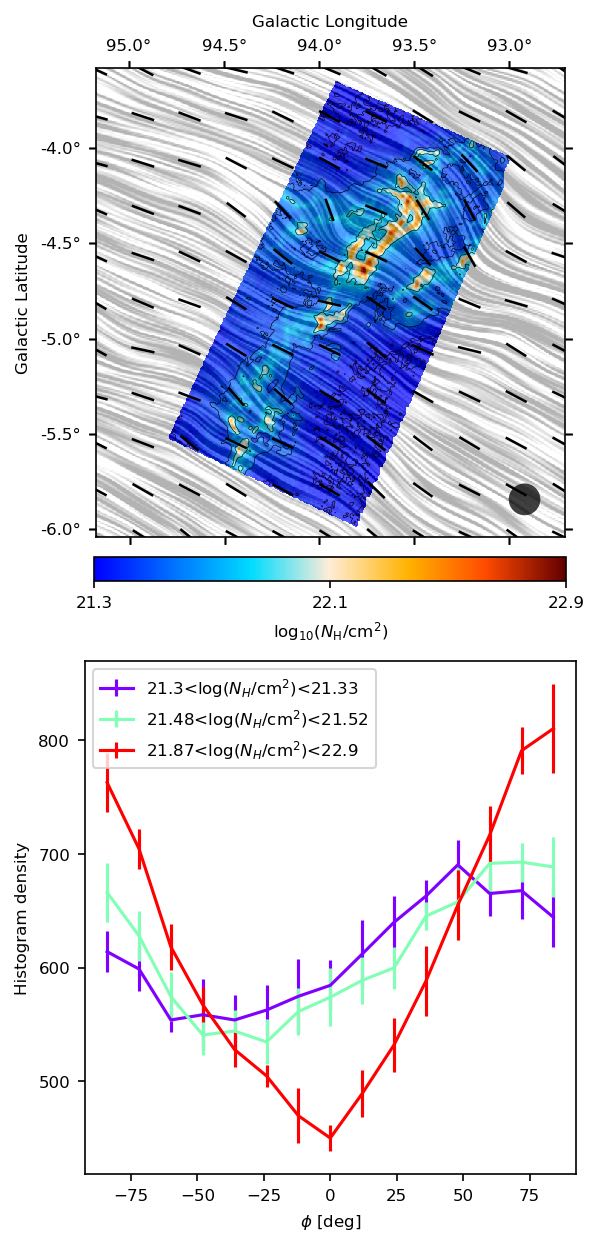}
}
\caption{Same as Fig.~\ref{fig:TaurusHROs} for the Perseus North (left) and Perseus South (center), and IC5146 (right) regions.}
\label{fig:Perseus-IC5146HROs}
\end{figure*}

\begin{figure*}[ht!]
\centerline{
\includegraphics[width=2.0in,angle=0,origin=c]{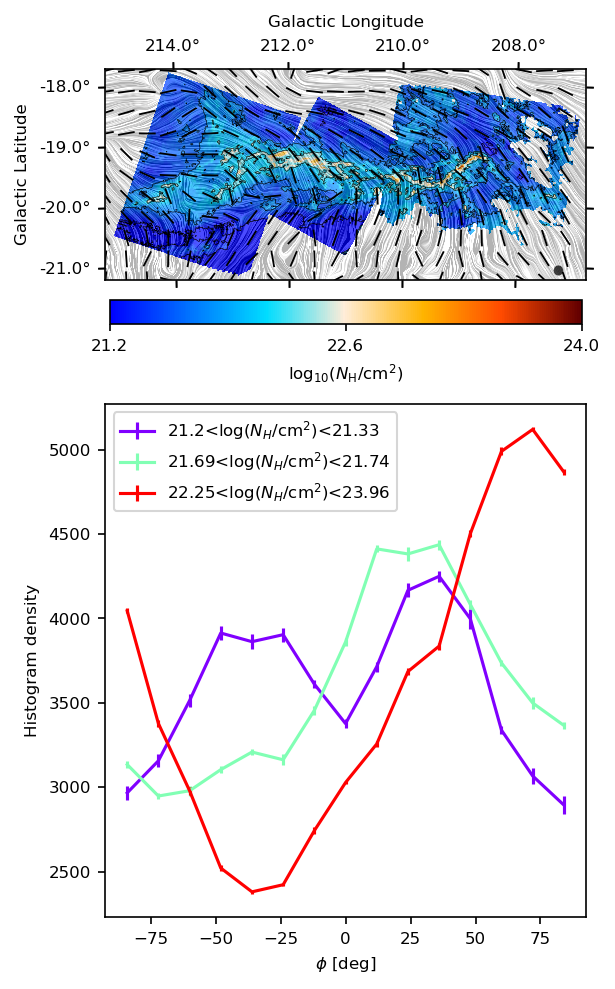}
\includegraphics[width=2.0in,angle=0,origin=c]{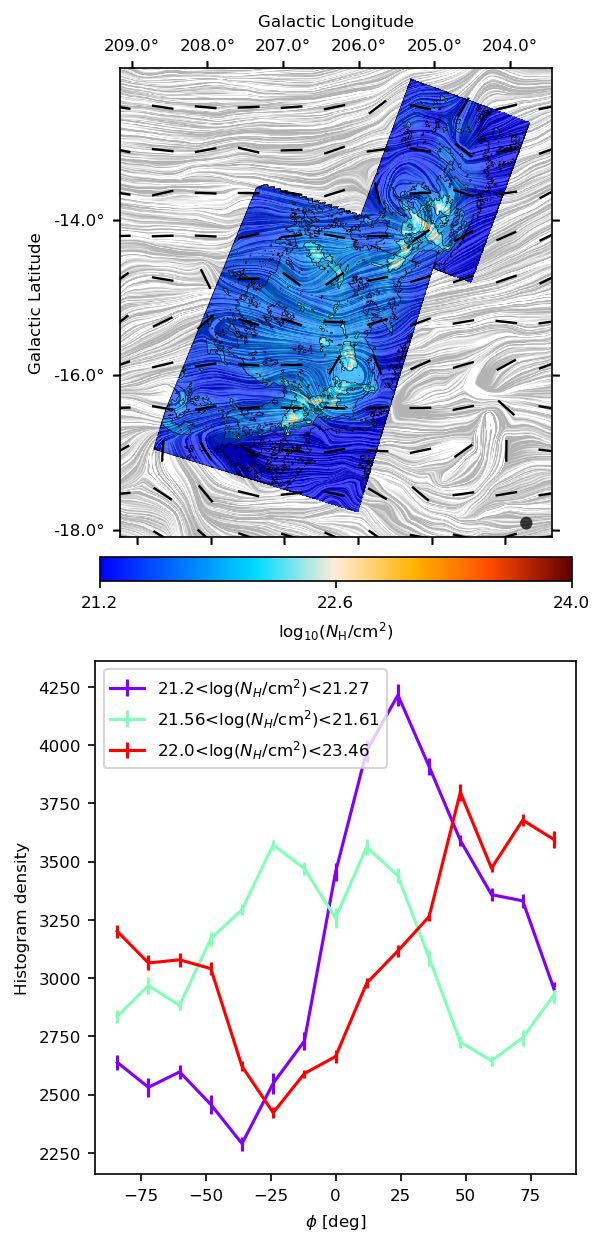}
\includegraphics[width=2.0in,angle=0,origin=c]{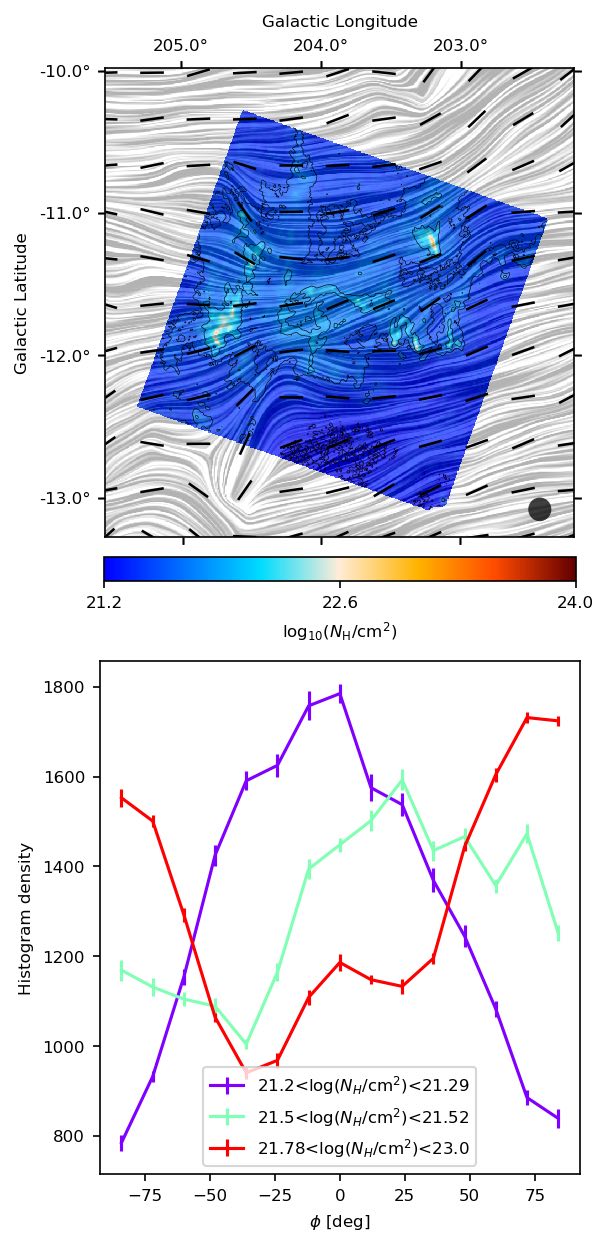}
}
\caption{Same as Fig.~\ref{fig:TaurusHROs} for the Orion~A (left), Orion~B (center), and L1622 (right) regions.}
\label{fig:OrionHROs}
\end{figure*}

\begin{figure*}[ht!]
\centerline{
\includegraphics[width=0.3\textwidth,angle=0,origin=c]{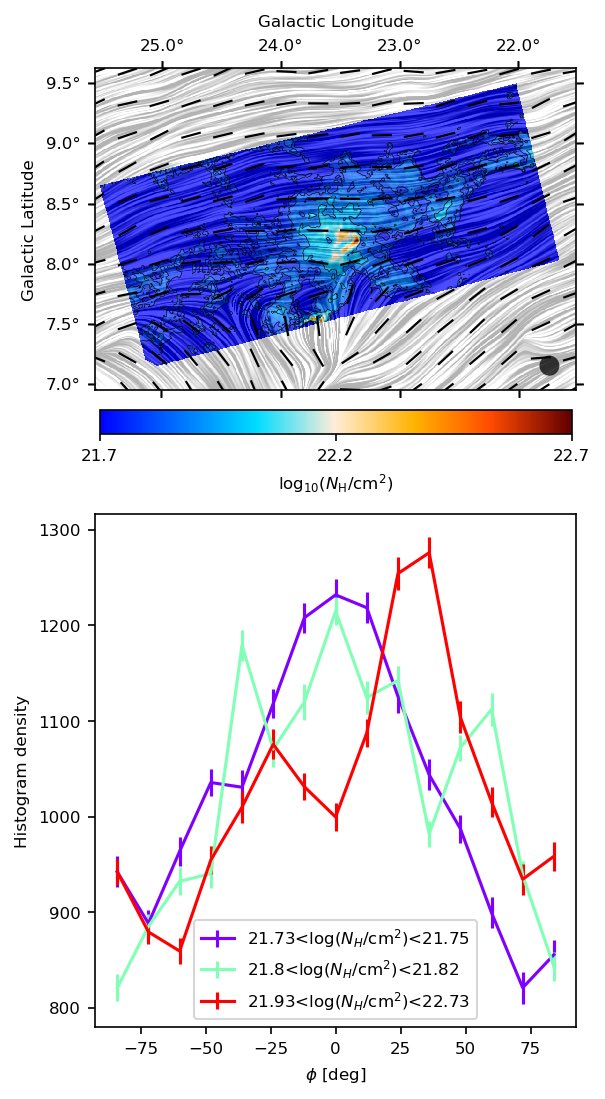}
\includegraphics[width=0.3\textwidth,angle=0,origin=c]{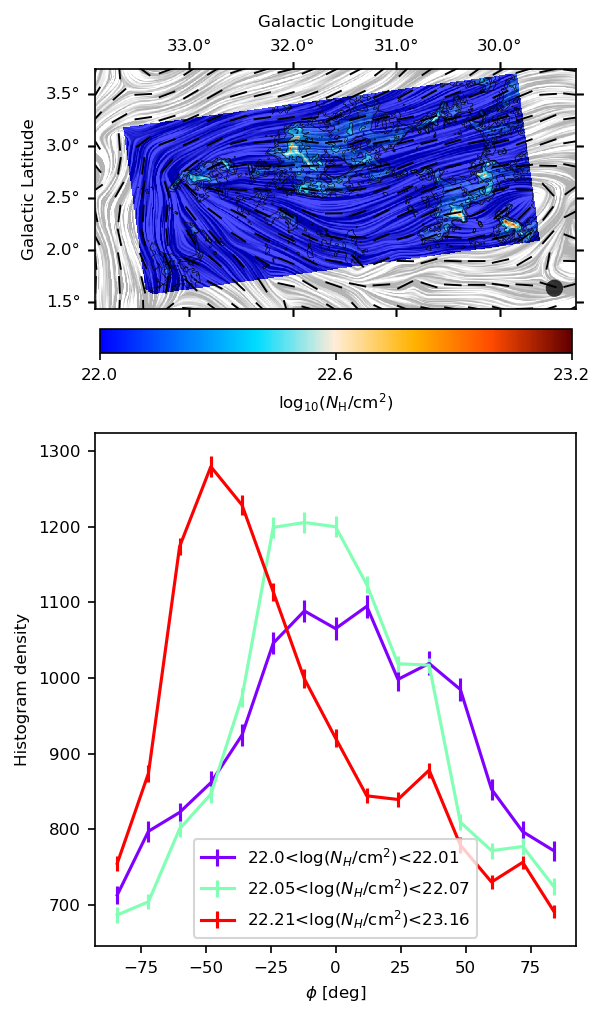}
}
\centerline{
\includegraphics[width=0.3\textwidth,angle=0,origin=c]{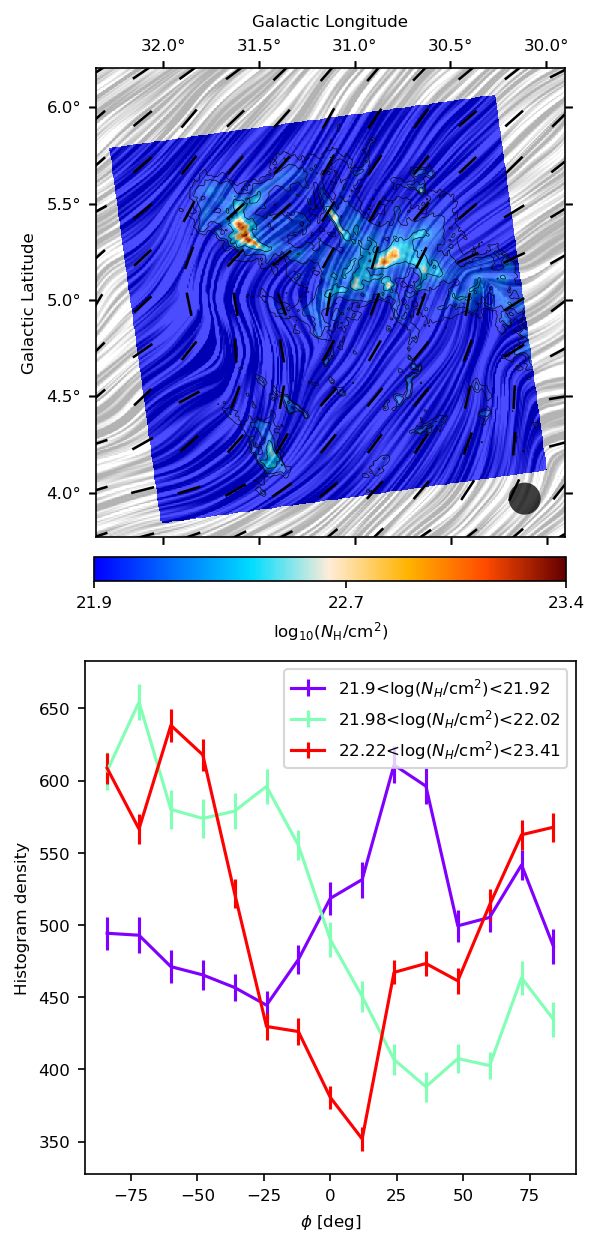}
\includegraphics[width=0.3\textwidth,angle=0,origin=c]{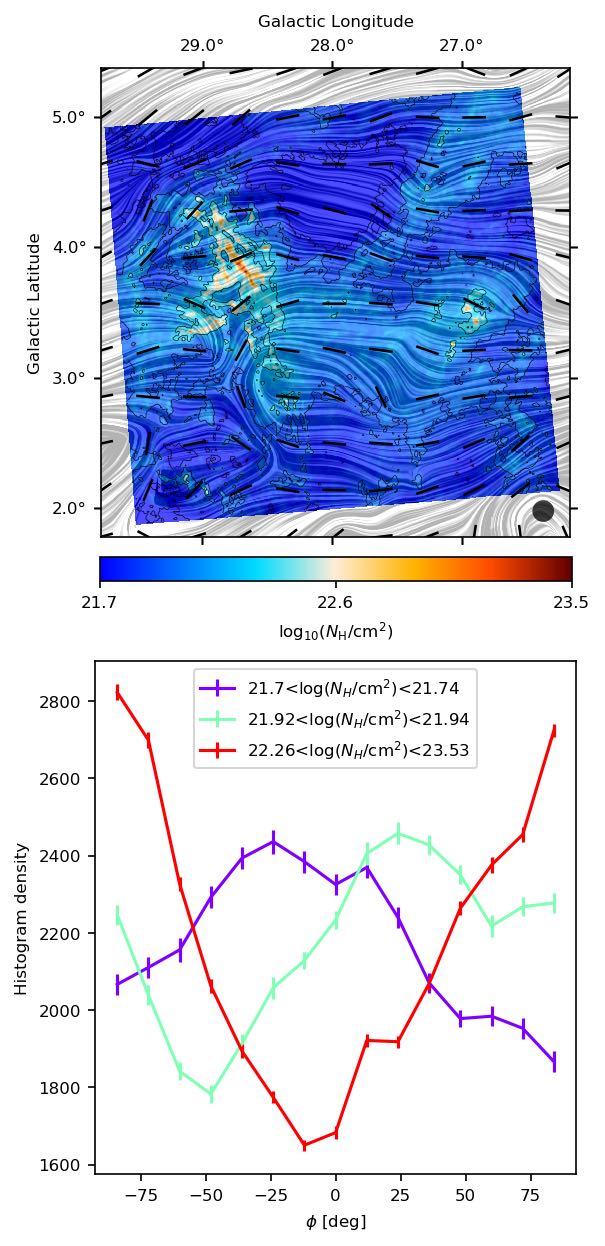}
}
\caption{Same as Fig.~\ref{fig:TaurusHROs} for the Aquila Rift subregions Serpens West, Serpens Main 1, Serpens Main 2, and Serpens (clockwise starting from the top left).}
\label{fig:AquilaRiftHROs}
\end{figure*}

\begin{figure*}[ht!]
\centerline{
\includegraphics[width=0.3\textwidth,angle=0,origin=c]{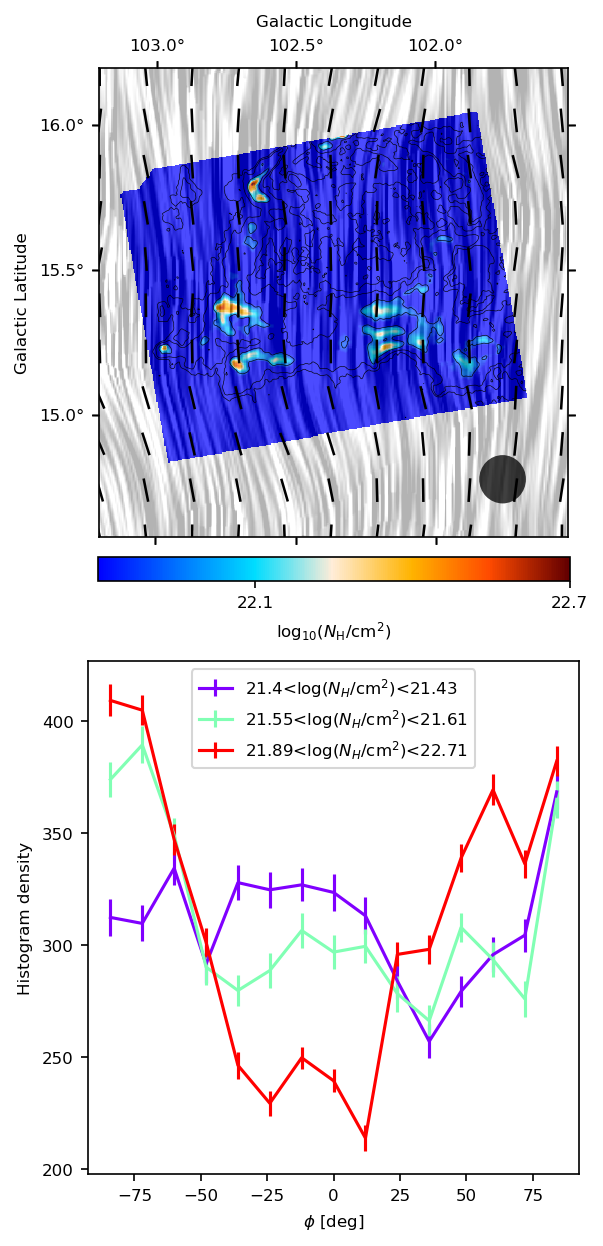}
\includegraphics[width=0.3\textwidth,angle=0,origin=c]{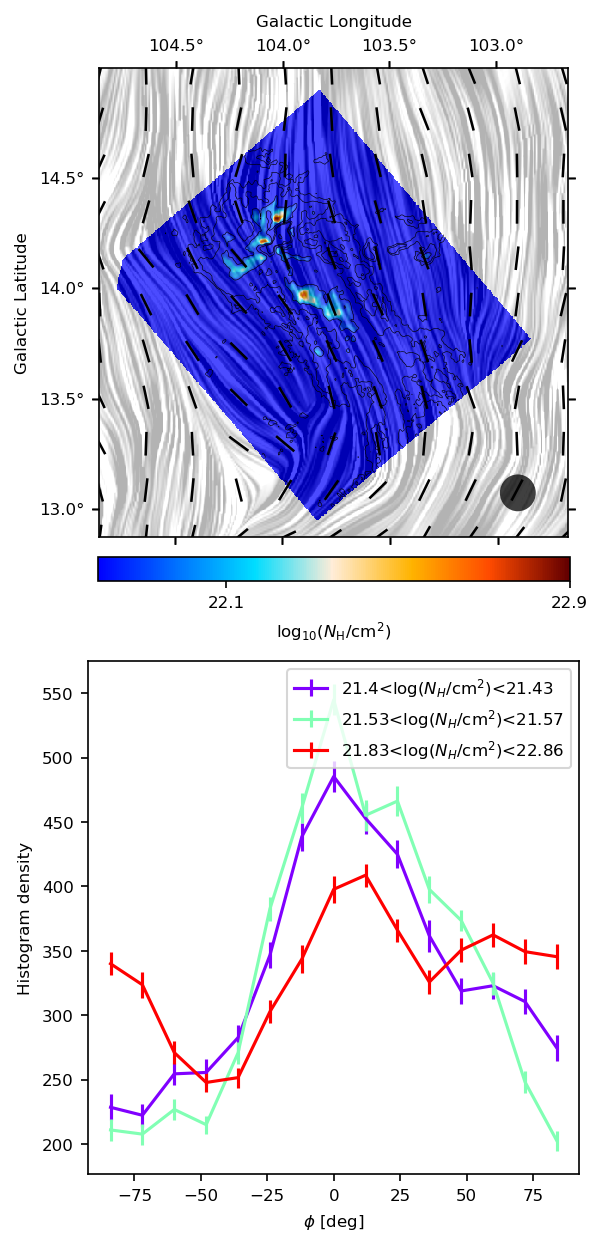}
}
\centerline{
\includegraphics[width=0.3\textwidth,angle=0,origin=c]{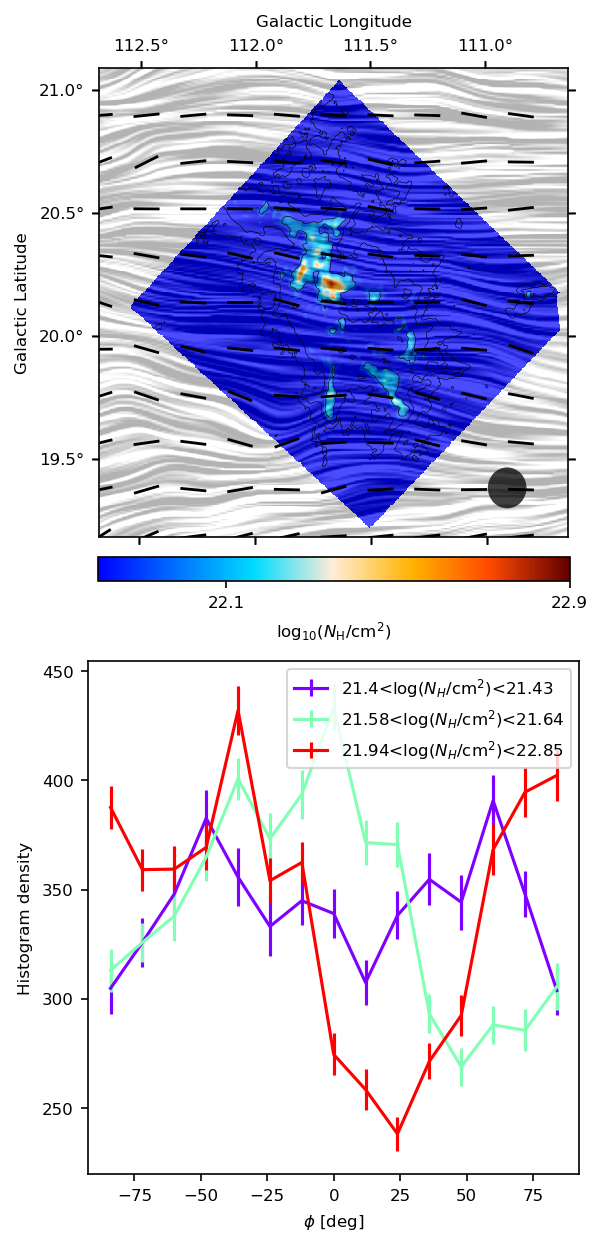}
\includegraphics[width=0.3\textwidth,angle=0,origin=c]{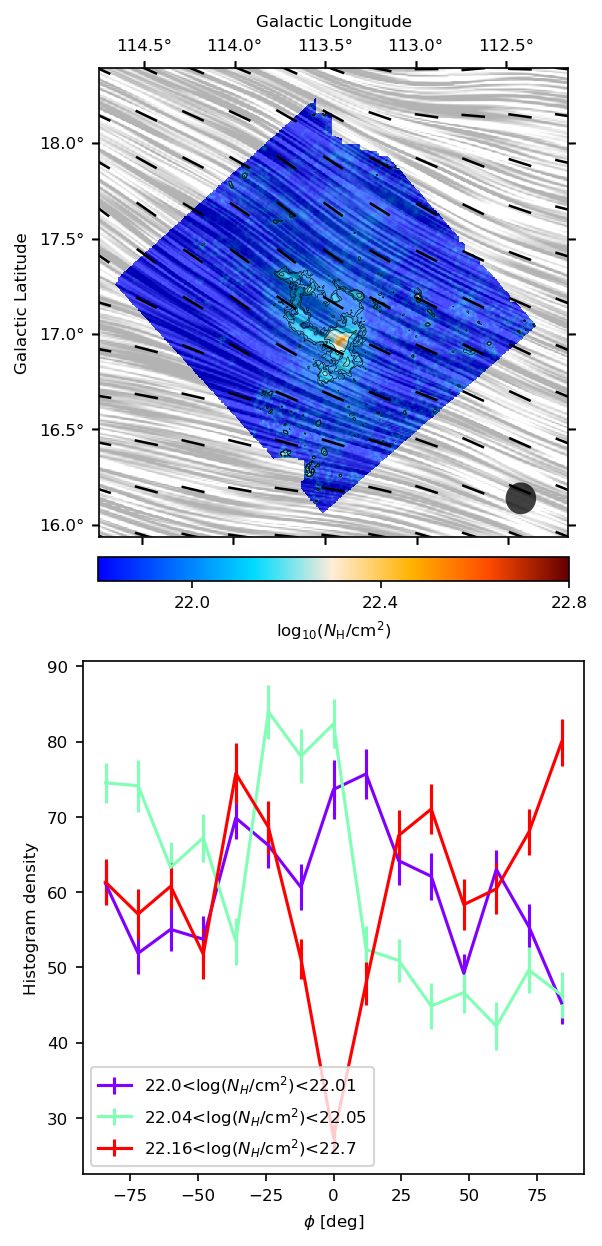}
}
\caption{Same as Fig.~\ref{fig:TaurusHROs} for the Cepheus subregions L1175, L1172, L1228, and L1241 (clockwise starting from the top left).}
\label{fig:CepheusHROs}
\end{figure*}

For the sake of sake of completeness, we also present maps of the gas column density (\nh) and the magnetic field projected onto the plane of the sky and integrated along the LOS (\bperp) represented by pseudo-vectors and by the drapery pattern that is produced using the line integral convolution method \citep[LIC,][]{cabral1993}.
Figure~\ref{fig:TaurusHROs}, Fig.~\ref{fig:OphiuchusHRO}, and Fig.~\ref{fig:LupusMaps} correspond to the Taurus, Ophiuchus, and Lupus regions, respectively.
Figure~\ref{fig:Cham-MuscaHROs} and Fig.~\ref{fig:CrAHROs} correspond to the Chamaeleon-Musca and the CrA regions.
Figure~\ref{fig:Perseus-IC5146HROs} corresponds to both the Perseus and IC5146 regions.
Finally, Fig.~\ref{fig:OrionHROs}, Fig.~\ref{fig:AquilaRiftHROs}, and Fig.~\ref{fig:CepheusHROs} correspond to the Orion, the Aquila Rift, and the Cepheus regions, respectively.

\section{Striation test}\label{app:striationsTest}
 
\begin{figure*}[ht!]
\centerline{
\includegraphics[width=0.35\textwidth,angle=0,origin=c]{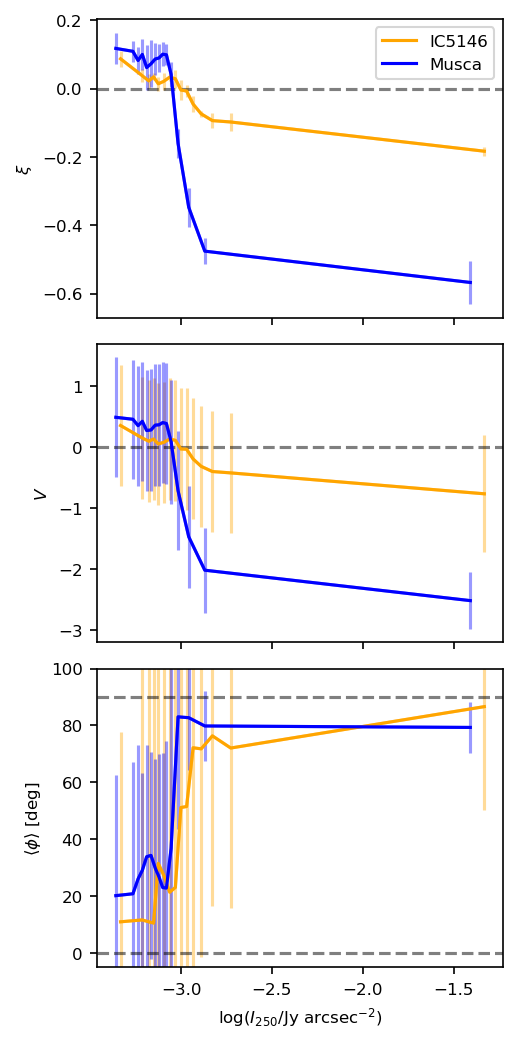}
}
\caption{Same as Fig.~\ref{fig:HOGpanel1} for the comparison between the observations of \Herschel\ observations at 250\micron\ and \bperp\ inferred from \Planck\ 353\,GHz polarization toward the Musca and IC 5146 regions.}
\label{fig:striationsHROs0}
\end{figure*}
 
\begin{figure*}[ht!]
\centerline{
\includegraphics[width=0.3\textwidth,angle=0,origin=c]{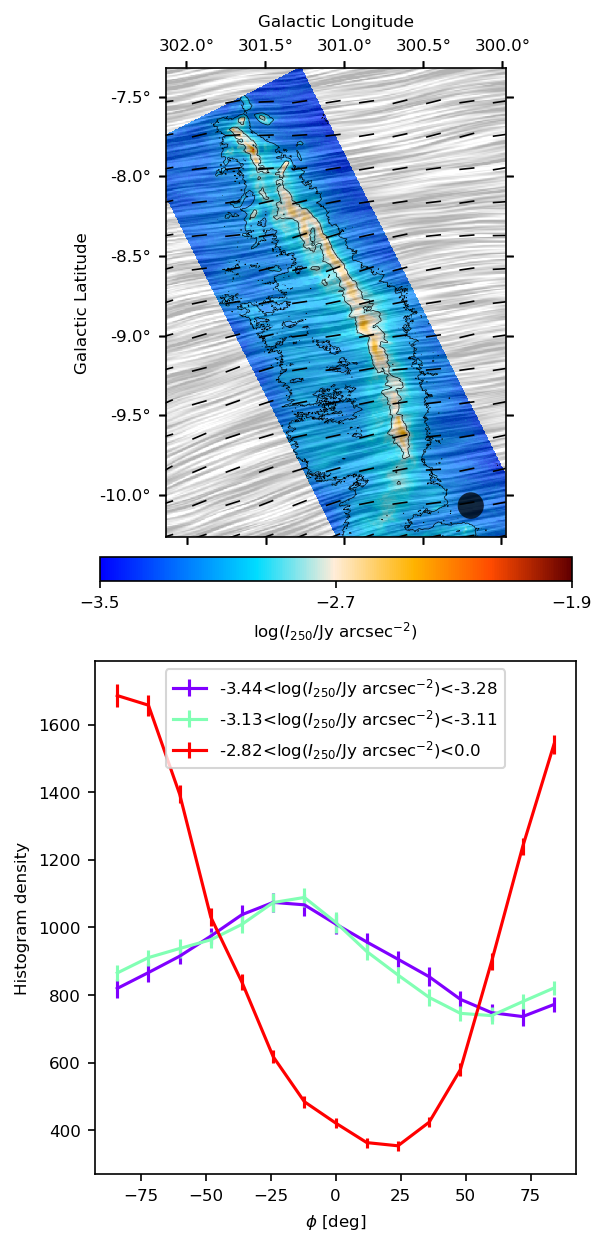}
\includegraphics[width=0.3\textwidth,angle=0,origin=c]{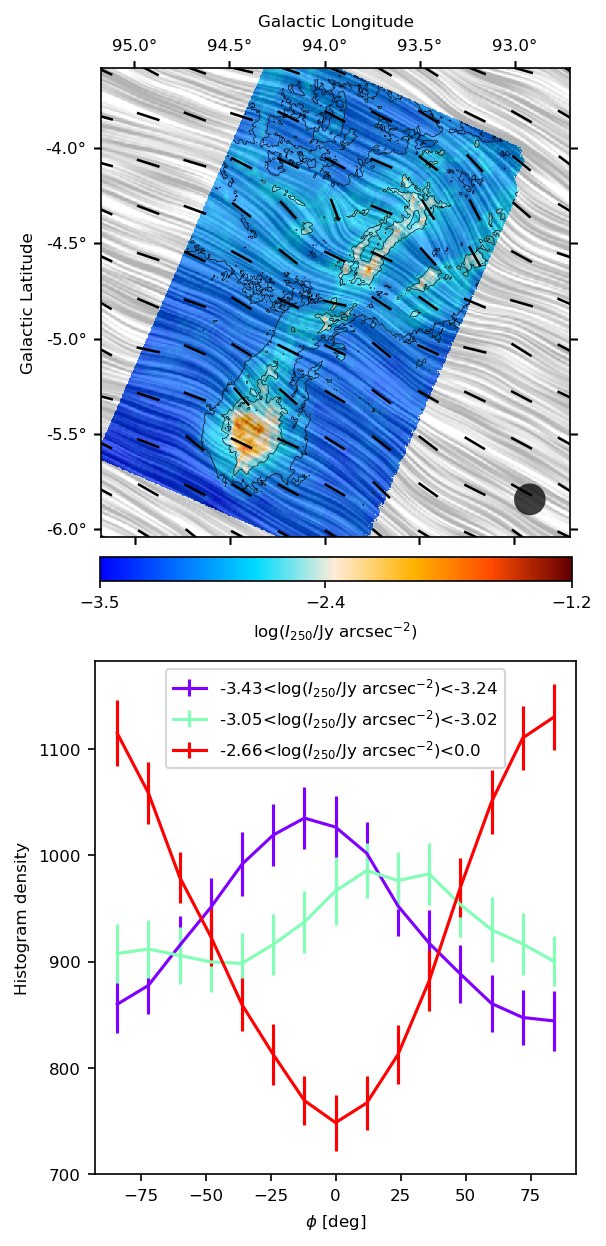}
}
\caption{\emph{Top.} Emission at 250\micron\ and magnetic field toward the Musca (left) and IC5146 (right) regions. 
The colors represent the emission observed in the \Herschel\ 250\micron\ band.
The drapery pattern represents the magnetic field orientation averaged along the LOS and projected onto the plane of the sky, \bperp, as inferred from the \Planck\ 353\,GHz data.
\emph{Bottom.} Histograms of relative orientations for three representative column density bins.}
\label{fig:striationsHROs1}
\end{figure*}

Because the derivation of column density (\nh) from the combination of \Herschel\ observations requires the degradation of the maps to the same common resolution, some of the striations or wispy elongated structures that appear abundantly around the dense filaments in the 250\micron\ maps are not present in the \nh\ maps.
In order to study the relative orientation between the magnetic field (\bperp) and the striations and establish a direct comparison with previous studies of these kind of structures \citep[namely,][]{cox2016}, in this appendix we quantify the relative orientation between the \Herschel\ 250\micron\ ($I_{250}$) observations and \bperp\ toward the Musca and IC5146 regions using the HROs method.

The relative orientation parameter ($\xi$) and the projected Rayleigh statistic ($V$), shown in Fig.~\ref{fig:striationsHROs0}, indicate that the structures sampled by the 250\micron\ observations show a clear transition from $V$\,$>$\,0 and $V$\,$<$\,0 with increasing $I_{250}$ toward both regions.
Moreover, \meanphi\ extends from less than 30\deg\ to roughly 90\deg\ with increasing $I_{250}$.
The significance of the results toward Musca is greater than those toward IC5146, as expected from the greater number of independent $\phi$ estimates toward that region.
However, the HROs presented in Fig.~\ref{fig:striationsHROs1} reveal that the transition from \nh\ and \bperp\ being mostly parallel to mostly perpendicular is clearer in IC5146.

\section{Column density slopes and relative orientation}\label{app:nhpdf}

\begin{figure*}[ht!]
\centerline{
\includegraphics[width=0.5\textwidth,angle=0,origin=c]{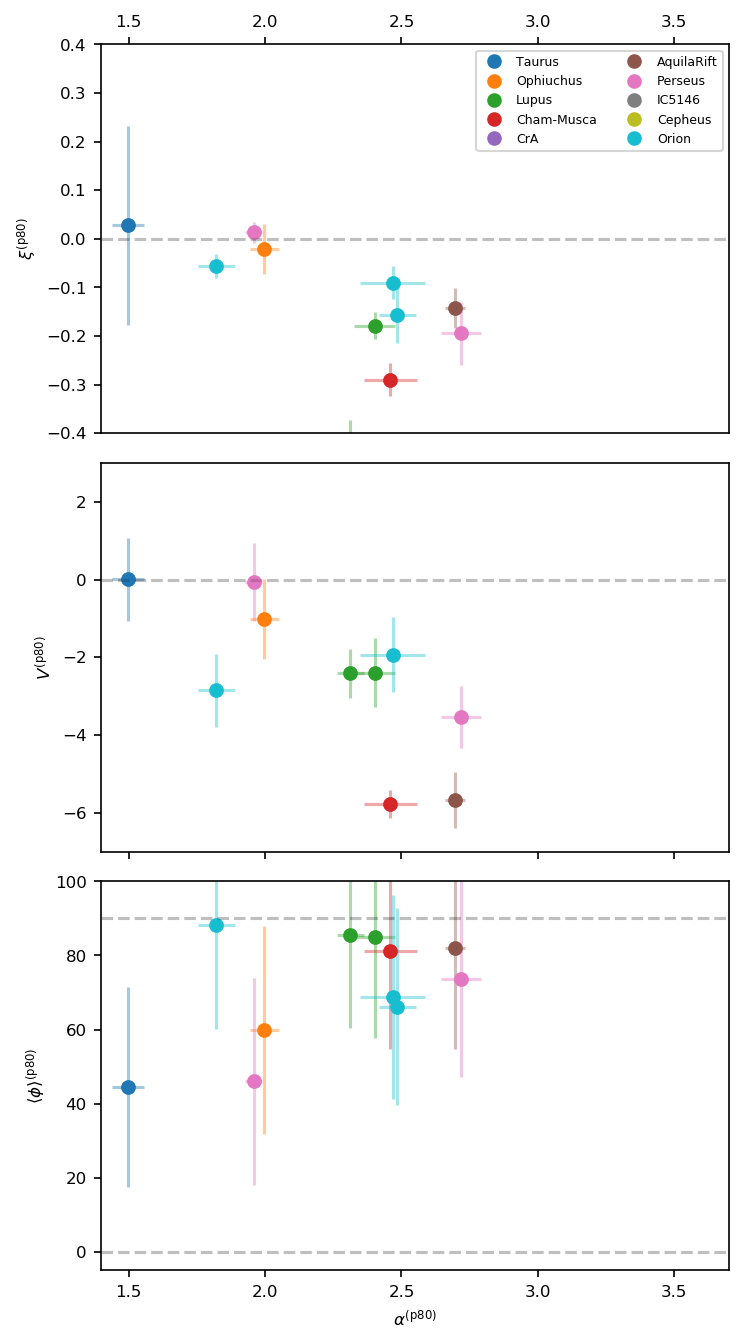}
\includegraphics[width=0.5\textwidth,angle=0,origin=c]{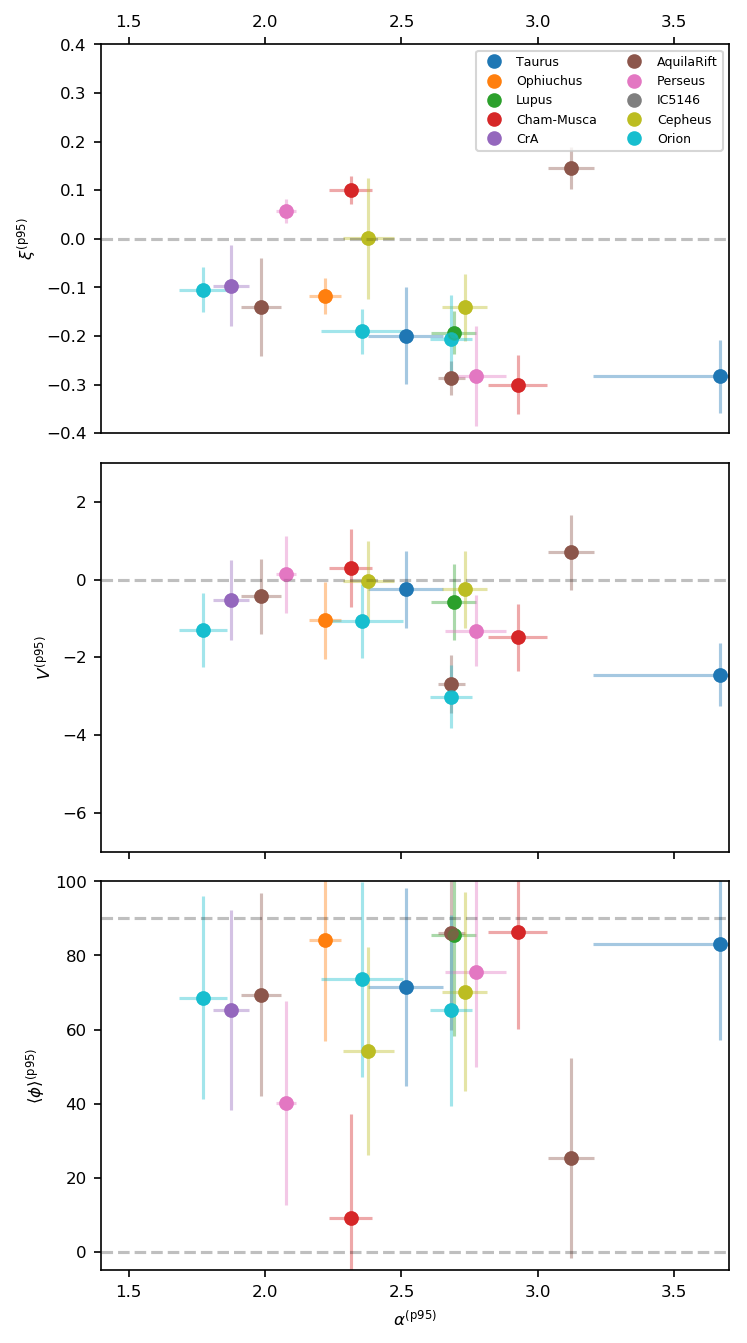}
}
\caption{Relative orientation parameter ($\xi$, top), projected Rayleigh statistic ($V$, middle), and mean relative orientation angle ($\left<\phi\right>$, bottom) as a function of the slope of the \nh\ distribution ($\alpha$, Eq.~\ref{eq:nhpdf}) within the 80th (left) and 95th (right) percentile of \nh.
All regions are selected using the criterion $|\left<\psi\right>^{({\rm in})}-\left<\psi\right>^{({\rm out})}|$\,$>$\,20\deg, where $\left<\psi\right>^{({\rm in})}$ and $\left<\psi\right>^{({\rm out})}$ are the mean \bperp\ orientation angles inside and outside the largest closed \nh\ contour, respectively.
}
\label{fig:GeneralTrendsNHPDFextra}
\end{figure*}

For the sake of completeness, we present the relation between the relative orientation parameter ($\xi$, top), projected Rayleigh statistic ($V$, middle), and mean relative orientation angle ($\left<\phi\right>$, bottom) as a function of the slope of the \nh\ distribution ($\alpha$) using a different selection than that discussed in Sec.~\ref{sec:DiscussionNHPDFs}.
Fig.~\ref{fig:GeneralTrendsNHPDFextra} shows the result of this relation for the regions within the 80th (left) and 95th (right) percentile of \nh.

\raggedright

\end{document}